\documentclass[10pt]{article}
\usepackage{graphicx}
\usepackage[all]{xy}
\usepackage{amsmath,amssymb}
\usepackage{geometry}
\def\colora{}
\def\colorb{}
\def\colorc{}
\def\colord{}
\def\colore{}

\def\p{{\boldsymbol p}}

\def\k{{\boldsymbol k}}

\def\x{{\boldsymbol x}}

\def\y{{\boldsymbol y}}

\def\B{{\boldsymbol B}}
\def\E{{\boldsymbol E}}

\def\z{{\boldsymbol z}}
\def\v{{\boldsymbol v}}

\def\bs{\boldsymbol}

\newcommand{\slB}{\raise.15ex\hbox{$/$}\kern-.53em\hbox{$B$}}
\newcommand{\slpartial}{\raise.15ex\hbox{$/$}\kern-.53em\hbox{$\partial$}}
\newcommand{\slA}{\raise.15ex\hbox{$/$}\kern-.7em\hbox{$A$}}
\newcommand{\slC}{\raise.15ex\hbox{$/$}\kern-.7em\hbox{$C$}}
\newcommand{\slL}{\raise.15ex\hbox{$/$}\kern-.53em\hbox{$L$}}
\newcommand{\slD}{\raise.15ex\hbox{$/$}\kern-.7em\hbox{$D$}}
\newcommand{\slcalD}{\raise.15ex\hbox{$/$}\kern-.7em\hbox{$\cal D$}}
\newcommand{\slP}{\raise.15ex\hbox{$/$}\kern-.53em\hbox{$P$}}
\newcommand{\slPP}{\raise.15ex\hbox{$/$}\kern-.53em\hbox{${\bs P}$}}
\newcommand{\slp}{\raise.15ex\hbox{$/$}\kern-.53em\hbox{$p$}}
\newcommand{\sll}{\raise.15ex\hbox{$/$}\kern-.53em\hbox{$l$}}
\newcommand{\slq}{\raise.15ex\hbox{$/$}\kern-.53em\hbox{$q$}}
\newcommand{\sla}{\raise.15ex\hbox{$/$}\kern-.53em\hbox{$a$}}
\newcommand{\slell}{\raise.15ex\hbox{$/$}\kern-.53em\hbox{$\ell$}}
\newcommand{\slv}{\raise.15ex\hbox{$/$}\kern-.53em\hbox{$v$}}
\newcommand{\slb}{\raise.15ex\hbox{$/$}\kern-.53em\hbox{$b$}}
\newcommand{\slk}{\raise.15ex\hbox{$/$}\kern-.53em\hbox{$k$}}
\newcommand{\slR}{\raise.15ex\hbox{$/$}\kern-.53em\hbox{$R$}}
\newcommand{\slQ}{\raise.15ex\hbox{$/$}\kern-.53em\hbox{$Q$}}
\newcommand{\slQQ}{\raise.15ex\hbox{$/$}\kern-.53em\hbox{${\bs Q}$}}
\newcommand{\slK}{\raise.15ex\hbox{$/$}\kern-.43em\hbox{$K$}}
\newcommand{\slKK}{\raise.15ex\hbox{$/$}\kern-.43em\hbox{${\bs K}$}}
\newcommand{\slSigma}{\raise.15ex\hbox{$/$}\kern-.53em\hbox{$\Sigma$}}
\newcommand{\slcalP}{\raise.15ex\hbox{$/$}\kern-.63em\hbox{$\cal P$}}

\def\build#1\over#2{\mathrel{\mathop{\kern 0pt#2}\limits_{#1}}}
\def\empile#1\over#2{\mathrel{\mathop{\kern 0pt#1}\limits_{#2}}}

\title{{\bf Some Aspects of the Theory\\ of Heavy Ion Collisions}}
\author{\sc Fran\c cois Gelis\\{\ }\\
  Institut de Physique Th\'eorique\\
  CEA/Saclay, Universit\'e Paris-Saclay\\
  91191, Gif sur Yvette, France
}

\begin{document}
\maketitle

\begin{abstract}
  We review the theoretical aspects relevant in the description of
  high energy heavy ion collisions, with an emphasis on the learnings
  about the underlying QCD phenomena that have emerged from these
  collisions.
\end{abstract}

\section{Introduction to heavy ion collisions}

\paragraph{Elementary forces in Nature} The interactions among the
elementary constituents of matter are divided into four fundamental
forces: gravitation, electromagnetism, weak nuclear forces and strong
nuclear forces. All these interactions except gravity have a well
tested a microscopic quantum description in terms of local gauge
theories, in which the elementary matter fields are spin-$1/2$
fermions, interacting via the exchange of spin-$1$ bosons. In this
framework, a special role is played by the Higgs spin-$0$ boson (the
only fundamental scalar particle in the Standard Model), whose
non-zero vacuum expectation value gives to all the other fields a mass
proportional to their coupling to the Higgs. The discovery of the
Higgs boson at the Large Hadron Collider in 2012 has so far confirmed
all the Standard Model expectations. In this picture, gravity has
remained a bit of an outlier: even though the classical field theory
of gravitation (general relativity) has been verified experimentally
with a high degree of precision (the latest of these verifications
being the observation of gravitational waves emitted during the merger
of massive compact objects - black holes or neutron stars), the quest
for a theory of quantum gravity has been inconclusive until now (and
possible experimental probes are far out of reach for the foreseeable
future).

\paragraph{Strong nuclear force} Quantum chromodynamics (QCD), the
microscopic theory that governs strong nuclear interactions, was
formulated in the early 1970's under the guidance of several
experimental clues. In particular, deep inelastic scattering of
electrons off proton targets indicated that the proton charge is
concentrated into smaller constituents (unresolved in the scattering)
of spin $1/2$ (this follows from the measured structure functions),
that interact weakly at high momentum transfer. These observations
paved the way towards a non-Abelian gauge theory with the property of
{\sl asymptotic freedom} \cite{Gross:1973id,Politzer:1973fx}, i.e. a
theory in which the coupling strength decreases at short
distance. Combined with some insights from hadron spectroscopy, this
led to an $SU(3)$ gauge theory, with spin-$1/2$ matter fields (the
quarks) in the fundamental representation. The fundamental property of
QCD that resolved the tension between the fact that quarks must
interact strongly enough to form bound hadronic states and the fact
that they appear to be weakly interacting in deep inelastic scattering
experiments is asymptotic freedom: namely, the property that the
running of the coupling due to quantum corrections is such that the
strong coupling constant becomes small at short distance and large on
distance scales compared to the size of a hadron.

We now know that there are six families of quarks: up, down, strange,
charm, bottom and top, ranging from nearly massless to about $175$~GeV
for the top quark \cite{Tanabashi:2018oca}. The nucleons that compose
the atomic nuclei of ordinary matter are built solely from the up and
down quarks, and the heavier quarks appear only in more massive
hadrons (at the exception of the top quark, whose lifetime is so short
that it decays before bound states can be formed). QCD has received
ample experimental support as the correct microscopic theory for
describing strong nuclear interactions. However, because of asymptotic
freedom, the most quantitative comparisons between theory and
experiments are based on hard processes (i.e., processes involving at
least one hard momentum particle in the final state). Although this is
sufficient to ascertain the fact that strong nuclear interactions are
indeed well described by QCD, these experiments leave unexplored
another important aspect of strong interactions, that has to do with
the rich properties of nuclear matter in extreme conditions of
temperature or density.

\paragraph{Asymptotic freedom, confinement and deconfinement} A
crucial property of QCD, consequence of asymptotic freedom, is color
confinement \cite{Wilson:1974sk}, namely the fact that isolated quarks
or gluons cannot exist but instead combine into bound states --the
hadrons-- in which their color charge is ``hidden''. Thus, trying to
pull a quark out of a hadron (for instance in a high energy
collision with another hadron) merely creates more
hadrons. Conversely, when one packs many hadrons in a small volume,
the average distance between their constituents decreases, and
therefore they interact more and more weakly\footnote{Let us clear out
  a possible misconception related to this: a small coupling constant
  in pairwise parton interactions does not necessarily imply that the
  system as a whole is weakly interacting. The latter is true only if
  the mean free path is large compared to the De Broglie wavelength of
  the constituents, and the inverse mean free path is the product of a
  cross-section by a density. In other words, in such a dense system,
  there may be strong collective effects despite a weak
  coupling.}. Given this, we may expect that the forces that bind
quarks inside hadrons eventually become weak enough to allow the
quarks to become unconfined, i.e., free to wander in the entire volume
of the system. This state of nuclear matter is called the {\sl
  quark-gluon plasma} (QGP). Note that this transition is
non-perturbative, since it happens at an energy scale where the
coupling constant is still too large to apply reliably perturbation
theory.

\begin{figure}[htbp]
  \centering
  \includegraphics[width=8cm]{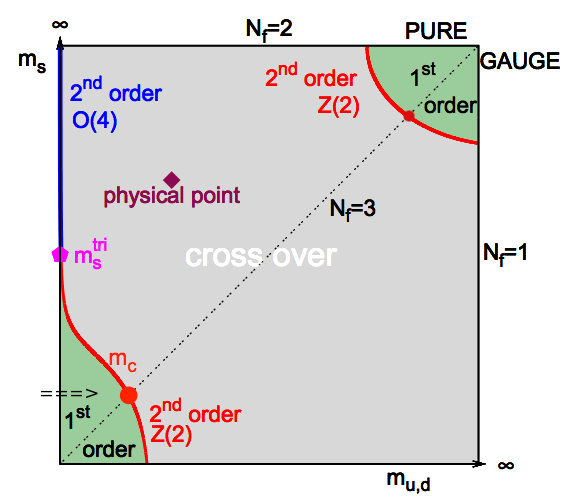}
  \caption{\label{fig:columbia}Nature of the transition at null
    chemical potential as a function of quark masses. From
    \cite{deForcrand:2017cgb}.}
\end{figure}

However, it is possible to formulate QCD non-perturbatively by
discretizing Euclidean space-time on a lattice. This setup provides a
way of computing certain observables without resorting to an expansion
in powers of the coupling. Some of the quantities that one may
calculate in lattice QCD are related to the confinement/deconfinement
transition: e.g., the expectation value of the trace of a Wilson loop
(that one may relate to the potential between a pair of infinitely
heavy quark and antiquark), or the entropy density (that measures the
number of active degrees of freedom in the system). Moreover, in
lattice QCD, one may vary several parameters of the theory, like the
number of quark families and their masses, in order to investigate the
role they play in the observed phenomena. Some of these results are
summarized in the plot of Figure \ref{fig:columbia}, taken from
\cite{deForcrand:2017cgb}. For instance, pure-glue QCD (equivalent to
QCD with infinitely massive quarks) has a first order deconfinement
transition at a temperature of the order of $270$~MeV. A first order
transition may also exist in the opposite limit, with massless
quarks. In this limit, the classical QCD Lagrangian also has a chiral
symmetry, spontaneously broken at low temperature (this transition is
also a first order transition).
The physical spectrum of light quarks in Nature lies in between these
two extreme situations, and there is now a consensus that this
physical point corresponds to a mere crossover transition, i.e., a
perfectly smooth (but rather rapid) transition from hadrons to
deconfined quarks, that takes place at a temperature around
$155$~MeV. Even in the absence of a genuine transition in the
thermodynamical sense, this transition exhibits signs of deconfinement
accompanied by a restoration of chiral symmetry (up to explicit
residual violations due to the non-zero quark masses).

\paragraph{Lattice QCD and baryon chemical potential} In statistical
equilibrium, the parameters that control the possible states of
nuclear matter are the temperature, the chemical potentials associated
to conserved quantities (the most important of which is the baryon
chemical potential $\mu_{_B}$), and possibly some external fields
(such as magnetic fields). However, a non-zero chemical potential is a
very serious obstacle for lattice QCD. Indeed, the chemical potential
turns the exponential of the action integrated over the quark fields
into a non-positive measure, that cannot be sampled by Monte-Carlo. At
very small $\mu_{_B}/T$, various lattice techniques
\cite{Hasenfratz:1983ba,Alford:1998sd,Hands:1999md,Fodor:2001au,Gupta:2004pk}
can be used to track the transition to non-zero values of $\mu_{_B}$,
but these methods loose their accuracy when the chemical potential
becomes of the order of the temperature. Note that a perturbative
approach becomes possible at very large $\mu_{_B}$ and/or large $T$,
since these parameters control the relevant scale for the running
coupling. At high $T$ and low $\mu_{_B}$, these analytic calculations
are consistent with lattice computations. At high $\mu_{_B}$ and very
low temperature, they indicate the presence of several color
super-conducting phases
\cite{Alford:1997zt,Berges:1998rc,Son:1998uk,Alford:2007xm} (in these
phases, the ground state of the system exhibits a non-zero quark-quark
condensate, very similar to Cooper pairing in BCS super-conductivity).

Outside the regions accessible to lattice simulations or to
perturbation theory, our knowledge of the phase diagram of nuclear
matter is mostly speculative. It is for instance expected that the
system is strongly interacting near the transition line, implying
small transport coefficients, consistent with the measurements of final
state correlations among the produced particles (see later the section
on hydrodynamics).

\paragraph{Heavy Ion Collisions} In the history of the early Universe,
the confinement transition was crossed when the Universe was about one
microsecond old, but as far as we know this did not leave any visible
imprint accessible to present astronomical observations. In the early
1980's emerged the idea to collide heavy nuclei in order to produce in
the laboratory nuclear matter at high temperature and density,
possibly sufficient to reach and go beyond the critical
line. Subsequently, several experiments have had all or part of their
scientific program devoted to the study of heavy ion collisions:
\begin{itemize}
\item \textbf{Bevatron} (Billions of eV Synchrotron) :\\
From 1954 to 1993 at Lawrence Berkeley National Laboratory, U.S 

\item \textbf{AGS} (Alternating Gradient Synchrotron) :\\
Since 1960 at Brookhaven National Laboratory, U.S\\
Now used as injector for RHIC

\item\textbf{SPS} (Super Proton Synchrotron) :\\
Since 1976 at CERN\\
Now the injector for the LHC

\item\textbf{SIS-18} (Schwer-Ionen-Synchrotron) :\\
Since 2001 at GSI

\item\textbf{RHIC} (Relativistic Heavy Ion Collider) :\\
Since 2000 at Brookhaven National Laboratory, U.S

\item\textbf{LHC} (Large Hadron Collider) :\\
Since 2009 at CERN
\end{itemize}
The first experimental hints of a deconfinement transition were
observed at the CERN SPS \cite{Heinz:2000bk,Abreu:2000ni,Andersen:1999ym,Agakishiev:1997au,Bearden:1999ck,Bachler:1999hu,Ambrosini:1999im,Aggarwal:1999hk}, that collided heavy ions
at a center of mass energy of $17$~GeV, and in the subsequent
experimental programs at higher energies (the RHIC at Brookhaven
National Laboratory, and the Large Hadron Collider at CERN) the focus
has shifted from assessing the production of a quark-gluon plasma
towards measuring quantitatively some of its properties
\cite{Adams:2005dq,Adcox:2004mh,Arsene:2004fa,Back:2004je,Aamodt:2010pa,Aamodt:2010jd,ALICE:2011ab,Abelev:2012ola,Aad:2010bu,ATLAS:2012at,Aad:2012gla,Khachatryan:2010gv,Chatrchyan:2011sx,CMS:2012qk}.

\begin{figure}[htbp]
  \centering
  \includegraphics[width=8cm]{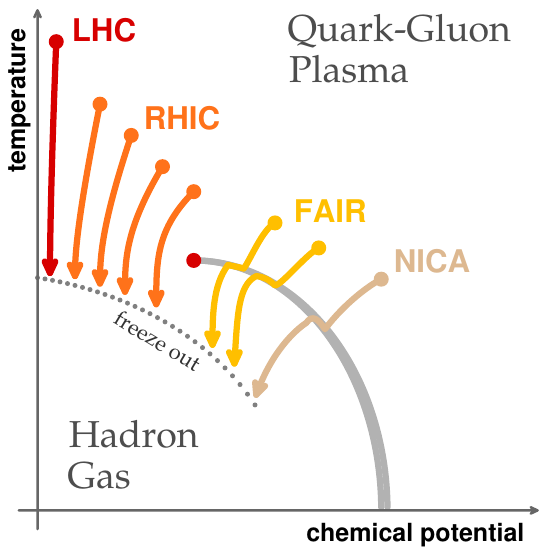}
  \caption{\label{fig:phase-diagram}Sketch of the phase-diagram of
    strongly interacting nuclear matter, and approximate reach of
    various experimental facilities.}
\end{figure}

\paragraph{Experimental handles} In heavy ion collisions, a few
experimental handles are available to vary the conditions in which the
quark gluon plasma may be formed. One of them is the atomic number of
the nuclei used in the collisions, whose main effect is to change the
volume of the interaction zone (but as we shall see in the next
section, this has also an incidence on the so-called saturation
momentum). When performing collisions with a given species of ions,
another variable that has a direct effect on the volume is the impact
parameter of each collision. Although the impact parameter is not
directly measurable, some observable quantities (such as the total
multiplicity in the final state, or the transverse energy) are
strongly correlated with the impact parameter. Finally, the collision
energy can in principle be varied (but of course, this is in practice
highly constrained by the accelerator design), which affects the
initial energy density (i.e., temperature) and the net baryon density
of the matter produced in a collision. This translates in different
reaches in the phase diagram for various heavy ion experiments, as
sketched in Figure \ref{fig:phase-diagram}.

\begin{figure}[htbp]
  \centering
  \includegraphics[width=13cm]{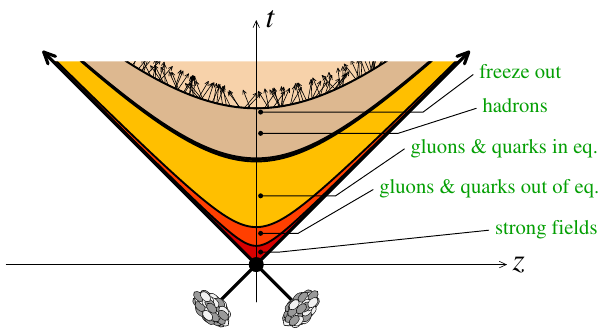}
  \caption{\label{fig:stages}Main stages of the evolution of the
    matter produced in a heavy ion collision. }
\end{figure}

\paragraph{Main stages of a heavy ion collision} From a theoretical
point of view, an ultrarelativistic collision between two nuclei can
be conveniently divided in several stages, sketched in the figure
\ref{fig:stages}.
Such a picture stems from the approximate boost invariance (i.e.,
independence on the spatial rapidity variable
$\eta_s\equiv \tfrac{1}{2}\ln((t+z)/(t-z))$) of such collisions, and the
successive stage are ordered by the proper time $\tau$. The collision
itself is very short due to the Lorentz contraction of the nuclei in
the laboratory frame. Just after the collision, the matter produced is
predominantly made of non-equilibrated gluons (they are not even
on-shell at the very beginning, and the system is better treated in
terms of fields rather than particles). This matter is strongly
interacting due to a large gluon occupation number, and evolves
towards equilibration (both kinetic and chemical, since
quark-antiquark pairs are produced in the process). In the subsequent
stages, the bulk evolution of the system is remarkably well described
by nearly ideal (i.e., with very small values of the viscous transport
coefficients) relativistic hydrodynamics. The expansion of the system
causes the system to cool down, and at some point the temperature
reaches the confinement temperature. In the framework of
hydrodynamics, as long as the system remains close to equilibrium, the
crossing of the confinement transition is rather transparent since it
is encoded in the equation of state. Soon after, the system becomes
dilute, the mean free path increases, and a description of its
expansion in terms of kinetic theory rather than hydrodynamics becomes
preferable. In such a description, the values of the various
cross-sections control when each type of reaction stops: the inelastic
processes stop first (chemical freezeout), soon followed by a kinetic
freezeout after which the momenta of the particles remain unchanged
(afterwards, all particles therefore fly on straight lines at constant
velocity until they hit a detector).

\section{Initial state, Color Glass Condensate}
\begin{figure}[htbp]
  \centering
  \includegraphics[width=5.5cm]{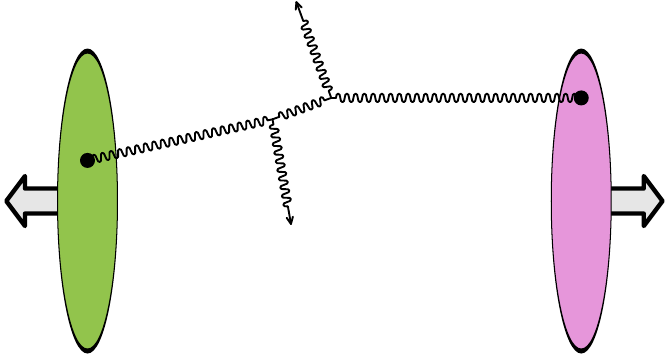}\hskip 10mm
  \includegraphics[width=5.5cm]{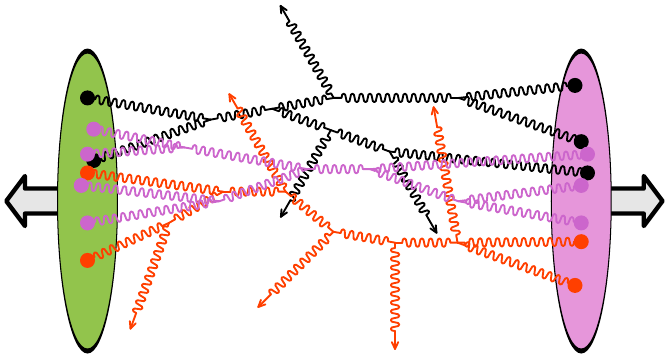}
  \caption{\label{fig:sat-nl}Left: scattering in the dilute
    regime. Right: multi-gluon scattering in the saturation regime.}
\end{figure}

\begin{figure}[htbp]
  \centering
  \includegraphics[width=8cm]{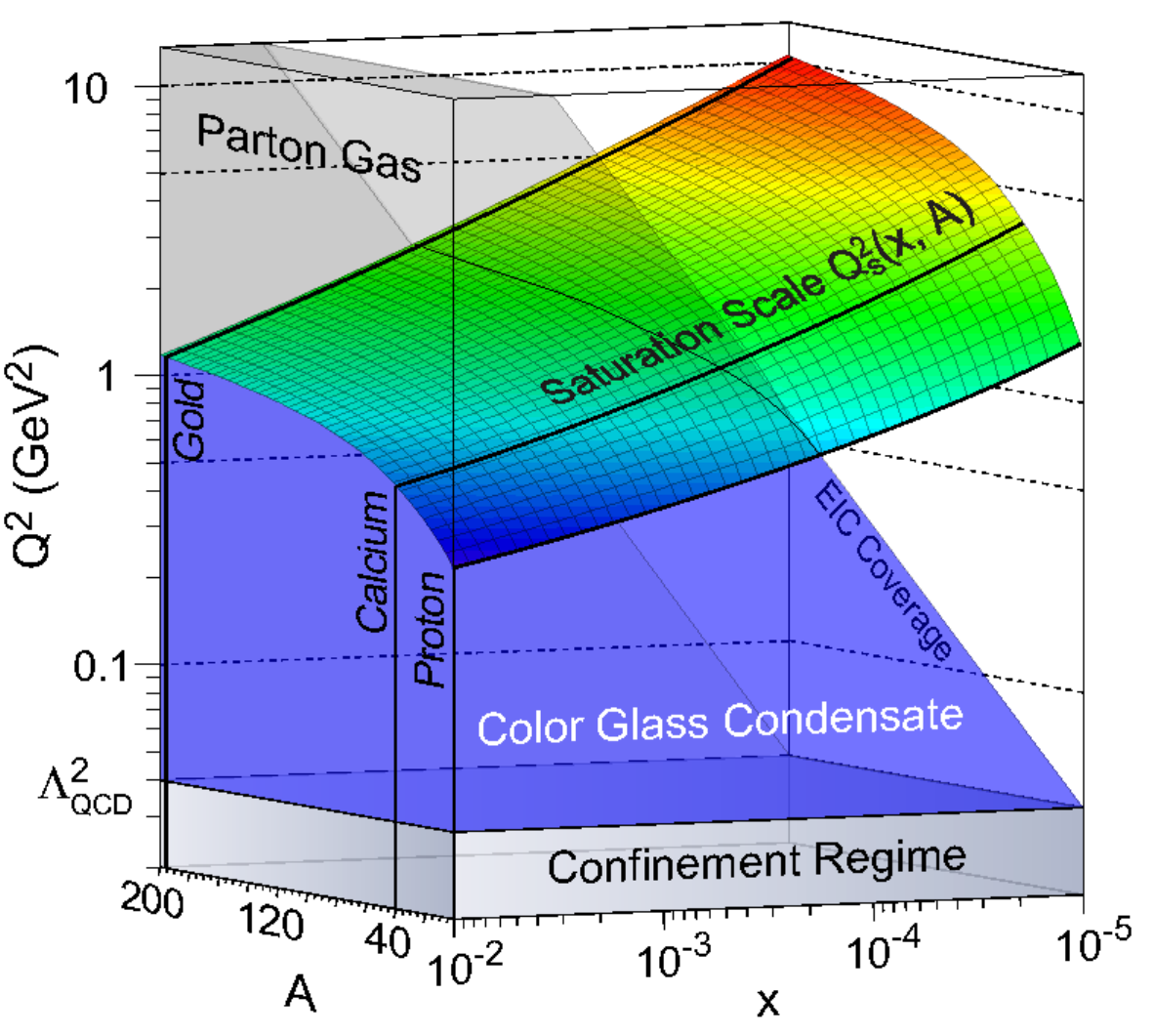}
  \caption{\label{fig:sat-domain}Saturation domain. From \cite{Deshpande:2009zz}.}
\end{figure}

\paragraph{Multiparton interactions and gluon saturation} Let us start
with the very first moments of a heavy ion collision. This is the
realm of the highest momentum scales of the entire collision, and one
may thus expect that this stage is amenable to a perturbative QCD
treatment. The situation is however more complicated. The main
difficulty is the fact that the typical transverse momentum of the
produced particles in such a collision is rather low (around $1$ GeV),
implying that high energy heavy ion collisions probe the partonic
content of the incoming nuclei at very small values of the
longitudinal momentum fraction $x$ (the fraction of the momentum of a
hadron carried by one of its constituents), i.e., in a regime where
the gluon density is large. The consequence of this is that processes
initiated by more than one parton in each nucleus become possible,
which invalidates the usual factorization schemes (based on single
parton densities), as illustrated in Figure \ref{fig:sat-nl}.
One may derive a simple criterion for the onset of these multi-parton
phenomena (known as {\sl gluon saturation}) by combining the estimated
gluon recombination cross-section and the number of gluons per unit of
transverse area
\cite{Gribov:1984tu,Mueller:1985wy,Blaizot:1987nc}. Gluon saturation
happens when the product of these two quantities is greater than one,
which can also be framed as an upper bound for the momentum transfer
$Q$ in deep inelastic scattering (the inverse of this scale plays the
role of a spatial resolution in such a scattering), $Q\lesssim Q_s$,
where $Q_s$ is the so-called saturation momentum. $Q_s$ depends both
on the atomic number of the nuclei, and on the collision energy via
the momentum fraction $x$,
\begin{equation}
Q_s^2\sim A^{1/3} x^{- \lambda}, 
\end{equation}
where the exponent $\lambda$ has been determined phenomenologically
from deep inelastic data and estimated to be $\lambda\approx
0.25$. The growth at small $x$ follows from the growth of the gluon
density, while the factor $A^{1/3}$ is a measure of the thickness of a
nucleus in the direction of the collision axis. The variations of the
saturation momentum as a function of $A$ and $x$ are shown in Figure
\ref{fig:sat-domain}.
Recalling that the typical value of $x$ scales as $p_\perp/\sqrt{s}$
where $p_\perp$ is the transverse momentum of produced particles and
$\sqrt{s}$ the energy in the center of mass of a nucleon-nucleon
collision, it appears that the bulk of particle production in heavy
ion collisions at the energy of the LHC is potentially affected by
gluon saturation, which calls for a theoretical framework going beyond
the usual collinear factorization. Indeed, the standard parton
distribution are single-parton densities and do not contain the
necessary information about the multi-parton initial states that
become important in the saturation regime.

\paragraph{Color Glass Condensate} Extending the framework of
collinear factorization by defining multi-parton densities in the same
way as the usual parton distributions is not practical. Instead, one
exploits the fact that gluon saturation is also a regime of large
gluon occupation number, which allows to treat the gluon field as
classical in a first approximation
\cite{McLerran:1993ni,McLerran:1993ka}.
\begin{figure}[htbp]
  \centering
  \includegraphics[width=\textwidth]{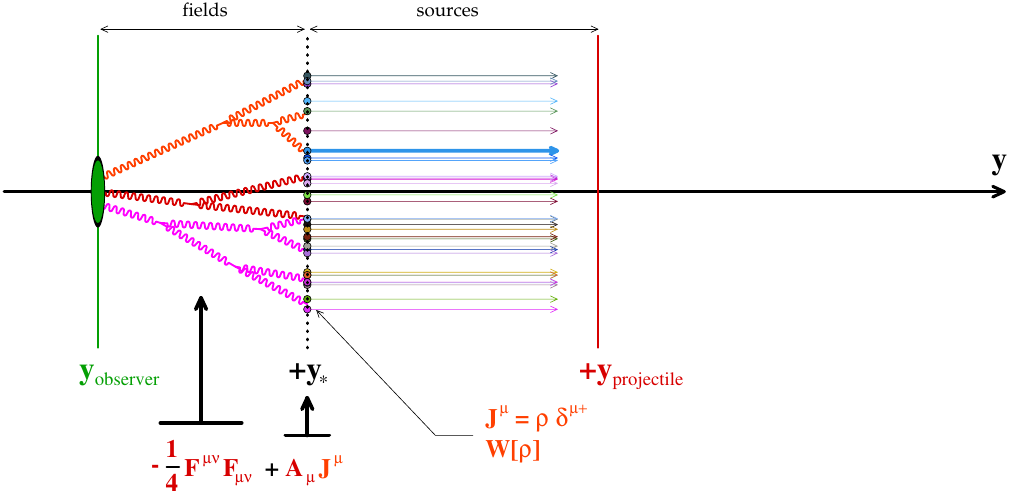}
  \caption{\label{fig:MVmodel}Illustration of the McLerran-Venugopalan model.}
\end{figure}
In such a description, the relevant information about an incoming
nucleus is the color current it carries, that acts as a source for
the color field. Moreover, this current is produced by the partons
that are comparatively fast (in the observer's frame), which implies
that it is nearly time independent thanks to time dilation. The
degrees of freedom in such a description are thus color currents (one
for each projectile) coupled to gluon fields, with an effective action
\begin{align}
  {\cal S}_{\rm eff}\equiv
  -\frac{1}{4} F^{\mu\nu}_a F_{\mu\nu\,a}+J^\mu_a A_{\mu\,a}. 
\end{align}
This setup, known as the {\sl Color Glass Condensate}
\cite{Iancu:2000hn,Iancu:2001ad,Ferreiro:2001qy,Iancu:2002xk,Weigert:2005us,Gelis:2010nm}
(CGC) is illustrated in Figure \ref{fig:MVmodel}. For a fast moving
projectile in the $+z$ direction, the color current has the form
\begin{align}
  J^\mu_a(x)=\delta^{\mu+}\rho_a(x^-,\x_\perp),
\end{align}
where $\rho_a(x^-,\x_\perp)$ is a function that describes the spatial
distribution of the color charges in the object under
consideration. In this expression, we have introduced the light-cone
coordinates, $x^\pm\equiv (t\pm z)/\sqrt{2}$ (the notation
$\delta^{\mu+}$ means that the only non-zero component of the current
is $J^+\propto J^-+J^3$). These coordinates are convenient when
discussing the kinematics of an object moving at the speed of light,
since $x^+$ acts as the time variable for this object (the fact that
the above current does not depend on $x^+$ simply reflects the fact
that this object is time-independent) and $x^-$ as a longitudinal
coordinate as measured in the rest frame of the object. The support of
the $x^-$ dependence of the current is very narrow and centered around
$x^-=0$, due to Lorentz contraction. Note also that the current must
be covariantly conserved, $[D_\mu, J^\mu]=0$. Since the covariant
derivative contains the color field, the color current may be affected
by its own radiated field (the light-cone gauge $A^-=0$ mitigates this
difficulty, since the current $J^+$ can be altered only by the field
$A^-$). The $\x_\perp$ dependence of $\rho_a(x^-,\x_\perp)$ reflects
the positions in the transverse plane of the color sources at the
instant of the collision (the duration of the collision, controlled by
the thickness of the Lorentz contracted nuclei, is much shorter than
the typical timescales of the internal motions of the constituents of
a hadron -- thus $J^\mu$ needs only to provide a snapshot of the
hadron content). But of course, the configuration of these charges is
not known event-by-event, and the best we may hope to know is a
statistical distribution of these densities, encoded in a functional
$W[\rho]$. This functional density is not something that we can
calculate perturbatively in QCD, since it depends on aspects such as
confinement, the nuclear wavefunction, etc...  For a large nucleus,
the McLerran-Venugopalan model \cite{McLerran:1993ni,McLerran:1993ka},
in which $W[\rho]$ is a Gaussian,
\begin{align}
W[\rho]
=
  \exp\Big(-\int d^2\x_\perp\frac{\rho_a(x^-,\x_\perp)\rho_a(x^-,\x_\perp)}{2\,\mu^2(x^-, \x_\perp)}\Big),
  \label{eq:gaussianMV}
\end{align}
is often employed due to its simplicity (in some cases, it even allows
analytical calculations). In this distribution, the mean value of the
charge distribution at a point $(x^-,\x_\perp)$ is zero, and
$\mu^2(x^-,\x_\perp)$ is a measure of its fluctuations. At lowest
order in the CGC effective theory, this parameter is a placeholder for
the value of the saturation momentum,
\begin{align}
  Q_s^2(\x_\perp)\propto g^2 \mu^2(\x_\perp)\ln\Big(\frac{\mu^2(\x_\perp)}{\Lambda_{_{QCD}}}\Big),\quad\mbox{with\ \ }\mu^2(\x_\perp)\equiv \int dx^-\;\mu^2(x^-,\x_\perp).
  \label{eq:Qs2mu2}
\end{align}
(This correspondence may be established by calculating the DIS
cross-section in the McLerran-Venugopalan model \cite{Lappi:2007ku}.)
Although a possible heuristic justification for this Gaussian model is
the central limit theorem, thanks to the fact that a large nucleus has
many constituents per unit of transverse area, one should keep in mind
that this distribution of $\rho_a$'s is not derived from first
principles in QCD (since doing so would require to control QCD in a
non-perturbative regime). Another reason why the Gaussian in
eq. (\ref{eq:gaussianMV}) does not have a very fundamental standing is
that this shape is not preserved when one includes one-loop
corrections: indeed, these corrections contain large logarithms of
energy that turn $W[\rho]$ into a non-Gaussian energy-dependent
distribution.

\paragraph{Power counting in the saturated regime} Let us now describe
how a typical CGC calculation is organized.  The color glass
condensate may be viewed as a Yang-Mills theory coupled to an external
source \cite{Gelis:2006yv,Gelis:2007kn}, which diagrammatically means
that all graphs contain only gluon propagators. Their endpoints can be
attached to the sources, to gluon vertices, or to the observable of
interest. In the saturation regime, the power counting for these
graphs is a bit peculiar due to the large gluon occupation number. The
3-gluon and 4-gluon vertices are respectively of order $g$ and $g^2$,
while the external source can be as large as $g^{-1}$ (this order of
magnitude is reached when the occupation number reaches its maximal
value, of order $g^{-2}$). Therefore, the order of magnitude of a
generic connected graph ${\cal G}$ is
\begin{align}
  {\cal G}\sim g^{-n_{_E}}g^{2n_{_L}} (gJ)^{n_{_J}},
\end{align}
where $n_{_E}$ is the number of external gluons, $n_{_L}$ the number
of loops and $n_{_J}$ the number of sources $J$ in the graph. We see
from this formula that when $J\sim g^{-1}$, the order of the graph
does not depend on the number of sources, implying that there is an
infinity of graphs contributing at each order (for instance, the {\sl
  leading order} is made of all the trees). The saturation regime is
therefore a strongly interacting non-perturbative situation, despite
the fact that the coupling constant may be small at high energy (the
typical scale at which the running strong coupling constant should be
evaluated is governed by the saturation momentum).

\paragraph{Leading order} At leading order, the infinite series of
tree diagrams that one needs to sum can always (for inclusive
observables) be expressed in terms of classical solutions of the
Yang-Mills equations,
\begin{align}
\big[D_\mu,F^{\mu\nu}\big]=J^\nu,
\end{align}
with retarded boundary conditions (the retarded nature of the boundary
condition follows from the fact that inclusive measurements do not put
any restriction on the final state). Since in the CGC the incoming
projectiles are completely encoded in the source $J^\mu$, the
classical initial condition is simply to have a null field (or more
generally a pure gauge) in the remote past, before the collision has
happened. 
\begin{figure}[htbp]
  \centering
  \includegraphics[width=5cm]{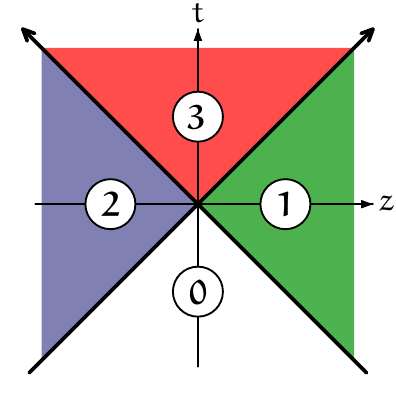}
  \caption{\label{fig:ST}Decomposition of space-time in four domains,
    when solving the classical Yang-Mills equations.}
\end{figure}
By causality, space-time is naturally divided into four domains shown
in Figure \ref{fig:ST}. Domain 0 corresponds to space-time points not
yet reached by any of the nuclei. Domain 1 is causally connected with
the right moving nucleus, but not with the left moving one. For domain
2, it is the opposite. And finally the domain 3 contains the outcome
of the collision. Given a null initial condition in domain 0, the
gauge field can be obtained analytically in the domains 1,2, and also
at the lower boundary of domain 3 (i.e., at a proper time
$\tau=0^+$). At later times in domain 3, no analytical solution is
known, but it is rather straightforward to solve the classical
Yang-Mills equations numerically \cite{Kovner:1995ja,Krasnitz:1998ns,Krasnitz:1999wc,Krasnitz:2000gz,Krasnitz:2001qu,Lappi:2003bi}.

\paragraph{Next to Leading Order} CGC calculations can in principle be
pushed to next-to-leading order, i.e., one-loop. The main difficulty
in doing this is that, like with the leading order, there is an
infinite set of diagrams contributing at NLO. These are all the
one-loop graphs embedded in the external classical gauge field
obtained at LO. For any inclusive observable, there exists an exact
relationship between the LO (tree level) and NLO (one loop) results,
that schematically reads \cite{Gelis:2008rw,Gelis:2008ad}
\begin{align}
{\cal O}_{_{\rm NLO}}
=
\frac{1}{2}\int\frac{d^3\k}{(2\pi)^3 2E_\k}\int\limits_{x^0=y^0=-\infty} d^3\x d^3\y\;
\Big[e^{+ik\cdot x}\frac{\delta}{\delta {\cal A}_{\rm ini}(x)}\Big]
\Big[e^{-ik\cdot y}\frac{\delta}{\delta {\cal A}_{\rm ini}(y)}\Big]
  {\cal O}_{_{\rm LO}},
  \label{eq:LO-NLO}
\end{align}
where ${\cal A}_{\rm ini}$ is the initial condition for the classical
field in the LO calculation.  
\begin{figure}[htbp]
  \centering
  \includegraphics[width=5.5cm]{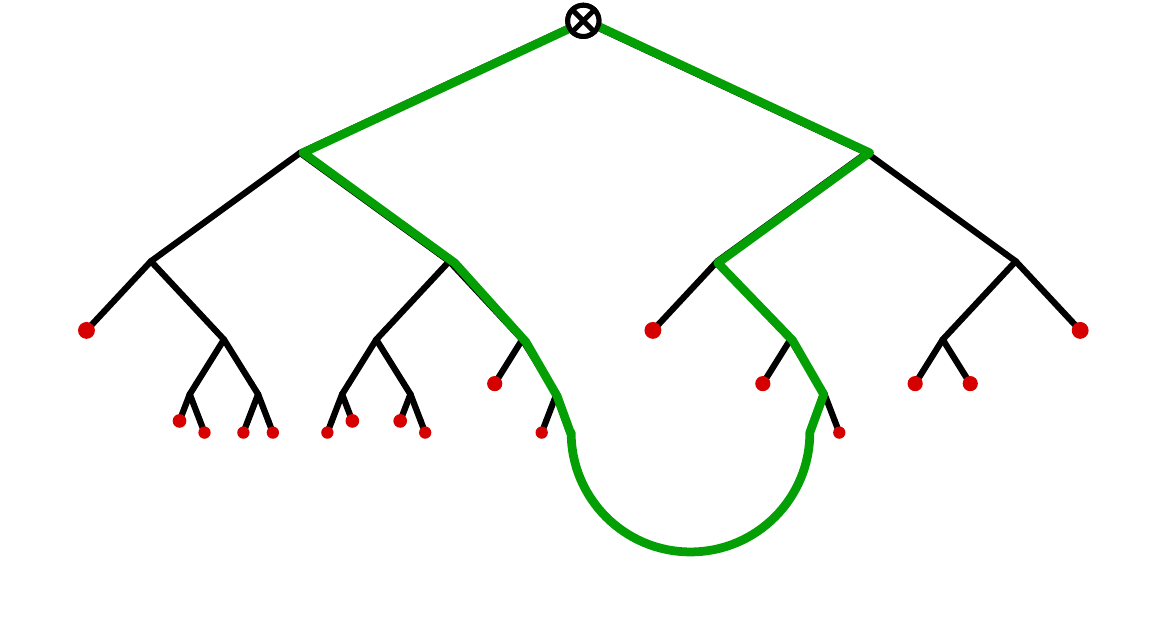}
  \includegraphics[width=5.5cm]{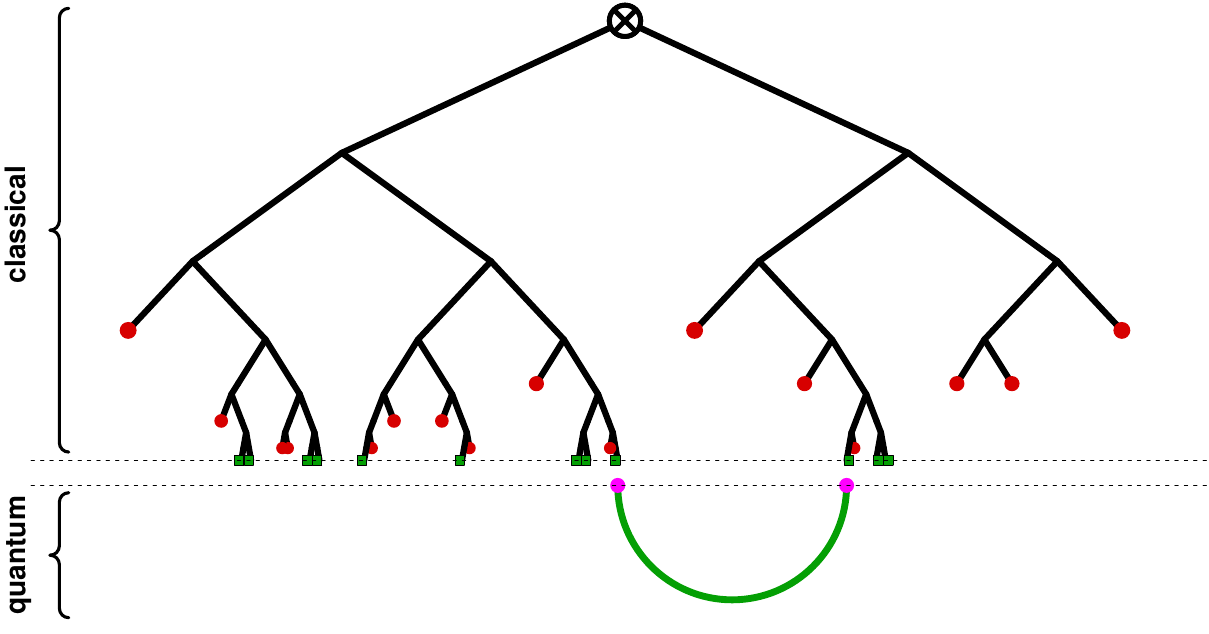}
  \caption{\label{fig:resum}Left: typical one-loop graph contributing
    to an inclusive observable. Right: interpretation of
    Eq.~(\ref{eq:LO-NLO}). In this graphical sketch, the horizontal
    direction represents space and the vertical direction represents
    time. The red dots represent
    the external source $J$.}
\end{figure}
This relationship indicates that the NLO may be obtained from the LO
simply by replacing at two points $x,y$ the classical initial
condition by plane waves $e^{+ik\cdot x}e^{-ik\cdot y}$, integrated
over all on-shell momenta $k$ (but the time evolution continues to be
governed by the classical equation of motion). This is consistent with
the fact that the first quantum correction (order $\hbar^1$) only
affects the initial state, and the first quantum effect that alters
the evolution arises at the order $\hbar^2$. In ordinary quantum
mechanics, this is best seen in the phase-space formulation (also
known as Moyal-Groenewold formulation), in which each quantum
operator ${\bs A}$ is replaced by its Wigner transform,
\begin{align}
  A(X,p)\equiv \int ds \; e^{i\tfrac{p\cdot s}{\hbar}}\,\big<X+\tfrac{s}{2}\big|{\bs A}\big|X-\tfrac{s}{2}\big>,
\end{align}
defined on the classical phase-space variables $(X,p)$. In this
mapping, the commutator of two such operators ${\bs A},{\bs B}$
becomes the {\sl Moyal bracket} of their Wigner transforms:
\begin{align}
  \{\!\{A, B\}\!\}(X,p)
  \equiv
  \frac{2}{i\hbar}
  \,
  A(X,p)
  \sin\Big(\tfrac{\hbar}{2}\big(\stackrel{\leftarrow}{\partial}_{p}\stackrel{\rightarrow}{\partial}_{_X}-\stackrel{\leftarrow}{\partial}_{_X}\stackrel{\rightarrow}{\partial}_{p}\big)\Big)
  B(X,p).
\end{align}
If we denote $W$ the Wigner transform of the density operator $\rho$
of a system, the von Neumann equation
$i\hbar \partial_t \rho=[H,\rho]$ becomes
\begin{align}
  \partial_t W = \{\!\{{\cal H},W\}\!\}=\underbrace{\{{\cal H},W\}}_{\mbox{\scriptsize Poisson bracket}}+{\cal O}(\hbar^2), 
\end{align}
where ${\cal H}$ is the Wigner transform of $H$ (i.e., ${\cal H}$ is
the classical Hamiltonian).

\paragraph{Large logarithms and JIMWLK evolution} When calculating
such one-loop corrections, an aspect of the CGC degrees of freedom
that we have ignored until now becomes important, namely that one must
introduce a cutoff to separate the color sources from the gauge
fields. For instance, such a cutoff should limit the range of
integration over the momentum $k$ in Eq.~(\ref{eq:LO-NLO}).  This
separation is based on the longitudinal momentum (or equivalently, the
rapidity $y\equiv\tfrac{1}{2}\ln((k_0+k_z)/(k_0-k_z))$), and is mandatory
when evaluating loops to avoid double countings. In practice, the
longitudinal component of the loop momentum must remain below the
cutoff, because all the higher momentum modes are already included in
the color current. This leads to all one-loop correction to be
sensitive (proportional to the logarithm of the cutoff on longitudinal
momentum) to the cutoff
\cite{Ayala:1995hx,JalilianMarian:1996xn,Iancu:2000hn,Iancu:2001ad,Ferreiro:2001qy,Gelis:2008rw,Gelis:2008ad,Gelis:2008sz}. However,
since this cutoff is an ad-hoc parameter of the CGC effective theory
rather than a physical parameter, it should not enter in physical
observables. This paradox is resolved by the fact that the cutoff
dependence is universal, in the sense that it depends on the nature of
the two colliding projectiles, but not on the inclusive observable one
is measuring. Therefore, it is possible to absorb the cutoff
dependence into a redefinition of the distributions $W[\rho]$ that
define the color source content of the projectiles, turning them into
cutoff dependent objects. For this to be feasible, one should perform
an average of the $\rho$-dependent observable, weighted by the
distributions of $\rho$'s of each projectile,
\begin{align}
  \big<{\cal O}\big>\equiv \int [D\rho_1 D\rho_2]\;
  W_1[\rho_1]W_2[\rho_2]\,{\cal O}(\rho_1,\rho_2).
\end{align}
The cutoff dependence of $W[\rho]$ is controlled by the so-called
JIMWLK equation, schematically of the form
\begin{align}
  \frac{\partial W[\rho]}{\partial \log\Lambda}
  =
  \frac{\delta}{\delta \rho_a}
  \chi_{ab}
  \frac{\delta}{\delta \rho_a}
  W[\rho],
\end{align}
where $\chi_{ab}$ depends on the LO classical field created by the
source $\rho$ (the possibility to transfer the cutoff dependence from
the observable to the distribution $W[\rho]$ is made possible by the
fact that the operator in the right hand side of the JIMWLK equation
is self-adjoint, via integration by parts). Thus, by evolving the
distributions $W[\rho]$ of each projectile to values of the
longitudinal momentum in the immediate vicinity of the scales relevant
for the observable of interest, one resums all the leading logarithms,
i.e., the powers $(g^2 \log\Lambda)^n$ where each logarithm of the
cutoff is accompanied by a factor $g^2$. This is very similar in spirit
to collinear factorization, the unphysical cutoff $\Lambda$ playing
the role of a factorization scale that should disappear from observables.

Since it is a functional equation, the JIMWLK equation is difficult to
solve, even numerically. The only known approach so far uses the fact
that the JIMWLK equation acts like a diffusion equation in the
functional space of the $\rho$'s (in a treatment more rigorous than
this general discussion, one would use Wilson lines built from the
$\rho$'s rather than the $\rho$'s themselves), and therefore can be
rewritten as a Langevin equation \cite{Blaizot:2002np}. Then, after
discretization of the transverse plane, this stochastic equation is
amenable to a numerical treatment in order to obtain an ensemble of
$\rho$'s evolved to the relevant value of the cutoff
\cite{Rummukainen:2003ns,Lappi:2011ju,Dumitru:2011vk}. Let us also
mention recent improvements: a modification of the Langevin equation
has been proposed to include the effects of a running coupling
\cite{Lappi:2014wya}, and the full NLO corrections to the JIMWLK
equation have also been evaluated
\cite{Balitsky:2013fea,Kovner:2013ona} (but not yet implemented in a
numerical code).

\paragraph{Balitsky-Kovchegov equation} To avoid this computationally
heavy approach, it is also possible to truncate the JIMWLK
equation. The first thing to note is that the functional form of the
JIMWLK equation is equivalent to an infinite sequence of equations for
the correlation functions of Wilson lines constructed from the
$\rho$'s
\begin{align}
  U(\x_\perp) \equiv {\rm T}\,\exp ig\int dx^-\,\tfrac{1}{{\bs\nabla}_\perp^2}\rho(x^-,\x_\perp). 
\end{align}
These equations are nested: the equation that drives the cutoff
dependence of the $2$-point correlation function depends on a
$4$-point function, etc... A possible approximation (that may be
justified in the limit of a large number of colors) consists in
factorizing this $4$-point function as a product of two $2$-point
functions, which has the effect of closing the evolution equation of
the latter. The resulting equation, known as the Balitsky-Kovchegov
equation \cite{Balitsky:1995ub,Kovchegov:1999yj}, reads
\begin{align}
  \frac{\partial T_{\x\y}}{\partial \log\Lambda}
  =\frac{\alpha_sN_c}{2\pi^2}\int d^2 \z_\perp\;
  \frac{(\x_\perp-\y_\perp)^2}{(\x_\perp-\z_\perp)^2(\y_\perp-\z_\perp)^2}\,
  \Big\{T_{\x\z}+T_{\z\y}-T_{\x\y}
  -T_{\x\z}T_{\z\y}\Big\},
\end{align}
where
\begin{align}
  T_{\x\y}
  \equiv
  1- N^{-1}\,{\rm tr}\big<U(\x_\perp)U^\dagger(\y_\perp)\big>
\end{align}
(with Wilson lines taken in the fundamental representation of
${su}(N)$). $T_{\x\y}$ is also proportional to the scattering
amplitude of a quark-antiquark dipole (at the transverse coordinates
$\x_\perp$ and $\y_\perp$, respectively) off a high energy nucleus. In
this equation, the first three terms, linear, correspond to the BFKL
equation, and the last term, non-linear, is a correction due to gluon
saturation, that becomes sizeable when the scattering amplitude
approaches the unitarity limit $T=1$. Since it is an equation for an
ordinary function, the BK equation is much easier to solve
numerically.  Note also that the previous equation has now been
improved by running coupling corrections
\cite{Kovchegov:2006vj,Gardi:2006rp}, by the full next-lo-leading log
corrections \cite{Balitsky:2008zza}, and by a resummation of collinear
logarithms \cite{Iancu:2015vea,Lappi:2016fmu}. These improvement has
allowed a successful phenomenology of small-$x$ phenomena in
deep-inelastic scattering and forward proton-proton or proton-nucleus
collisions based on the BK evolution equation
\cite{Albacete:2014fwa,Altinoluk:2014eka,Lappi:2015fma,Iancu:2015joa}.

\section{Pre-equilibrium evolution just after the collision}

\paragraph{Energy-momentum tensor at Leading Order} As we have seen in
the previous section, large nuclei at high energy may be described by
using the CGC framework, in which the large momentum degrees of
freedom are treated as random color currents coupled to the color
field. At leading order in the strong coupling constant, all
expectation values are given by tree diagrams, whose sum is the
classical solution of Yang-Mills equations with a null retarded
boundary condition. Having in mind a description of the subsequent
stages of the collision in terms of relativistic hydrodynamics, it is
therefore natural to calculate the components of the energy momentum
tensor. In the CGC framework, the dominant contribution comes from the
gluons (the valence quarks have a very small contribution at small
$x$, and the sea quark distribution is suppressed by a power of
$\alpha_s$ compared to that of the gluons). In a classical field, they
are given by the following formulas
 \begin{align}
&
T^{00}_{_{\rm LO}}
=
\frac{1}{2}\big[\underbrace{{\bs E}^2+{\bs B}^2}_{\mbox{\scriptsize class. fields}}\big],
\qquad
T^{0i}_{_{\rm LO}}=\big[{\bs E}\times {\bs B}\big]^i,
\nonumber\\
&
T^{ij}_{_{\rm LO}}
=
\frac{\delta^{ij}}{2}\big[{{\bs E}^2+{\bs B}^2}\big]
-\big[{\bs E}^i{\bs E}^j+{\bs B}^i{\bs B}^j\big],
\end{align}
in terms of the chromo-electric and chromo-magnetic fields (note that
this tensor is traceless in classical Yang-Mills theory -- a non-zero
trace would arise from loop corrections via the $\beta$-function, and
from explicit quark masses when quarks are taken into account). Let us
first mention the very special configuration of the $\E$ and $\B$
fields just after the collision: at $\tau=0^+$, these two fields are
both parallel to the collision axis \cite{Lappi:2006fp}, which leads
to the following form of $T^{\mu\nu}_{_{\rm LO}}(\tau=0^+)$,
\begin{align}
T^{0i}_{_{\rm LO}}=0,\quad T^{11}_{_{\rm LO}}=T^{22}_{_{\rm LO}}=T^{00}_{_{\rm LO}},\quad T^{33}_{_{\rm LO}}=-T^{00}_{_{\rm LO}}. 
\end{align}
In other words, the matter is produced at rest, with a negative
longitudinal pressure (i.e., the system resists longitudinal
expansion). Such a negative pressure means that this system should not
be viewed as a collection of on-shell particles, but rather as fields. 
\begin{figure}[htbp]
  \centering
  \includegraphics[width=8cm]{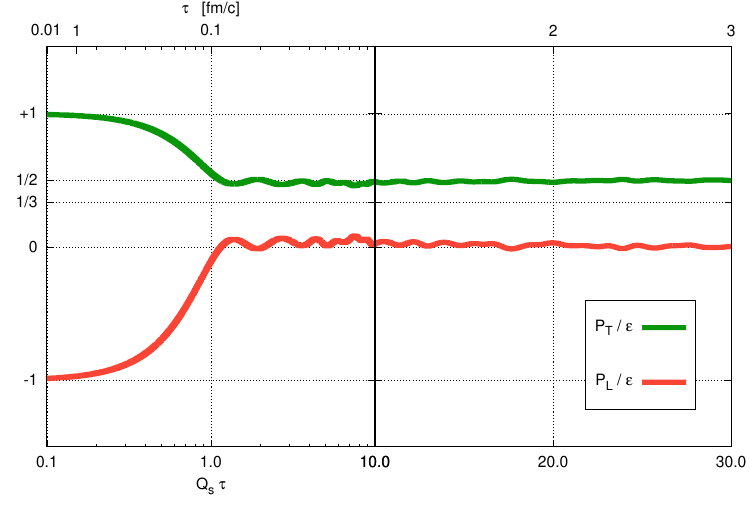}
  \caption{\label{fig:TLO}Time evolution at Leading Order of the ratios $T^{11}/T^{00}$ and $T^{33}/T^{00}$.}
\end{figure}
At later times (see Figure \ref{fig:TLO}), the longitudinal pressure
progressively builds up to become positive at a time around
$Q_s \tau\sim 1$, which is indeed the time at which the color fields
may be interpreted as nearly on-shell gluons
\cite{Fukushima:2011nq}. However, it is also clear from this plot that
the ratio of longitudinal to transverse pressure remains very
small. In fact, at leading order, this ratio decreases as $\tau^{-2}$,
which is characteristic of a free streaming system (i.e., a system
whose self-interactions are too weak to compete with the longitudinal
expansion).

\paragraph{Next to Leading Order, Instabilities} Such a behavior of
the longitudinal pressure is not consistent with hydrodynamical
evolution, where the ratio $P_{_L}/P_{_T}$ would instead increase to
eventually approach unity. For an underlying QCD description to allow
a smooth matching to a subsequent hydrodynamical expansion, there
should be a range of times where the two descriptions lead to similar
behaviors. It turns out that higher order corrections in the CGC
description are potentially more important than the power counting
suggests. Indeed, the power counting correctly states that loop
corrections (i.e. corrections beyond the classical field
approximation) are suppressed by additional powers of the coupling
constant, but it implicitly assumes that the coefficients in this
power expansion remain of order one at all times. It is this
assumption that turns out to be incorrect, because the classical
solutions of Yang-Mills equations are subject to instabilities that
make them exponentially sensitive to their initial conditions
\cite{Muller:1992iw,Biro:1993qc,Kunihiro:2010tg,Romatschke:2005pm},
combined with the fact that one-loop corrections can be expressed in
terms of small perturbations to the initial condition of the LO
classical field.
\begin{figure}[htbp]
  \centering
  \includegraphics[width=8cm]{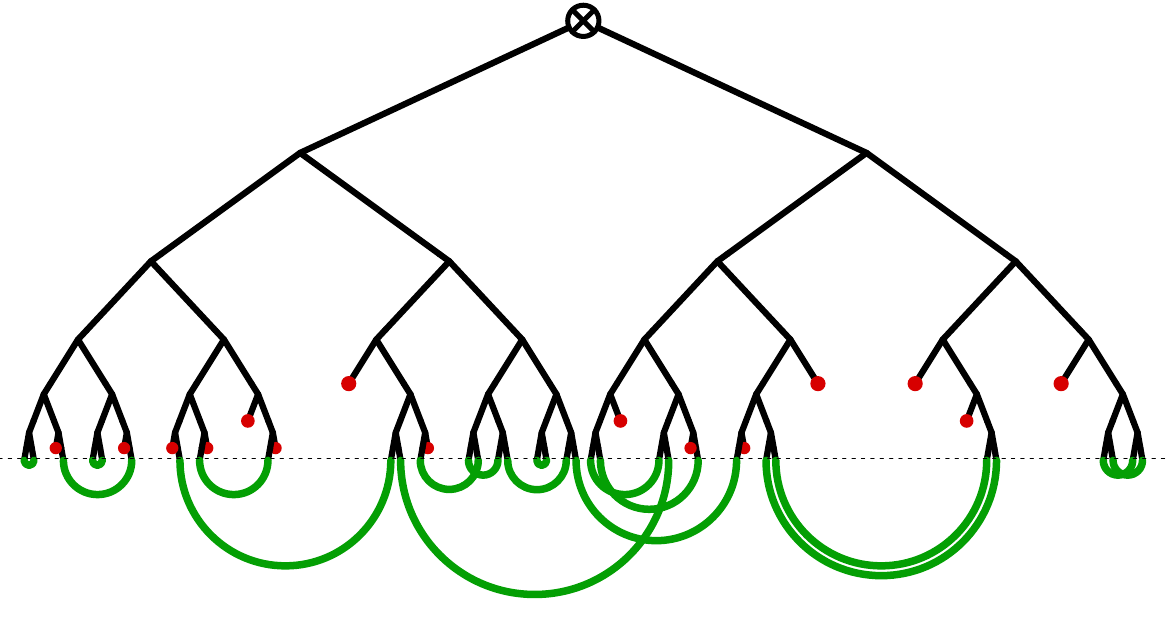}
  \caption{\label{fig:resum1} Graphs that have the leading time
    behavior in the presence of instabilities. }
\end{figure}

\paragraph{Resummation, Classical Statistical Approximation} By a more
careful power counting that keeps track of these terms that have an
exponential growth with time \cite{Gelis:2012ri}, it is possible to determine the set of
graphs that provide the leading contributions at large time. These
graphs are shown in the Figure \ref{fig:resum1} (and in the left part
of Figure \ref{fig:resum}, we show for comparison a typical
next-to-leading order contribution). It turns out that the sum of this
infinite set of higher-loop order graphs can be obtained by letting
the initial value of the gluon field fluctuate around its classical
value, with a Gaussian distribution which is completely determined by
the one-loop result. Schematically, this resummed result reads:
\begin{align}
  {\cal O}_{\rm resummed}
  =
  \int \big[D{\colorc a}\big]\;\exp\Big[-\frac{1}{2\,{\colorb\hbar}}\int_{\x,\y}
  {\colorc a(\x)}{\colora{\bs\Gamma}_2^{-1}(\x,\y)}{\colorc a(\y)}\Big]\;{\cal O}_{_{\rm LO}}[{\cal A}_{\rm in}+{\colorc a}].
  \label{eq:resum}
\end{align}
Such a resummation is of course only a part of the full answer (with
an appropriate choice of the variance ${\bs\Gamma}_2$, one may obtain
the complete LO and NLO results, but only a subset of the higher
orders), known in the literature as the {\sl Classical Statistical
  Approximation} (CSA). We have written the explicit dependence of the
Gaussian distribution with respect to $\hbar$ to emphasize the quantum
nature of these fluctuations of the field.  In the special case of
heavy ion collisions, the variance of the Gaussian fluctuations,
${\bs\Gamma}_2$, can be determined analytically at the time
$\tau=0^+$, i.e., just after the collision
\cite{Epelbaum:2013waa}. Note that in the vacuum, this variance reads
(this is just a sketch, that ignores the complications due to gauge
fields)
\begin{align}
  \big<a(x)a(y)\big>={\bs\Gamma}_2(x,y)=\int \frac{d^3\k}{(2\pi)^3 E_\k}\;\frac{1}{2}\,e^{ik\cdot(x-y)}.
  \label{eq:variance}
\end{align}
The factor $1/2$ in the integrand can be interpreted as the zero point
occupation of the vacuum (at one-loop, one may show that the only
quantum effect is the fact that the ground state is not empty but
subject to zero point fluctuations -- the resummation considered here
is an approximation in which this  is extended to higher-loop orders).

Once the variance ${\bs\Gamma}_2$ is known, Eq.~(\ref{eq:resum}) can
be evaluated numerically on a lattice in a straightforward way, since
it simply amounts to reproducing the leading order classical CGC
computation with a fluctuating initial condition. The Gaussian
integral in Eq.~(\ref{eq:resum}) can be estimated by Monte-Carlo
sampling. These simulations, with the variance given in
Eq.~(\ref{eq:variance}), lead to an increase of the ratio of
longitudinal to transverse pressure \cite{Gelis:2013rba}. However, the
interpretation of this result is obscured by the fact that this setup
has no continuum limit when the lattice spacing goes to zero. This can
be understood easily from the fact that Eq.~(\ref{eq:variance})
corresponds to a flat momentum distribution of gluons, that extends to
arbitrarily large momenta (only cut-off by the inverse lattice
spacing).

\paragraph{CSA with a compact fluctuation spectrum} An
alternative to the variance given in Eq.~(\ref{eq:variance}) that does
not have this ultraviolet problem would be to replace the factor $1/2$
by another gluon distribution that has a fast enough fall-off at large
momentum,
\begin{align}
  \big<a(x)a(y)\big>_{\rm alt}=\int \frac{d^3\k}{(2\pi)^3 E_\k}\;f_0(k)\,e^{ik\cdot(x-y)}.
  \label{eq:variance-alt}
\end{align}
Although such a distribution cannot be derived from first principles,
unlike Eq.~(\ref{eq:variance}), a handwaving argument in favor of it
is that after a time of order $Q_s^{-1}$ the gluons produced in a
collision are nearly on-shell with a compact distribution that extends
up to momenta $k\sim Q_s$ (with an occupation number of order $g^{-2}$
within this support). With such a spectrum of initial field
fluctuations, the behavior of the ratio of pressures $P_{_L}/P_{_T}$
is at odds with what was obtained with Eq.~(\ref{eq:variance}),
everything else being equal: with Eq.~(\ref{eq:variance-alt}), one has
$P_{_L}/P_{_T}\sim \tau^{-2/3}$, showing no sign of isotropization
\cite{Berges:2013eia,Berges:2013fga}. In this scenario, it is argued
  that isotropization is delayed until the gluon occupation number
  becomes of order one, which would happen eventually at a time
  $Q_s \tau \sim \alpha_s^{-3/2}$.

  Going beyond the classical statistical approximation in a field
  theoretical framework is possible with the two-particle irreducible
  formalism \cite{Luttinger:1960ua,Baym:1961zz,Berges:2004yj}. This
  formalism amounts to a self-consistent determination of the
  propagator (which in a many-body context also contains the
  information about the particle distribution) by resumming a
  self-energy --itself a function of the propagator-- on the
  propagator. The 2PI framework can be renormalized
  \cite{vanHees:2001ik,Berges:2005hc}, thereby avoiding the issues
  with the CSA and zero point vacuum fluctuations, and can thus be
  used to track the real-time evolution of a system starting from any
  quantum state. The main drawback of this approach is that it is very
  demanding in terms of computational resources, especially in the
  case of an expanding system like the one formed in a heavy ion
  collision. At the time of this writing, there has only been one
  ``proof of concept'' implementation for an expanding system
  \cite{Hatta:2011ky}, in which the questions related to
  isotropization were not investigated.

\paragraph{Kinetic theory and Boltzmann equation} A less costly
alternative is {\sl kinetic theory}, that one may obtain from the 2PI
approach provided one makes two additional approximations:
\begin{itemize}
\item Quasi-particle approximation: this amounts to assuming that the
  propagator describes on-shell infinitely long lived particles. With
  this assumption (that can only be valid in a system where the mean
  free path is much larger than the De Broglie wavelength of the
  particles), the only unknown in the propagator is the particle
  distribution $f(x,\p)$.
\item Gradient approximation: this amounts to assuming that the
  spatial variations of the system due to its off-equilibriumness
  occur only on time and distance scales much larger than the De
  Broglie wavelength of the particles. 
\end{itemize}
With these two approximations, the Kadanoff-Baym equation of motion of
the 2PI formalism reduces to a much simpler Boltzmann equation,
schematically of the form
\begin{align}
  \big(\partial_t+\v_{\p_1}\cdot{\bs\nabla}_\x)\,f_1
  &= C_{\p_1}[f]\\
  &=\int_{\p_{2,3,4}}|M(12\to 34)|^2\, \delta(p_1+p_2-p_3-p_4)\nonumber\\
  &\qquad\times\big\{f_3 f_4(1+f_1)(1+f_2)
  -f_1 f_2(1+f_3)(1+f_4)\big\}. 
\end{align}
The first line is the generic Boltzmann equation obtained when using
these two approximations, with a {\sl collision term} local in $x$
that can a priori contain arbitrary orders in the distribution $f$. In
the second line, we have written specifically the Boltzmann equation
that includes only $2\to 2$ elastic reactions (but this truncation is
an extra approximation that goes beyond the quasi-particle and
gradient approximations).

\begin{figure}[htbp]
  \centering
  \includegraphics[width=8cm]{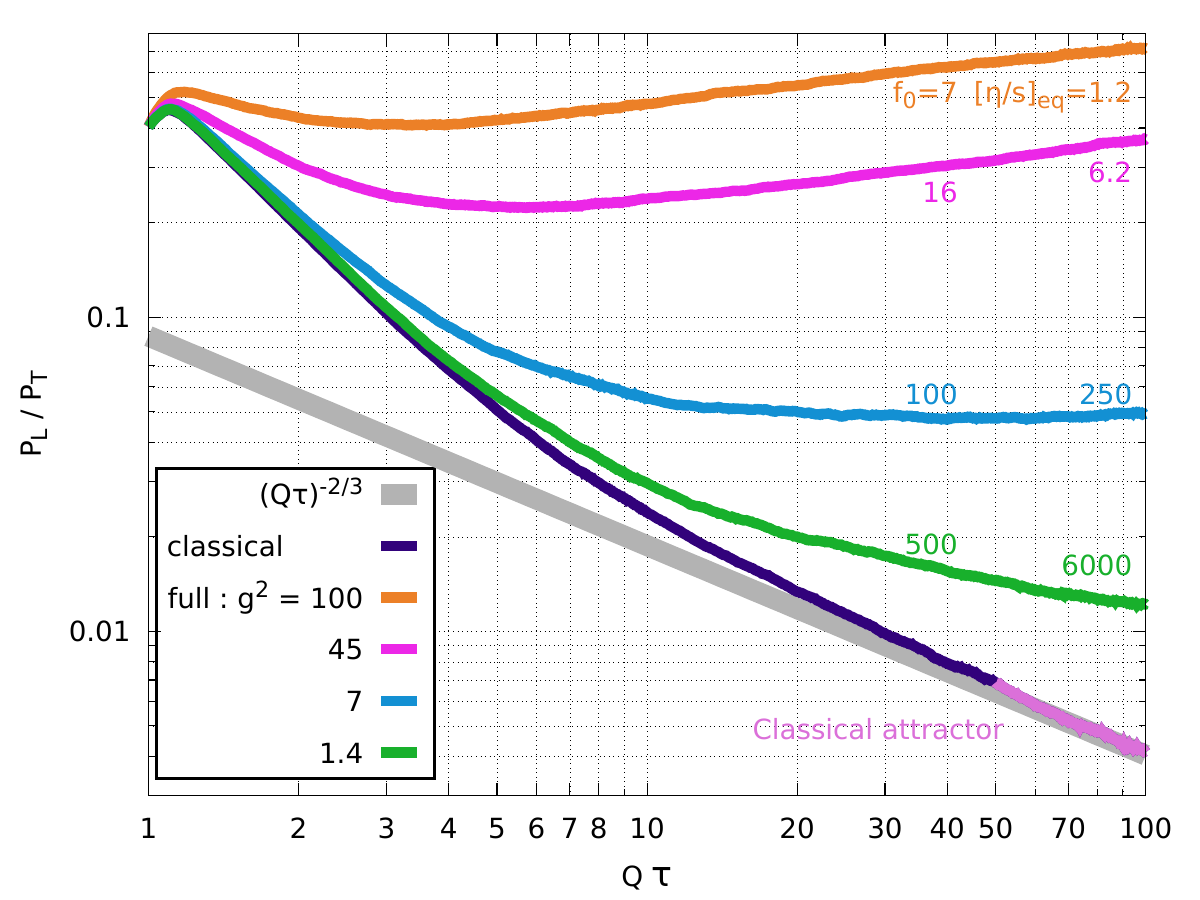}
  \caption{\label{fig:Kin-scal-evol}Comparison of the classical and
    full evolution of $P_{_L}/P_{_T}$ in kinetic theory for a
    longitudinally expanding system. The dark purple curve is the
    kinetic theory analogue of the classical statistical approximation
    based on eq.~(\ref{eq:variance-alt}).}
\end{figure}

\paragraph{Testing semi-classical approximations via kinetic theory}
Besides being a tool to study the evolution of the particle
distribution, the Boltzmann equation can also provide some insights
about the limitations of the classical statistical approximation,
because identical classical approximations may be applied to the right
hand side of the Boltzmann equation. To that effect, the
correspondence is the following \cite{Mueller:2002gd,Jeon:2004dh}
\begin{itemize}
\item[1.] CSA with Eq.~(\ref{eq:variance-alt}) $\Longleftrightarrow$ keep only the terms cubic in $f$,
\item[2.] CSA with Eq.~(\ref{eq:variance}) $\Longleftrightarrow$ replace $f\to f+\tfrac{1}{2}$ in the previous approximation (note that in this approximation, the collision integral has the correct cubic and quadratic terms, and also some spurious linear terms not present in the exact collision term).
\end{itemize}
With the approximation 2 of the collision integral, one can for
instance reproduce quantitatively the ultraviolet sensitivity (i.e.,
the lack of continuum limit) of the CSA when zero point fluctuations
are included \cite{Epelbaum:2014mfa}. With the approximation 1, the
Boltzmann equation (for an expanding system) leads to the behavior
$P_{_L}/P_{_T}\sim \tau^{-2/3}$. It is also possible to understand by
kinetic arguments why this approximation misses important physics for
a longitudinally expanding system \cite{Epelbaum:2015vxa}. In such a
system, isotropization results from a competition between the
expansion of the system, that drives the particles towards a more a
more anisotropic distribution, and collisions that tend to
redistribute the directions of the velocities (for collisions that are
sufficiently isotropic, which is the case for a dense system due to a
strong Debye screening).
\begin{figure}[htbp]
  \centering
  \includegraphics[width=4cm]{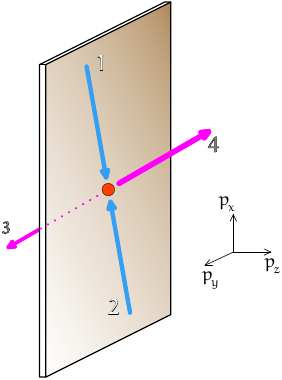}
  \caption{\label{fig:22scat}Typical $2\to 2$ scattering in a highly anisotropic system.}
\end{figure}
However, as shown in Figure \ref{fig:22scat}, when two particles
(labeled $1$ and $2$) from a very anisotropic distribution scatter
out-of-plane, momentum conservation implies that the two final state
particles (labeled $3$ and $4$) end in the empty region of
phase-space. Such a scattering is forbidden when one keeps only the
cubic terms in the collision integral, because these terms correspond
to stimulated emission, that can only happen when one of the produced
particles goes in an already populated region. The large-angle
scattering process of Figure \ref{fig:22scat} is allowed only by the
terms quadratic in $f$, that are not present in the approximation of
Eq.~(\ref{eq:variance-alt}).  The difference between a semi-classical
approximation such as Eq.~(\ref{eq:variance-alt}) and one that
includes the zero-point fluctuations can then be seen on the time
evolution of the ratio $P_{_L}/P_{_T}$, as shown in Figure
\ref{fig:Kin-scal-evol} in the case of a scalar theory (see
\cite{Epelbaum:2015vxa} for details): in this figure, we see that the
classical approximation leads to a decrease of this ratio as
$\tau^{-2/3}$, while with the full collision term it first decreases
(while the expansion outpaces the scatterings) and then increases
(when the expansion rate has become low enough) to approach unity.
It is only for very small couplings (i.e., unrealistically large
values of the ratio of the shear viscosity to entropy, $\eta/s$ -- see
the next section for a discussion of the range of values of this ratio
expected in heavy ion collisions) that this classical approximation
agrees with the evolution driven by the full collision term long
enough to reach the asymptotic scaling regime.

\begin{figure}[htbp]
  \centering
  \includegraphics[width=8cm]{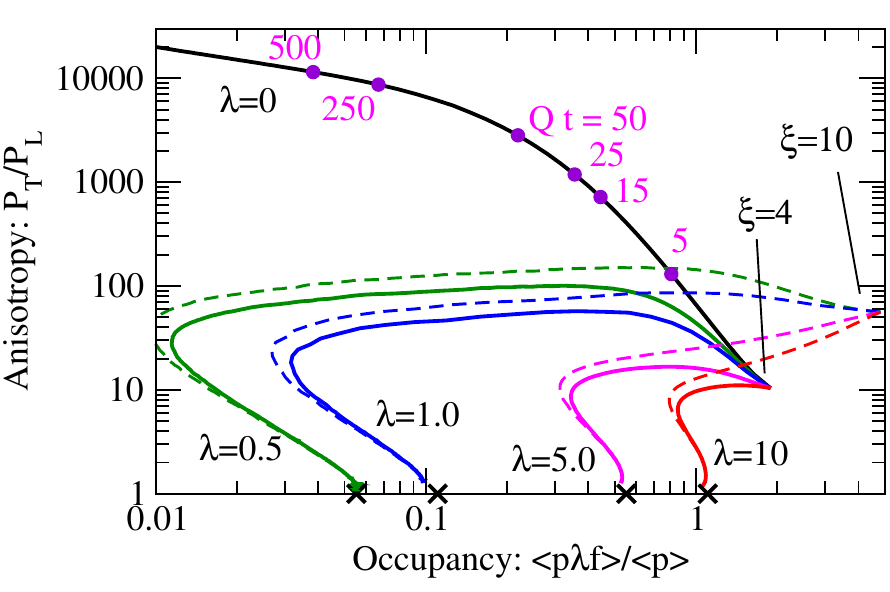}
  \caption{\label{fig:kinevol}Isotropization in Yang-Mills theory in a
    kinetic description. In this figure $\lambda\equiv g^2 N_c$ is the
    't Hooft coupling.  The solid black curve is the kinetic theory
    analogue of the classical statistical approximation based on
    eq.~(\ref{eq:variance-alt}). From \cite{Kurkela:2015qoa}.}
\end{figure}

\begin{figure}[htbp]
  \centering
  \includegraphics[width=8cm]{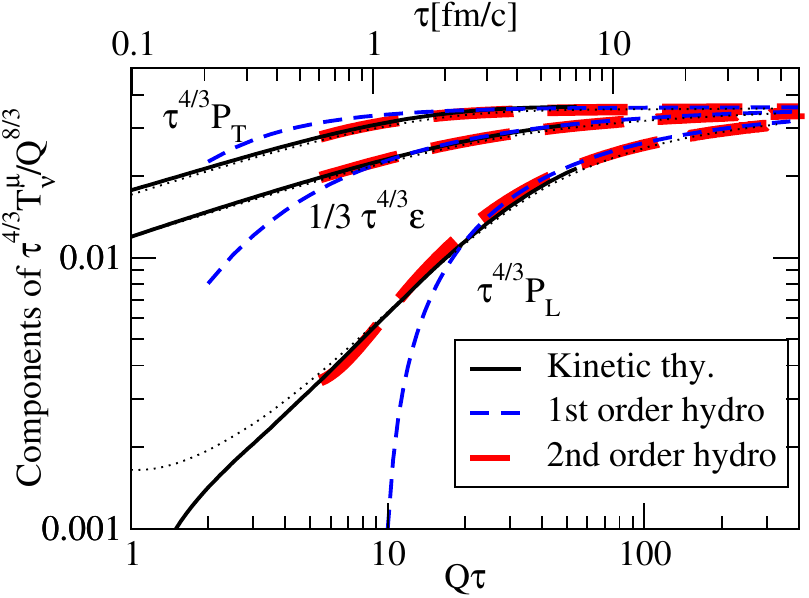}
  \caption{\label{fig:kinhydro} Comparison of kinetic theory with second order hydrodynamics. From \cite{Kurkela:2015qoa}.}
\end{figure}

\paragraph{Kinetic approach in Yang-Mills theory} A similar
computation has also been performed in the more realistic setting of
Yang-Mills theory \cite{Arnold:2002zm,Kurkela:2015qoa}, with similar results as
shown in Figure \ref{fig:kinevol}. There also, one sees the classical
approximation depart from the full evolution fairly quickly. For
instance, the curve $\lambda=0.5$, i.e. $\alpha_s=0.02$ for $N_c=2$,
deviates from the classical approximation around $Q_s\tau\approx 2$.
Moreover, this happens much earlier than the presumed range of
validity $Q_s\tau\lesssim\alpha_s^{-3/2}\approx 350$ predicted within
the classical approximation itself (in fact, we see from this plot
that the point $Q_s\tau =350$ on the classical evolution is orders of
magnitude off the correct trajectory).
In addition, this computation has shown a very good agreement with
second order hydrodynamics already at times where isotropization is
still far from being achieved, thereby providing a justification for
the applicability of hydrodynamics as early as $\tau\approx 0.6 $~fm/c
(compare the red and solid black curves in Figure \ref{fig:kinhydro}).

\paragraph{Fixed points of kinetic evolution} Kinetic theory has also
been used to obtain results about the fate of a system that undergoes
longitudinal expansion
\cite{Blaizot:2017lht,Blaizot:2017ucy,Blaizot:2019scw,Blaizot:2020gql},
with only generic assumptions about the strength of the collisions.
The starting point of this study is the Boltzmann equation in the {\sl
  relaxation time approximation},
\begin{align}
    \Big(\partial_\tau -\frac{p_z}{\tau}\Big)\, f(\tau,\p)
  =-\frac{f-f_{\rm eq}}{\tau_{_R}}. 
\end{align}
In this equation, $\tau_{_R}$ is a relaxation time that controls how
fast the particle distribution relaxes to its local equilibrium
value. This parameter may be chosen in various ways:
\begin{itemize}
\item $\tau_{_R}=\infty$~:~~ for a collisionless system,
\item $\tau_{_R}\sim \epsilon^{-1/4}$~:~~ for a ``conformal'' system,
  i.e., a system where the collision rate scales as the inverse temperature,
\item $\tau_{_R}={\rm const}$~:~~ for a fixed collision rate (although
  this is not very realistic with expansion).
\end{itemize}
\begin{figure}[htbp]
  \centering
  \includegraphics[width=8cm]{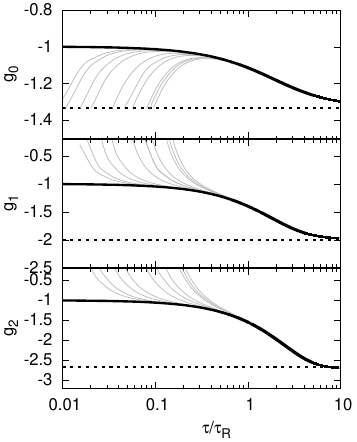}
  \caption{\label{fig:kinattract}Kinetic evolution of the exponents
    $g_n$ for many distinct initial conditions. The dotted line
    indicates the interacting fixed point. From \cite{Blaizot:2017ucy}.}
\end{figure}
Then, one may define the following moments,
\begin{align}
  L_n\equiv \int_\p \p^2 \; P_{2n}(p_z/p)\;f(\tau,\p),\quad g_n\equiv \tau \partial_\tau \ln L_n. 
\end{align}
(In this definition, $P_{2n}(z)$ is the order-$2n$ Legendre
polynomial. Note that $L_0=\epsilon = P_{_L}+2P_{_T}$,
~~$L_1= P_{_L}-P_{_T}$.)  The quantity $g_n$ gives the exponent in the
power law behavior of the corresponding moment. The Boltzmann equation
can then be transformed into an infinite set of linear equations that
govern the evolution of the moments $L_n, L_{n+1}$ and $L_{n-1}$, in
which all but one  coefficients depend only on the left hand side of
the Boltzmann equation. Then, it was observed that these coupled
equations admit two (and only two) fixed points:
\begin{itemize}
\item A {\sl free-streaming fixed point}, obtained for $\tau_{_R}=\infty$,
  where all the $L_n$ behave as $\tau^{-1}$ (i.e., all the $g_n$ go to
  $-1$), with fixed ratios among them. In particular, one has
  $L_1/L_0\to -\tfrac{1}{2}$, and $P_{_L}/P_{_T}$ decreases
  like $\tau^{-2}$.
\item An {\sl interacting fixed point}, obtained for
  $\tau_{_R}\sim \epsilon^{-1/4}$, where $g_0\to -\tfrac{4}{3}$, $g_1\to
  -2$. At this fixed point, the system has a locally isotropic
  particle distribution, and an energy density that decreases like
  $\tau^{-4/3}$.
\end{itemize}
Note that in a scaleless system where the collision rate scales as
$\epsilon^{-1/4}$ (this is the conformal case mentioned above), the
ratio $\tau/\tau_{_R}$ always goes to infinity as $\tau\to \infty$,
and the system therefore always converges to the interacting fixed
point. The figure \ref{fig:kinattract} shows the evolution of the
first three $g_n$'s for an ensemble of initial conditions, as a
function of $\tau/\tau_{_R}$. After a short transient evolution, all
these evolutions coalesce on a universal ``attractor'' (indicated by
the thick black curve), on which the exponents are the free streaming
ones or the interacting ones depending on the value of the ratio
$\tau/\tau_{_R}$.

\section{Hydrodynamical evolution; Late stages}

The next, and in fact main, stage of the bulk evolution of the system
formed in the collision of two heavy ions is a rather long period of
relativistic hydrodynamical expansion. Hydrodynamics
\cite{Teaney:2009qa,Romatschke:2009im,Romatschke:2017ejr,Florkowski:2017olj,Gale:2013da}
is a coarse grained description whose starting point is the local
conservation laws, for energy and momentum and for any other conserved
charge (such as baryon number),
\begin{align}
  \partial_\mu T^{\mu \nu}=0,\quad
  \partial_\mu J^\mu_{_B}=0,\quad\cdots
\end{align}
However, these equations are too general to constrain uniquely the
evolution of the system. For this, it is necessary to express the
energy-momentum tensor and the other conserved currents in terms of a
small number of quantities.

\paragraph {Perfect fluid} The simplest case is that of a perfect,
i.e., non dissipative, fluid. In this case, the energy-momentum tensor
depends only on the local energy density $\epsilon$, pressure $p$ and fluid
$4$-velocity vector $u^\mu$,
\begin{align}
  T^{\mu\nu}&\empile{=}\over{{\rm perfect}}(\epsilon+p) u^\mu u^\nu -p\,g^{\mu\nu}\nonumber\\
            &= \epsilon\,u^\mu u^\nu+ p\,\Delta^{\mu\nu}\quad \mbox{with\ \ }\Delta^{\mu\nu}\equiv u^\mu u^\nu-g^{\mu\nu},
              \label{eq:idealT}
\end{align}
where $g^{\mu\nu}$ is the Minkowski metric tensor with $(+,---)$
signature. The tensor $\Delta^{\mu\nu}$ is a projector on the local
rest frame of the fluid, $\Delta^{\mu\nu} u_\nu=0$, in terms of which
one may define time and spatial derivatives,
$D\equiv u^\mu \partial_\mu$ and
$\nabla^\mu \equiv \Delta^{\mu\nu}\partial_\nu$, in the fluid rest
frame. The equations of ideal hydrodynamics therefore read
\begin{align}
  D\epsilon = -(\epsilon+p)\nabla_\mu u^\mu,\quad D u^\mu = -(\epsilon+p)^{-1}\,\nabla^\mu p. 
\end{align}
These equations are the relativistic analogue of Euler fluid equations.
The first equation indicates that the local variation of the energy
density is proportional to the variation of the volume of fluid cells,
since $dV/V = dt \, \nabla_\mu u^\mu$. The second equation relates the
acceleration of the fluid to the gradient of its pressure. Note that
the first equation implies that entropy is conserved (indeed this
equation is equivalent to $d(\epsilon V)+pdV=TdS=0$).

\paragraph{Boost invariant ideal flow} In heavy ion collisions at
ultrarelativistic energies, the longitudinal momenta of the produced
particles are typically much larger than their transverse
momenta. This leads to a strong correlation between the longitudinal
momentum of a particle and its longitudinal position in coordinate
space. More precisely, one has
\begin{align}
  y\equiv \frac{1}{2}\ln\Big(\frac{p^0+p^3}{p^0-p^3}\Big)\approx
  \eta_s \equiv \frac{1}{2}\ln\Big(\frac{x^0+x^3}{x^0-x^3}\Big). 
\end{align}
If the energy density of the fluid at some initial proper time
$\tau_0$ is independent of the spatial rapidity $\eta_s$ (this is
approximately the case in the Color Glass Condensate framework, since
the rapidity dependence comes from the JIMWLK evolution of the
distributions of color sources, which is significant only on scales
$\delta\eta_s\sim \alpha_s^{-1}$), the subsequent hydrodynamical
evolution of the fluid is boost invariant. In this case, the evolution
of a perfect fluid is governed by a single equation,
\begin{align}
  \frac{d\epsilon}{d\tau}=-\frac{\epsilon+p}{\tau}.
  \label{eq:BIhydro}
\end{align}
(If the fluid is not locally isotropic, the pressure in the right hand
side should be replaced by the longitudinal pressure.) In a conformal
theory (i.e., a theory with only massless particles and no running
coupling), the energy-momentum tensor is traceless and one has
$p=\epsilon/3$ in equilibrium. Therefore, this leads to
\begin{align}
  \epsilon, p &\sim \tau^{-4/3},\nonumber\\
  T&\sim \epsilon^{1/4}\sim \tau^{-1/3},\nonumber\\
  s&\sim T^3 \sim \tau ,\nonumber\\
  sV&\sim s\tau\sim \mbox{const},
\end{align}
where $s$ is the entropy density. Recall that the assumption of boost
invariance, and therefore these scaling laws, are only true as long as
the longitudinal expansion dominates over the transverse one (this is
hidden in the assumption that $V\sim \tau$), and are therefore
expected to change when the proper time becomes comparable to the
diameter of the colliding nuclei. Note also an important fact, equally
valid for solving Eq.~(\ref{eq:BIhydro}) or the general hydrodynamical
equations: in order to close the system of equations and obtain a
solution, it is necessary to use an {\sl equation of state} that
relates for instance the pressure to the energy density.

\paragraph{Viscous corrections}
In order to go beyond the simple description in terms of a perfect
fluid, one should first alter Eq.~(\ref{eq:idealT}) by writing
\begin{align}
  T^{\mu\nu}=T^{\mu\nu}_{\rm perfect}+\pi^{\mu\nu}+\Pi\,\Delta^{\mu\nu},
\end{align}
where we have split the deviation from the perfect fluid into a
traceless tensor $\pi^{\mu\nu}$ and a term $\Pi \Delta^{\mu\nu}$ that
has a non-zero trace. The equations of motion are
$\partial_\mu T^{\mu\nu}=0$, combined with an equation of state and
constituent equations that express $\pi^{\mu\nu}$ and $\Pi$ in terms
of gradients. In a system which is not too far from local equilibrium,
these expressions may be expanded in powers of the
gradients\footnote{This expansion should be regarded as rather formal,
  as it may not lead to a convergent series \cite{Heller:2016rtz}.},
and at lowest order one may write
\begin{align}
  \pi^{\mu\nu}&=-\eta \,\sigma^{\mu\nu}\quad\mbox{with\ \ } \sigma^{\mu\nu}\equiv \nabla^\mu u^\nu+\nabla^\nu u^\mu-\frac{2}{3}\Delta^{\mu\nu}\,(\nabla_\rho u^\rho),\nonumber\\
  \Pi &= -\zeta\,(\nabla_\rho u^\rho).
        \label{eq:1storderexp}
\end{align}
The coefficients $\eta$ and $\zeta$ (respectively, the shear and bulk
viscosities) describe how the stress tensor responds to a small
gradient of the fluid velocity. The resulting hydrodynamical equations
are the relativistic analogue of the Navier-Stokes equations.

When applied to a boost invariant system, the resulting hydrodynamical
equations lead to
\begin{align}
  \frac{d\epsilon}{d\tau}=-\frac{\epsilon+p-\tfrac{4}{3}\tfrac{\eta}{\tau}}{\tau},
  \label{eq:BI1storder}
\end{align}
resulting in a slower decrease of the energy density compared to the
case of a perfect fluid. This equation indicates that the first order
gradient expansion it was obtained from is legitimate as long as
$\eta\ll \tau(\epsilon+p)$. Using the thermodynamic relation
$\epsilon+p=sT$, this condition can be turned into
\begin{align}
\frac{\eta}{s}\ll \tau T,
\end{align}
where the left hand side of the inequality is a local property of the
fluid while the right hand side is a property of the flow itself. From
kinetic theory, the ratio $\eta/s$ may be estimated to be of the order
of $\lambda T$ where $\lambda$ is the mean free path. Thus, the
inequality also reads $\lambda \ll \tau$, implying that the system
cannot be described by hydrodynamics at times that are smaller than
the time between two successive scatterings of a particle. Conversely,
the hydrodynamical description improves as $\tau$ increases. In a
scale invariant system, we may estimate the relative magnitude of the
first viscous correction in the right hand side of
Eq.~(\ref{eq:BI1storder}) as follows,
\begin{align}
  \frac{\eta}{\tau(\epsilon+p)}\sim \frac{T^3}{\tau T^4}\sim \tau^{-2/3},
\end{align}
where in the last step we use the behavior of $T$ from ideal
hydrodynamics.

\paragraph{Causality, Second order hydrodynamics}
The first order gradient expansion in Eq.~(\ref{eq:1storderexp}) leads
to some pathologies in a relativistic context, because the correction
to the stress tensor follows instantaneously any modification to the
velocity field. This causality violating behavior eventually leads to
numerical instabilities when solving the corresponding hydrodynamical
equations. A possible practical strategy to fix this problem is to
modify Eqs.~(\ref{eq:1storderexp}) into relaxation equations, in order
to introduce a delay between changes of the gradients and the
resulting variation of the stress tensor. There is no unique way of
doing this, but on timescales longer that the ad-hoc relaxation time,
all these models lead to identical physical predictions.

At a more fundamental level, the modifications introduced by turning
Eqs.~(\ref{eq:1storderexp}) into relaxation equations can be motivated
from the study of second order terms in the gradient expansion
\cite{Israel:1979wp,Denicol:2010xn}. For instance, the second gradient
order in $\pi^{\mu\nu}$ contains a term of the form
$\eta \tau_\pi D \sigma^{\mu\nu}$, where $\tau_\pi$ has the dimension
of a time. If we take this second order expansion as is, it displays
similar causality issues as the first order one. These problems may be
avoided by replacing $\sigma^{\mu\nu}$ in the time derivatives that
appear at second order by the first order relationship between
$\sigma^{\mu\nu}$ and $\pi^{\mu\nu}$, thereby producing a term in
$-\tau_\pi D \pi^{\mu\nu}$. By doing this, the second order
constituent relation becomes a differential equation in time, with a
relaxation time $\tau_\pi$. The benefit of this point of view compared
to the more phenomenological one described before is that it allows to
relate the relaxation time to the underlying microscopic theory in an
unambiguous fashion.

\paragraph{Equation of state}
A key ingredient in order to turn the equations of hydrodynamics into
a closed set of equations is an equation of state that relates for
instance the entropy density to the temperature, or equivalently the
pressure and the energy density. 
\begin{figure}[htbp]
  \centering
  \includegraphics[width=8cm]{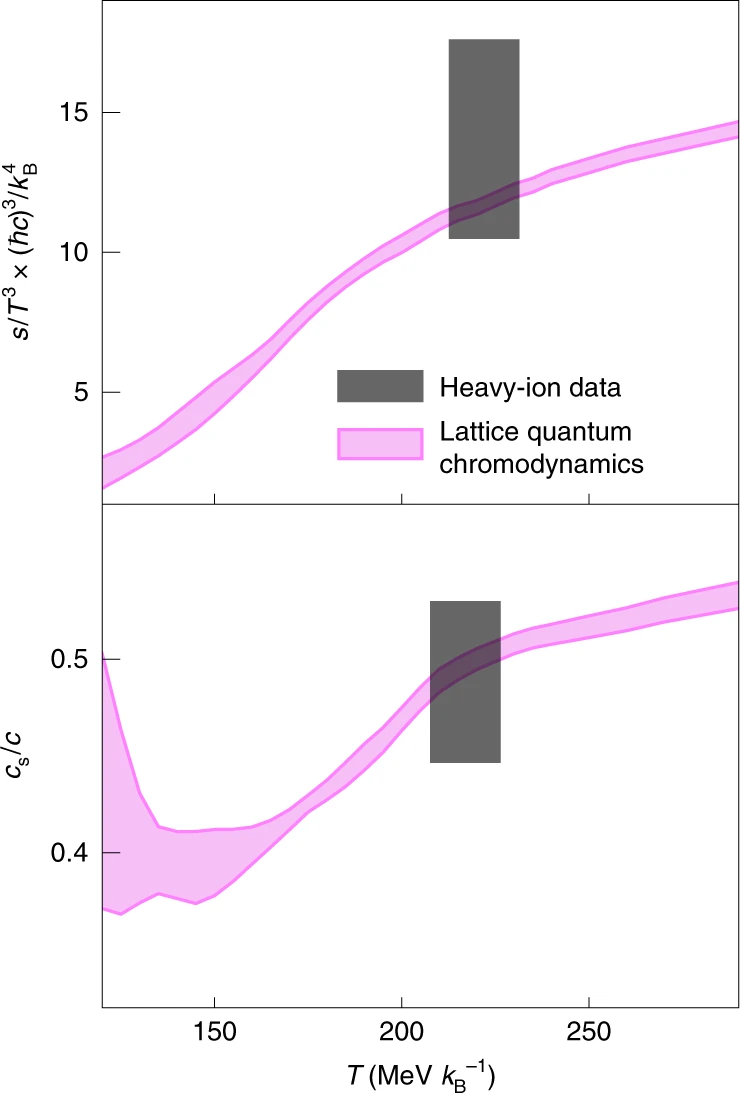}
  \caption{\label{fig:eos}Top: equation of state obtained from lattice QCD
    at zero baryon chemical potential. Bottom: speed of sound as a
    function of temperature. Boxes: extraction from LHC data (see
    text). From \cite{Gardim:2019xjs}.}
\end{figure}
Perturbation theory, improved by the resummation of hard thermal
loops\cite{Braaten:1989mz}, allows to obtain robust results for temperatures only a couple
of times above the critical temperature \cite{Blaizot:1999ip,Blaizot:1999ap,Blaizot:2000fc,Blaizot:2003iq,Andersen:1999sf,Andersen:2000yj,Andersen:2011sf,Haque:2014rua}, but becomes less and less
reliable as $T_c$ is approached from above. A non-perturbative
first-principles alternative is lattice QCD
\cite{Philipsen:2012nu}. At null baryon chemical potential, the
advances in computing hardware and algorithms are by now allowing to
perform unquenched (i.e., with virtual quark loops) simulations with
realistic quark masses. At low temperatures compared to the
deconfinement transition, these computations may be continued by an
equation of state based on a gas of hadron resonances.

This approach works well only at zero baryon chemical potential, where
it has shown that the deconfinement transition is a continuous
crossover rather than a discontinuity for realistic quark masses. At
$\mu_B>0$, the determinant resulting from integrating out the fermion
fields is complex valued, leading to a {\sl sign problem} that
precludes any direct approach based on a Monte-Carlo sampling. When
$\mu_B/T$ is small enough, various workarounds are possible:
reweighting, analytic continuation from calculations at imaginary
$\mu_B$ (for which there is no sign problem), that allow to make an
incursion into the territory of positive chemical potentials
\cite{Bazavov:2017dus,Borsanyi:2019hsj}. It is expected that the
crossover at small $\mu_B$ becomes a first order phase transition at
larger $\mu_B$, the beginning of the transition line being a second
order critical point (the red dot in Figure \ref{fig:phase-diagram}).
However, the quest in lattice simulations for such a second order
critical point has remained rather inconclusive until now.

There has been a recent attempt to extract directly some information
about the equation of state from LHC heavy ion data
\cite{Gardim:2019xjs}. In this work, the authors used a hydrodynamical
simulation to estimate the effective temperature $T_{\rm eff}$ and
effective volume $V_{\rm eff}$ of a hypothetical homogeneous fluid
that would evolve into a system with the same energy and entropy as
the QGP at the time where the particles decouple (see the discussion
of {\sl freeze-out} later in this section). They observed that this
effective temperature is related to the mean transverse momentum
$\big<p_\perp\big>$ of the final state particles by
$T_{\rm eff}\approx 3.07\,\big<p_\perp\big>$ (with a proportionality
coefficient roughly independent of the equation of state and transport
coefficients). Given ALICE data in the bin of centrality $[0,5]\%$,
one gets $T_{\rm eff}=222\pm 9$~MeV.  The total entropy is inferred
from the number of produced charged particles,
$S\approx 6.7\,N_{\rm ch}$, while the effective volume $V_{\rm eff}$
also comes from the hydrodynamical simulation, giving an entropy
density $s=20\pm 5$~fm${}^{-3}$, and $s/T_{\rm eff}^3=14\pm 3.5$, in
agreement with the lattice equation of state. In particular, this is
much higher than the value $\sim 3$-$4$ expected in the confined
phase, suggesting a large number of degrees of freedom consistent with
deconfinement. By repeating this analysis at two collision energies
($2.76$ and $5.02$~TeV), one may also estimate the speed of sound via
\begin{align*}
  c_s^2(T_{\rm eff})=\frac{sdT}{Tds}\Big|_{T_{\rm eff}}
  =
  \frac{d\ln\big<p_\perp\big>}{d\ln (dN_{\rm ch}/d\eta)}
  =0.24\pm 0.04,
\end{align*}
again in agreement with lattice computations.

\paragraph{Transport coefficients}
Among the transport coefficients that enter in hydrodynamics, the one
that has received most interest is the shear viscosity $\eta$. As we
mentioned earlier, the ratio $\eta/s$ is the ratio of the mean free
path to the quantum wavelength of the particles. This allows to make
some simple estimates in several limits. Firstly, in the perturbative
limit (weak coupling, and low enough particle density), this leads to \cite{Jeon:1994if,Jeon:1995zm,Arnold:2000dr,Arnold:2003zc}
\begin{align}
\frac{\eta}{s}\sim \frac{1}{\alpha_s^2 \ln(\alpha_s^{-1})} \gg 1. 
\end{align}
(Although we do not write the prefactor here, it can be determined in
the weak coupling limit.)  Another limit is that of a strongly coupled
plasma. Although this limit is not accessible in QCD, this calculation
is possible in a supersymmetric cousin of QCD thanks to the AdS/CFT
correspondence, leading the following result \cite{Policastro:2001yc}
\begin{align}
\frac{\eta}{s}=\frac{1}{4\pi}. 
\end{align}
Note that such a constant value, independent of the coupling, is
consistent with the fact that quantum mechanics prevents this ratio
from becoming arbitrarily small since the quantum wavelength is a
lower value for the mean free path (but such an argument does not give
the value of the constant ratio one would reach).

Out of equilibrium, there is another interesting situation, where the
coupling constant is weak but the gluon occupation number is large, possibly
as large as $\alpha_s^{-1}$. In this case, the scattering rate should
contain a factor $f(1+f)$ where $f$ is the occupation number of the
scattering centers (when the occupation number is small, only a factor
$f$ is necessary, leading to the usual formula
$\lambda^{-1}\sim n \sigma$). Although the scattering cross-section is
proportional to $\alpha_s^2$ (up to logarithms), the factor $f(1+f)$
leads to a mean free path that does not contain any power of the
coupling constant. This situation of weak coupling but high density,
which is relevant in the very early stages of a heavy ion collision,
illustrates the fact that a strong coupling is not the only
possibility for having a small ratio $\eta/s$: more generally, the
system should be {\sl strongly interacting} for this to be true.

Outside of the above limiting cases, one may also consider lattice
QCD.  Unlike the equation of state, the transport coefficients remain
very difficult to even estimate in lattice QCD. Thanks to Green-Kubo's
formulas, transport coefficients may be expressed in terms of the
Fourier transform of a retarded current-current (the current should be
the one that couples to the quantity whose transport one is interested
in, e.g., a charge current for an electrical conductivity) correlation
function at zero momentum,
\begin{align}
\sigma \propto \lim_{\omega\to 0} \frac{\rho(\omega,\k=0)}{\omega}
\end{align}
This formula expresses the transport coefficient in terms of the slope
at zero energy of the corresponding spectral function
$\rho(\omega,\k)$. However, on the lattice, one has only a direct
access to the imaginary time version of these correlation functions.
This imaginary time correlator admits a spectral representation
involving the relevant spectral function,
\begin{align}
  G(\tau,\k)=\int d\omega\; K_{_T}(\tau,\omega)\,\rho(\omega,\k),
  \label{eq:spectral-rep}
\end{align}
where $K_{_T}(\tau,\omega)$ is a known temperature-dependent kernel.
In a lattice approach, the left hand side of this equation would be
obtained (with statistical errors, and a finite number of values of
$\tau$ and $\k$) from numerical simulations, and one would then try to
invert this integral relationship to obtain the spectral function
$\rho(\omega,\k)$. The difficulty with this is that it is a very
ill-posed problem when the left hand side is only imperfectly known,
implying that a direct inversion is unfeasible. Attempts at
constraining the spectral function in this way have been made by using
Bayes theorem in order to find the most likely spectral function
compatible with Eq.~(\ref{eq:spectral-rep}) and with a set of prior
assumptions about its shape (a minimal assumption would be that it is
positive) \cite{Asakawa:2000tr,Yamazaki:2001er,Sasaki:2005ap,Morningstar:2001je,Burnier:2014gta}.

More recently, a hybrid approach was proposed
\cite{Christiansen:2014ypa}, that combines a skeleton loop expansion
for the correlator $\big<[\pi_{ij}(x),\pi_{ij}(0)]\big>$ that enters
in Green-Kubo's formula, truncated at two loops, and a
non-perturbative input for the couplings and the propagator that
enters in this expansion obtained from the Euclidean functional
renormalization group \cite{Berges:2000ew}.
\begin{figure}[htbp]
  \centering
  \includegraphics[width=8cm]{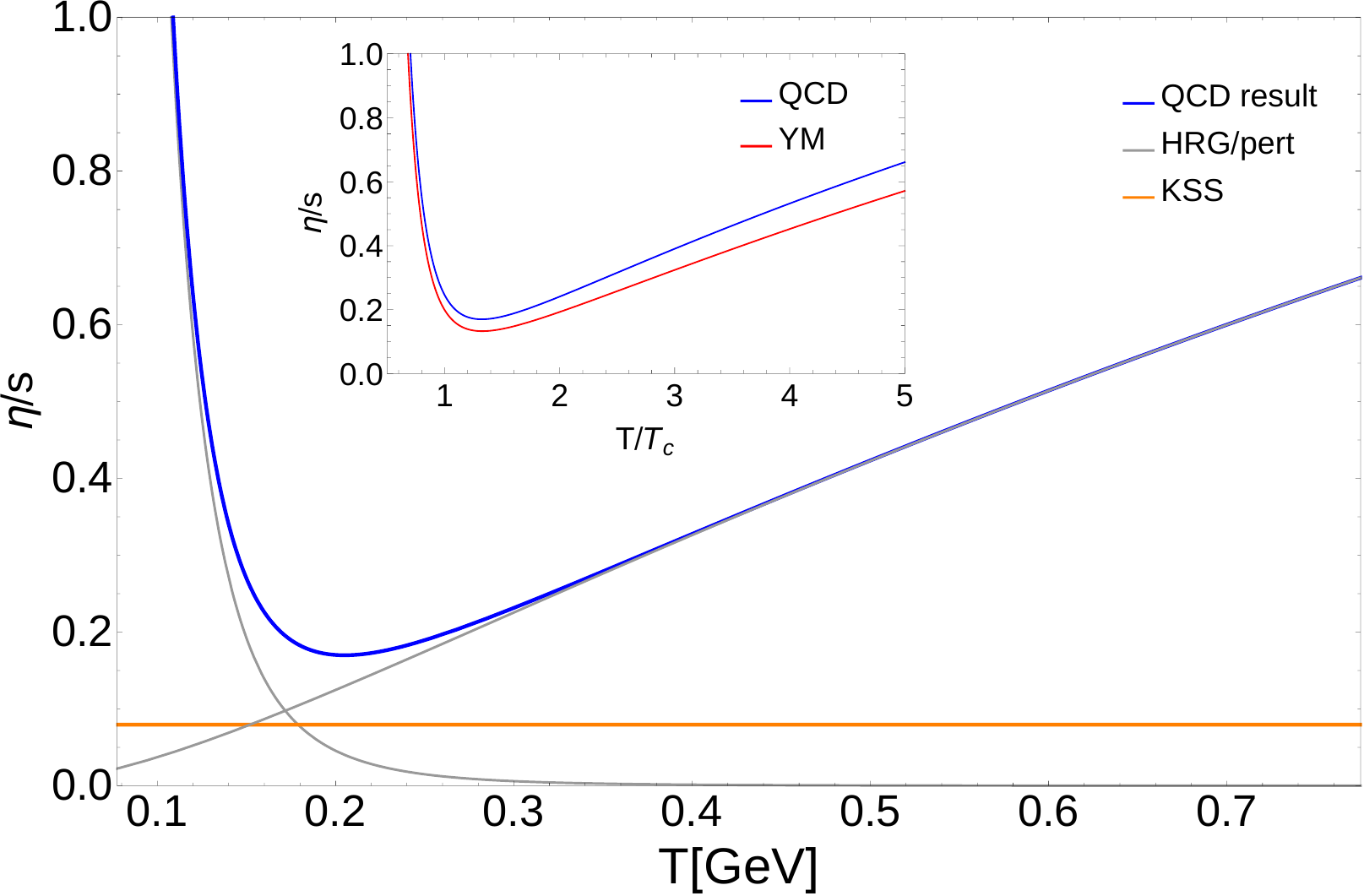}
  \caption{\label{fig:viscosity}Theoretical evaluations of the viscosity to entropy ratio as a function of temperature. From \cite{Christiansen:2014ypa}.}
\end{figure}
The results obtained with this approach are shown in
Fig. \ref{fig:viscosity}, for a pure Yang-Mills theory (the result
shown for QCD with quarks is an estimate based on the pure YM result,
not an ab initio calculation). Interestingly, this calculation
suggests that $\eta/s $ has a rather pronounced temperature
dependence, with a minimum around $1.25\,T_c$ (the location of this
minimum in units of the critical temperature is almost identical in
Yang-Mills theory and in QCD). Moreover, the value of $\eta/s$ at this
minimum is only slightly above the value $1/(4\pi)$ obtained in the
strong coupling limit in the AdS/CFT approach for $N=4$ super-Yang-Mills theory.

\paragraph{Freeze-out}
Since at the end of the day experiments observe particles, it is
necessary to convert the objects (the fluid energy density and its
velocity) evolved by the hydrodynamical equations into distributions
of particles. In fact, the need for a description in terms of
particles arises much earlier than the time at which particles are
detected because the density decreases as the system expands, and
therefore the mean free path increases, leading to a situation where
the conditions of applicability of hydrodynamics are not met anymore.

\begin{figure}[htbp]
  \centering
  \includegraphics[width=8cm]{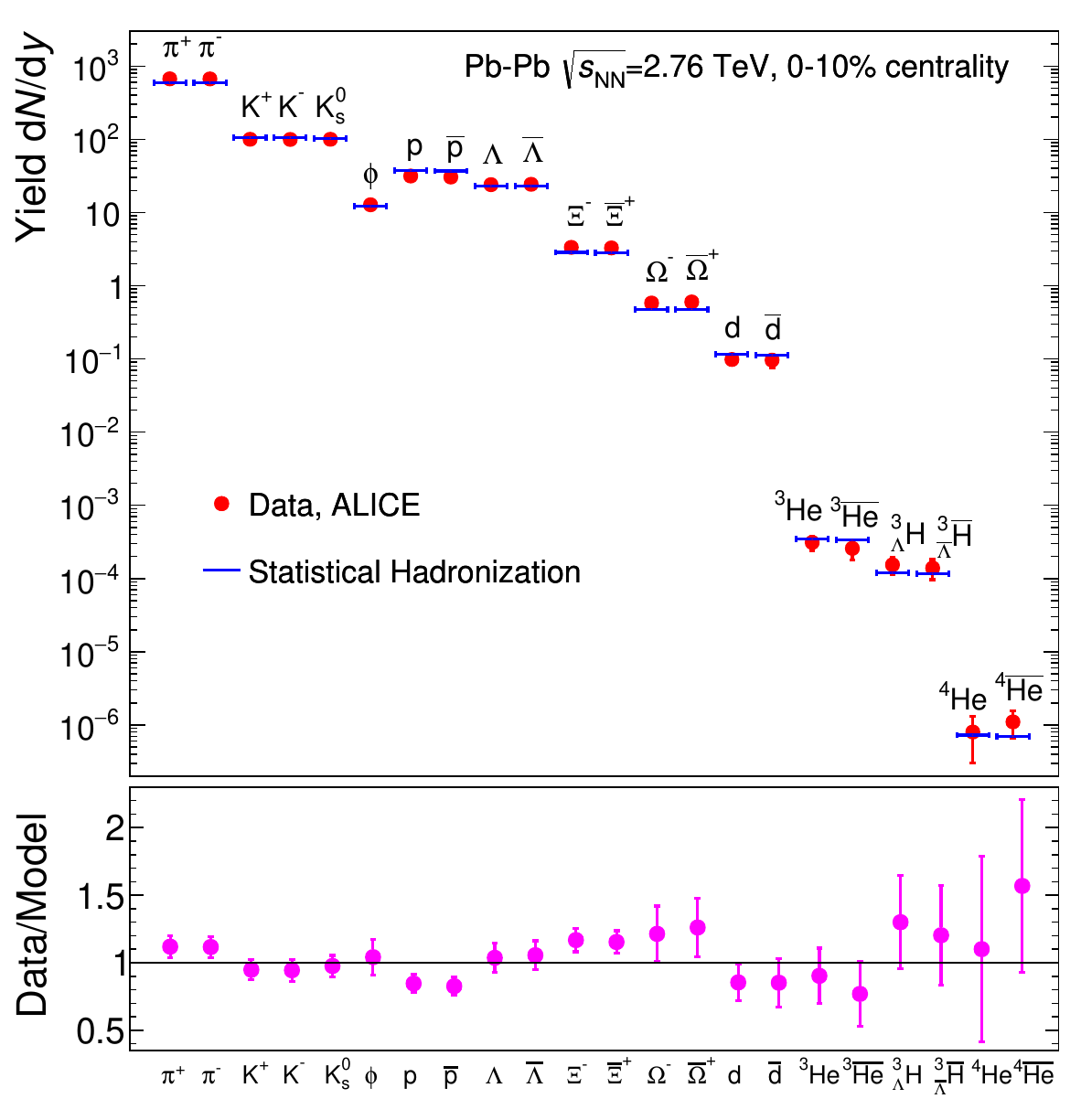}
  \includegraphics[width=9cm]{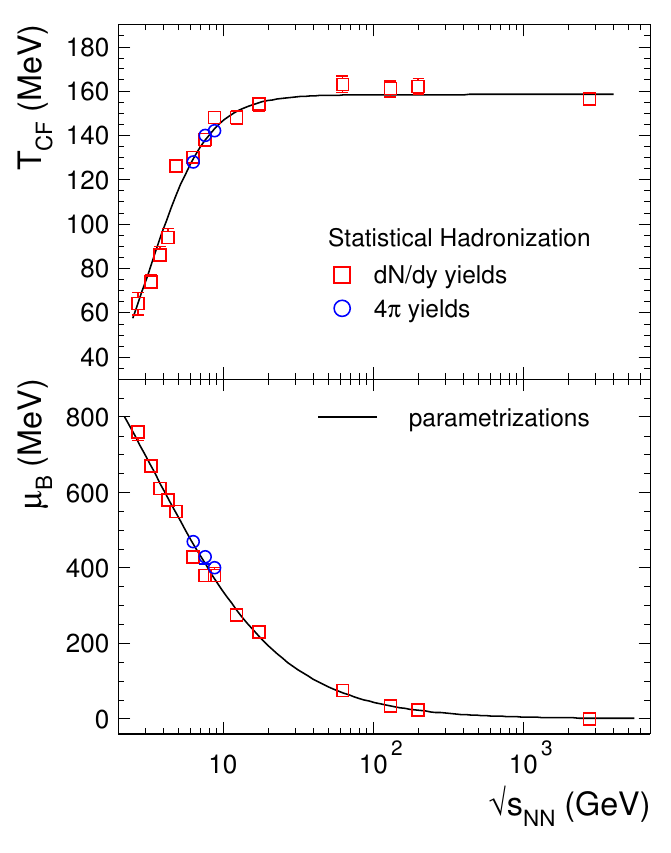}
  \caption{\label{fig:chem_freeze}Top: measured hadron yields compared
    to the statistical hadronization model output. Bottom: temperature
    and chemical potential at chemical freeze-out, extracted from
    collisions at various energies. From \cite{Andronic:2017pug}.}
\end{figure}

On should distinguish between a {\sl chemical freeze-out}, at which
the inelastic collisions become rare (after that, the chemical
composition of the system is frozen), and a {\sl kinetic freeze-out}
where the elastic collisions also stop and the momentum distributions
freeze. Experimentally, the temperature and chemical potential at the
chemical freeze-out are well constrained from the ratios of abundances
of various species of particles, as shown in the figure
\ref{fig:chem_freeze}
\cite{Andronic:2003zv,Andronic:2006ky,Stachel:2013zma,Andronic:2017pug}
(see also \cite{Fries:2003vb,Fries:2003kq} for a somewhat related
approach to hadron formation).
In the simplest version of this model, one assumes that all particle
distributions are the equilibrium ones at a common temperature and
chemical potential. A volume is also necessary to obtain absolute
yields, but this parameter drops out in ratios of the yields of
various particle species.

The conversion from a fluid to free particles should be done at the
kinetic freeze-out, by choosing a locally space-like ``surface''
$\Sigma$, and by using the {\sl Cooper-Frye formula},
\begin{align}
  E_\p \frac{dN}{d^3\p}=\frac{1}{(2\pi)^3}\int_\Sigma d^3S_\mu\; P^\mu\, f(P\cdot u),
\end{align}
where $f$ is the local distribution function. For a fluid in local
thermal equilibrium, $f$ is the Bose-Einstein of the Fermi-Dirac
distribution, evaluated at the local fluid temperature. However, when
the fluid is viscous, there should also be deviations from local
thermal distributions,
\begin{align}
  f(p)=f_{\rm eq}(p)+\delta f(p). 
\end{align}
The form of the deviation $\delta f(p)$ is related to the transport
coefficients (such as the shear viscosity) and therefore depends on
the microscopic interactions in the fluid.

In this approach, the freeze-out conditions (temperature and chemical
potential) are a priori free parameters that may be adjusted to best
fit the spectra of produced particles. A more sophisticated
alternative would be to convert the fluid into particles at an earlier
time, and continue the evolution with kinetic equations
\cite{Hirano:2004en,Hirano:2005xf,Monnai:2009ad}. In this fancier approach, the
freeze-out temperature is controlled by the values of the various
cross-sections used in the kinetic description (this would even allow
different species to decouple at different times), and is no longer an
ad hoc input of the model.

\paragraph{Flow anisotropies}
Experimentally, several predictions of hydrodynamical models may be
compared with data. One of them is the transverse momentum spectra of
the produced particles, that are sensitive to the temperature at which
the freeze-out occurs (and to whether the particle distributions are
the equilibrium ones or not). 

Another main class of observables directly related to the
hydrodynamical expansion of the quark gluon plasma consists in
measuring angular correlations among the detected particles
\cite{Ollitrault:1992bk,Dusling:2007gi,Song:2007ux,Luzum:2008cw,Voloshin:2008dg,Alver:2010gr,Alver:2010dn,Teaney:2010vd,Muller:2012zq,Heinz:2013th}. The
quantities used to express these correlations are the so-called $v_n$,
defined as the Fourier coefficients of the azimuthal distribution of
particles. These flow coefficients may be measured as a function of
transverse momentum, of the centrality of the collisions, of the
species of particles. The hydrodynamical expansion provides a
one-to-one mapping between the spatial anisotropy of the initial
distribution of energy density (and possibly its initial flow) and the
final momentum anisotropy, that may be understood as an effect of
pressure gradients (the fluid is accelerated in the direction of the
pressure gradient). Moreover, the conversion of spatial
inhomogeneities into momentum space anisotropies depends quite
sensitively on the transport coefficients, mostly the shear viscosity:
if the ratio $\eta/s$ is too large, this conversion is very
ineffective and hydrodynamics cannot explain the rather large values
of the measured $v_n$s.
\begin{figure}[htbp]
  \centering
  \includegraphics[width=8cm]{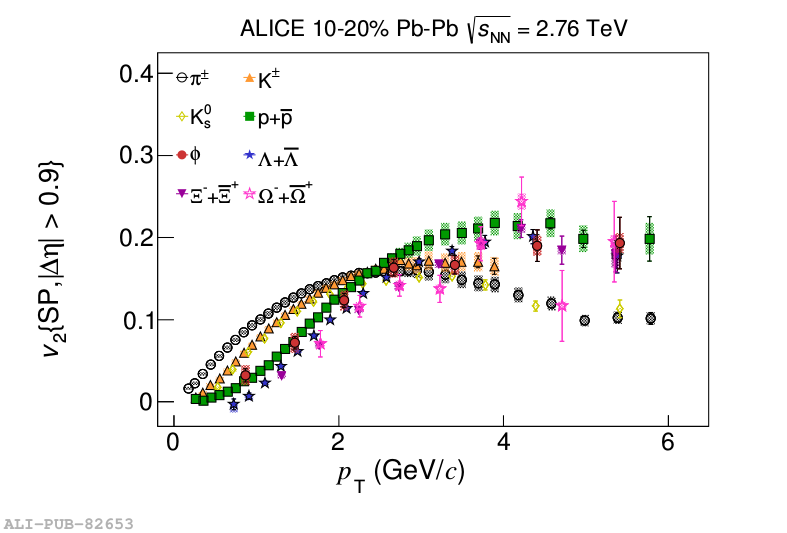}
  \caption{\label{fig:v2}Second Fourier coefficient of the azimuthal distribution for identified particles. From \cite{Abelev:2014pua}.
  }
\end{figure}

\begin{figure}[htbp]
  \centering
  \includegraphics[width=12cm]{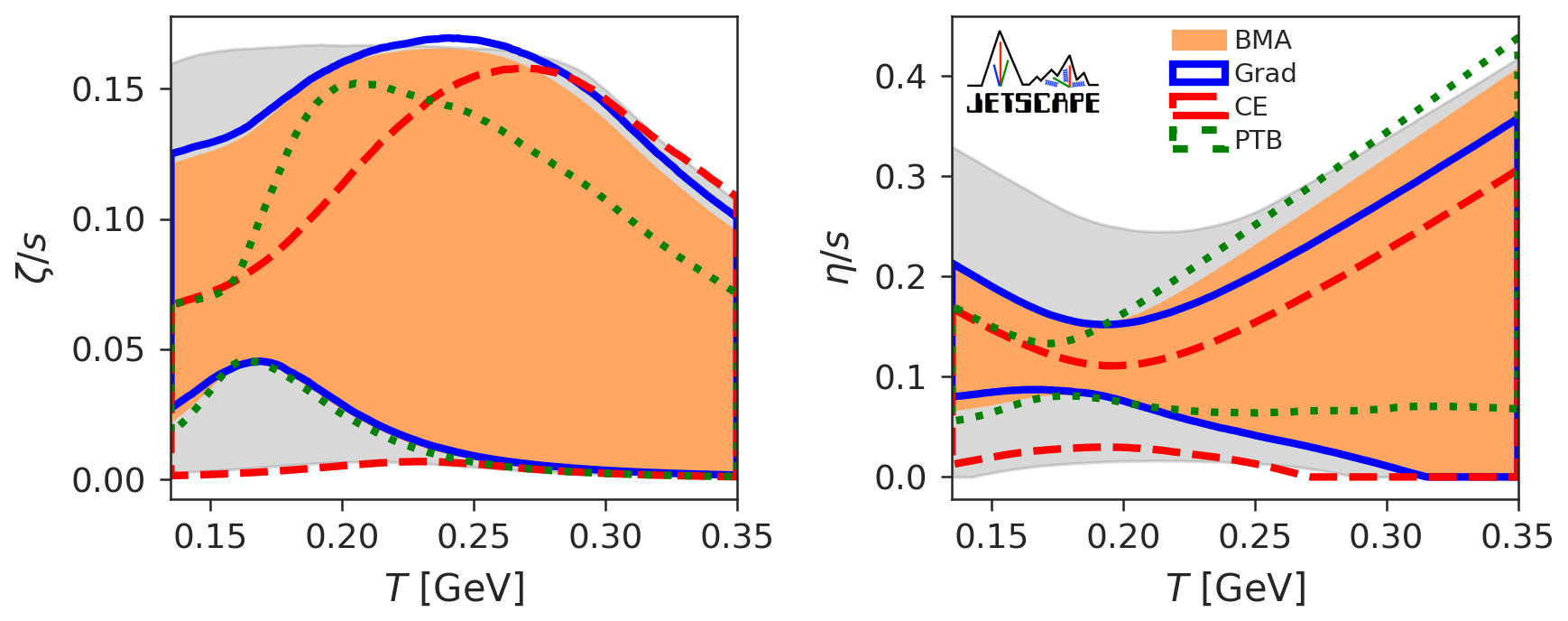}
  \caption{\label{fig:visco}Bayesian extraction of the shear and bulk
    viscosities from Pb-Pb collisions at $2.76$~TeV. Gray: 90\%
    confidence interval for the prior. Red, Green, Blue, Orange: 90\%
    confidence intervals for the posteriors in three models, and their
    average. From \cite{JETSCAPE:2020shq}.}
\end{figure}

Recently, the JETSCAPE collaboration has used a Bayesian approach in
order to extract the most likely values of the shear and bulk
viscosities as a function of temperature \cite{JETSCAPE:2020shq,JETSCAPE:2020mzn}, as
shown in Figure \ref{fig:visco}.
One can see that the constraints provided by the data restricts the
range of values of the viscosities, down to fairly low values after
normalization by the entropy density, compared to the prior
distribution used in this analysis. Although these results are very
promising, one should keep in mind that, besides the transport
coefficients one wants to extract, the model has other unknowns, especially in the
modelling of the initial condition and in the details of the
implementation of the freeze-out.

\paragraph{Hydrodynamics from kinetic theory}
In local thermal equilibrium, the hydrodynamical description has six
unknowns (the local baryon density $n_{_B}$, energy density $\epsilon$
and pressure $P$, plus the three independent components of the fluid
velocity $u^\mu$), and the conservation equations for the baryonic
current and the energy-momentum tensor provide five equations. The
system is closed by the equation of state that relates
$n_{_B}, \epsilon$ and $P$.

Away from this ideal situation, there are nine additional unknowns
($\Pi$, five independent components of $\pi^{\mu\nu}$, and three
components of the mismatch between the energy flow vector and the
baryon number flow vector). In addition to the equation of state, we
thus need $14$ equations to close the system, that are made by the
conservation equations ($5$ equations), and by $9$ constitutive
equations relating the stresses to the gradients. But these equations
can also be obtained from the underlying microscopic dynamics, in the
form of a Boltzmann equation $p^\mu \partial_\mu f = C_\p[f]$. Indeed,
by weighting the Boltzmann equation by $1$, $p^\nu$, $p^\nu p^\lambda$
and integrating over $\p$, we get
\begin{align}
  &\partial_\mu \int_\p p^\mu\;f= \int_\p C_\p[f]=0,\\
  &\partial_\mu \int_\p p^\mu p^\nu \;f =\int_\p p^\nu \,C_\p[f]=0,\\
  &\partial_\mu \int_\p p^\mu p^\nu p^\lambda \; f
    =
    \int_\p p^\nu p^\lambda\,C_\p[f].
\end{align}
The right hand side of the first two equations is identically zero
given the symmetry properties of the collision integral, for any
distribution $f$. This set of $5$ equations is in fact the five
conservation equations. The last equation forms a set of $10$
independent equations (given the symmetry of $p^\nu p^\lambda$). Note
that by summing over $\nu=\lambda$, one recovers the conservation
equation of particle number. Therefore, this set of equations contains
only $9$ novel equations, i.e., precisely the number needed to close
our macroscopic description. This approach, supplemented by an
expansion of $f$ around the equilibrium distribution $f_0$, is known
as Grad's $14$-moment method. It allows to relate the hydrodynamical
description (in particular the transport coefficients \cite{Denicol:2010xn,Betz:2010cx,Denicol:2012cn,Denicol:2012es,Muronga:2001zk,Muronga:2003ta}) to the
underlying microscopic interactions encoded in the collision integral.

\paragraph{Hydrodynamics far from equilibrium}
For a long time, it has been thought that the hydrodynamical
description is a near-equilibrium effective description (this point of
view stems in part from the fact that one may obtain hydrodynamics
from a truncated gradient expansion). However, it was realized
recently that this gradient expansion may be a non-convergent series
with a null radius of convergence \cite{Heller:2013fn,Heller:2016rtz}. This
observation suggests to reconsider the conditions of applicability of
the hydrodynamical description (indeed, if the radius of convergence
is zero, it cannot serve as a parameter that defines what ``close
enough to equilibrium'' means). Another, observational, reason for
reassessing the applicability of hydrodynamics as an effective
macroscopic description is that in heavy ion collisions it appears to
perform much better than what one may naively expect by viewing it as
the result of an expansion around equilibrium.

A step towards a better understanding of these questions has been the
discovery (so far, in simple cases, where the dynamics and the flow
are sufficiently symmetric) of attractors towards which hydrodynamical
solutions evolve, even while gradients are still sizeable
\cite{Denicol:2017lxn,Heller:2011ju,Strickland:2018ayk,Strickland:2019jut}. In other
words, these solutions quickly reach a universal behavior independent
of the details of their initial conditions, long before the system is
in a state of isotropic local equilibrium. On these attractors, the
dissipative currents behave in a universal way in terms of the
(possibly still large) gradients, which shed another light on the
constitutive relations, indicating that their validity may not be
conditioned by a gradient expansion.

\paragraph{Hydrodynamics in small systems}
Another pressing question, closely related to the previous point, is
to determine what is the ``minimal size'' of a system that may be
described by hydrodynamics. Indeed, as the system becomes smaller, the
collectivity (that may be quantified by the ratio of the system size
by the mean free path, a gross measure of the number of collisions per
particle) is expected to decrease, leading to a situation less
favorable for the applicability of hydrodynamics.

Experimentally, this question was put forward by the observation that
certain correlation patterns (e.g., the ``ridge'' long range rapidity
correlation observed in the two-particle spectrum) are seen in
nucleus-nucleus collisions \cite{Ray:2011zza}, proton-nucleus
\cite{Abelev:2012ola} and even proton-proton collisions
\cite{Khachatryan:2010gv} (provided one triggers on high multiplicity
final states in the latter cases). In nucleus-nucleus collisions, the
ridge is interpreted as the result of the collective radial motion of
the produced matter
\cite{Voloshin:2003ud,Shuryak:2007fu,Dumitru:2008wn}, i.e., as
flow. Moreover, in this case, there is only a modest contribution to
the correlation provided by the CGC initial condition.

\begin{figure}[htbp]
  \centering
  \includegraphics[width=8cm]{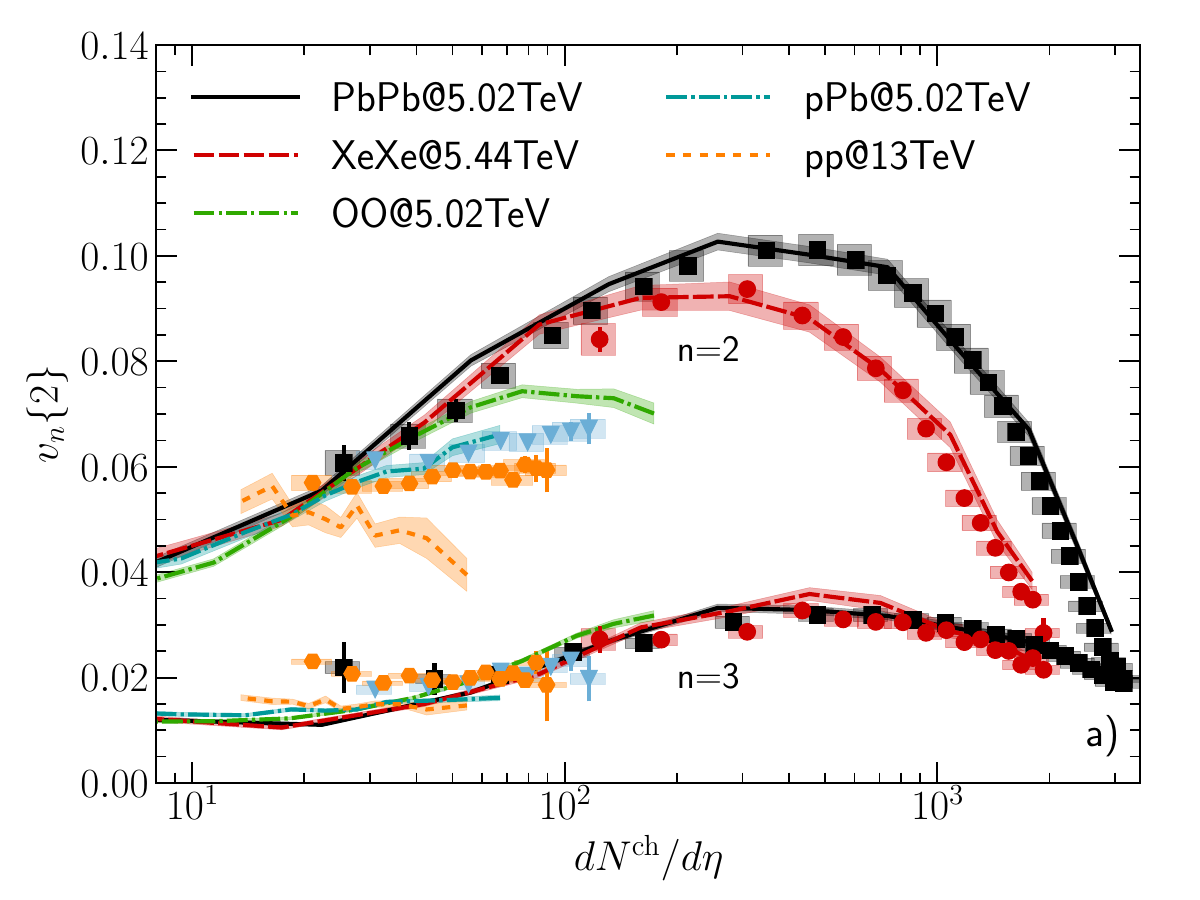}
  \caption{\label{fig:small}Points: $v_2$ and $v_3$ flow coefficients
    in collisions of nuclei of various sizes, down to proton-proton
    collisions (ALICE collaboration). Bands: hydrodynamical model
    predictions. From \cite{Schenke:2020unx}. See also \cite{Acharya:2019vdf,Sievert:2019zjr}.
  }
\end{figure}

In smaller systems (see Figure \ref{fig:small} for some recent data),
the interpretation of the ridge correlation has been the subject of
intense debates between two extreme positions: that the correlation
can be entirely explained by collective flow
\cite{Bozek:2012gr,Bozek:2013uha} (i.e., final state interactions), or
entirely due to initial state correlations
\cite{Dumitru:2010iy,Dusling:2012iga,Dusling:2012wy,Dusling:2013oia}.
By now, it seems that a consistent description of the flow
observables in systems for which $dN_{\rm ch}/d\eta\ge 10$ calls for a
dominance of the final state interactions over the initial state
correlations.  At low multiplicity, the observed flow is probably the
result of the combination of initial state correlations and final
state interactions \cite{Schenke:2019pmk} (it has been observed in a
realistic hydrodynamical model that the final state flow is less
correlated with the initial geometry but more correlated with the
initial momentum anisotropy at low multiplicity, and that this trend
is reversed at high multiplicity).

\section{Tomography and  hard probes}
Until now, we have discussed mostly the bulk properties of the matter
produced in heavy ion collisions, from its formation to its subsequent
evolution. There, the quark-gluon-plasma or its precursor was the core
subject of the study, with flow observables a tool for accessing
(almost) directly some of their properties. Now, we turn to a class of
observables in which this matter plays the role of a substrate that
modifies them compared to a situation where the final state would be
mostly empty. Here, the strategy is to measure these observables both
in nucleus-nucleus collisions and in collisions of smaller projectiles
--such as proton-proton collisions-- where the formation of a
quark-gluon plasma is not expected, and where the observables are
supposedly well understood.  The comparison between the two (with an
appropriate rescaling to account for the different sizes of the two
systems) provides informations on the properties of the matter
produced in heavy ion collisions. A better control is provided by
observables that are perturbative (meaning that they are characterized
by a hard scale, allowing to use QCD in a regime of weak coupling), so
that they can at least be computed with some degree of accuracy in
proton-proton collisions. However, even in this case, one should keep
in mind that their modifications by the surrounding medium may involve
much smaller momentum scales that render a perturbative treatment
difficult or even impossible in nucleus-nucleus collisions.

\paragraph{Basics of QCD radiation}
The most important feature of QCD in this context is the fact that
gluons are massless, and that three flavors of quarks (u,d, and s)
have masses lower or comparable to the non-perturbative scale
$\Lambda_{\rm QCD}$. For this reason, gluon radiation is enhanced for
soft gluons and for gluons emitted collinearly to their parent. For
instance, the emission probability of a gluon off a parton goes like
\begin{align}
  dP\propto \alpha_s \, \frac{d^2\k_\perp}{k_\perp^2 }\frac{dz}{z},
  \label{eq:softrad}
\end{align}
where $\k_\perp$ is the transverse momentum of the gluon (relative to
the direction of the momentum of its parent) and $z$ is the fraction
of its longitudinal momentum relative to that of its parent. The
unwritten prefactor contains a color factor that depends on whether the
emitter is a quark or a gluon. This probability diverges at small
$k_\perp$ and at small $z$. For radiation in vacuum (e.g., in the
final state of a proton-proton collision), the consequences of these
divergences are well understood:
\begin{itemize}
\item After one resums the soft radiation, the probabilities for
  partonic final states with a prescribed number of (massless) gluons
  is actually zero. This is of course largely irrelevant, since
  because of confinement the gluons are not the objects that are
  eventually detected.
\item The observables that match the most closely the concept of
  perturbative parton are the so-called {\sl jets}. Loosely speaking,
  a jet is a collimated beam of particles produced by the splitting of
  a common ancestor (quark or gluon). At a more operational level,
  defining jet cross-sections requires that one defines a procedure
  for deciding when two distinct partonic final states correspond to
  the same configuration of jets. For instance, two graphs that differ
  by a loop correction, or by an extra soft or collinear gluon, should
  contribute to the same jet final state. The jet definition is not
  unique (in particular, because there is no unique way of defining
  ``soft'' or ``collinear''), but they all share a crucial property:
  they lead to finite cross-sections when all mass scales are sent to
  zero (observables that have this property are said to be {\sl
    infrared and collinear safe}) -- in other words, all the logarithms
  that one would get by integrating eq.~(\ref{eq:softrad}) down to
  $k_\perp=0$ or $z=0$  cancel in these observables. The
  theoretical definition of what one means by a jet in turn defines
  the procedure (the so-called ``jet algorithm'') for extracting jet
  cross-sections from experimental data: given a final state made of
  detected particles, the jet algorithm defines how they should be
  clustered into jets.
\item Jet cross-sections are calculable without any non-perturbative
  input, but the price to pay for this is to give up on the idea of
  saying something about individual particles in the final state. It
  is also possible to consider cross-sections for producing a certain
  hadron with a given momentum, but their calculation requires to
  introduce {\sl fragmentation functions}, that may be viewed as the
  inclusive probability that a certain quark or gluon turns into this
  hadron (plus any number of additional particles that we do not care
  about).  This time, the collinear logarithms from
  eq.~(\ref{eq:softrad}) do not cancel and they must be resummed,
  which introduces a scale dependence into the fragmentation
  function. This scale dependence is perturbative (governed by the
  DGLAP evolution equation
  \cite{Gribov:1972ri,Altarelli:1977zs,Dokshitzer:1977sg}, now known
  up to three-loop accuracy
  \cite{Moch:2004pa,Vogt:2004mw,Vermaseren:2005qc}), but the initial
  condition of this evolution is non-perturbative and must be
  extracted from experimental input.  The hadron production
  cross-section is obtained as the convolution of a partonic
  cross-section with the fragmentation function, evaluated at a
  certain scale (called the {\sl factorization scale}). This scale is
  not a physical parameter, but rather a remnant of the truncation of
  the perturbative series at a finite order (the residual scale
  dependence decreases by going to higher loop order, and a fully
  non-perturbative calculation would have no such scale at all).
\end{itemize}

\paragraph{Production of hard probes in nucleus-nucleus collisions}
When extending these ideas to nucleus-nucleus collisions, a generic
assumption is that the {\sl production} of the hard object proceeds in
the same perturbative way as in proton-proton collisions, the only
change being a change of the parton distribution functions that
describe the initial state. In the light of the earlier discussion of
gluon saturation, it is clear that for this to be true the observable
of interest should probe these distributions away from the non-linear
saturation regime, in order to be dominated by processes that probe a
single parton in each projectile. This is a reasonable assumption at
high virtuality $Q^2$, since the non-linear corrections are typically
suppressed as powers of $Q_s^2/Q^2$. Another thing to keep in mind is
that the direct measurements of {\sl nuclear parton distributions} are
scarce. An approximate treatment consists in viewing a nucleus as an
incoherent superposition of protons and neutrons. The parton
distributions of protons are very well known. For the neutrons, one
usually treats the neutron (udd) as the isospin partner of the proton
(uud). This implies for instance that the u-quark distribution in a
neutron is the same as the d-quark distribution in a proton,
etc... But note that the measurement of the structure function $F_2$
in deep inelastic scattering off a proton does not allow to
disentangle the u and d quark distributions. To separate them, one
also needs DIS measurements with deuteron (plus the assumption that
the binding of the deuteron is weak enough compared to the relevant
virtuality scales so that its parton distributions are just the sum of
the proton and neutron ones).

\paragraph{Gluon formation time} The discussion of parton and jet
energy loss requires to introduce the concept of {\sl formation time}
of a radiated gluon. Consider the emission of a gluon of momentum $k$
off a colored particle (quark or gluon) of momentum $p+k$. According
to the uncertainty principle, the virtuality of the line of momentum
$p+k$ just before the gluon emission vertex defines the gluon
``formation time''
\begin{align}
  t_f^{-1}
  \equiv
  {E_\p+E_\k-E_{\p+\k}}
  \empile{\approx}\over{k\ll p}
  \frac{(p+k)^2}{2E_\p}
  =
  E_k(1-\cos\theta)
  \empile{\approx}\over{\theta\ll 1}
  \tfrac{1}{2}E_\k\theta^2.
  \label{eq:tf}
\end{align}
This time can also be interpreted as the time necessary for the
wave-packet of the new gluon to separate sufficiently from that of the
emitter. Indeed, the transverse separation between the emitter and the
gluon grows with time according to
$\Delta r_\perp\approx \theta t$. On the other hand, the wavelength
of the gluon, projected on the plane orthogonal to the emitter, reads
\begin{align}
  \lambda_\perp
  =
  \frac{1}{E_\k\sin\theta}
  \empile{\approx}\over{\theta\ll 1}
  \frac{1}{E_\k \theta}.
\end{align}
We see that the formation time is also given by the condition
$\Delta r_\perp\Big|_{t_f}= \lambda_\perp$.

\begin{figure}[htbp]
  \centering
  \includegraphics[width=8cm]{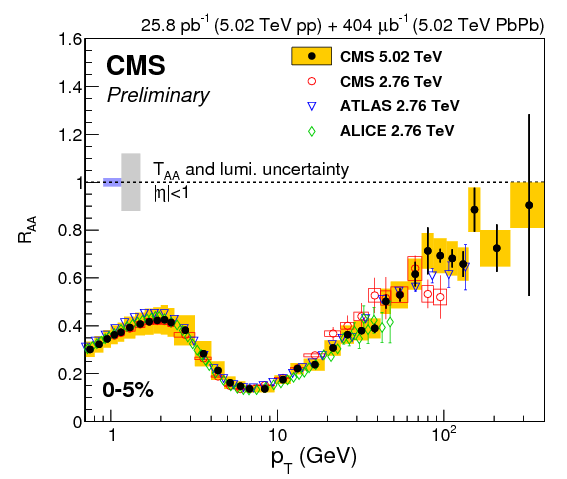}
  \caption{\label{fig:raa}Suppression factor of charged hadrons in nucleus-nucleus collisions. See \cite{Foka:2016zdb,Abelev:2012hxa}.
  }
\end{figure}

\paragraph{Parton energy loss}
The simplest of the observables sensitive to the medium modification
of parton splitting consist in comparing inclusive hadron spectra
measured in nucleus-nucleus collisions with the same spectra measured
in proton-proton collisions at the same energy. However, a direct
comparison of these two spectra is no meaningful because these yields
come from system of vastly different volumes. In order to account for
this trivial geometrical effect, it is customary to define the
so-called ``nuclear modification factor'',
\begin{align}
  R_{AA}\equiv \frac{\frac{dN}{d^3\p}\Big|_{AA}}{N_{\rm coll}\times\frac{dN}{d^3\p}\Big|_{pp}},
\end{align}
where $N_{\rm coll}$ is the number of binary nucleon-nucleon
collisions. For hard objects whose production is obtained as an
incoherent sum of nucleon-nucleon collisions, scaling by this factor
effectively eliminates the trivial volume dependence (in other words,
the ratio $R_{AA}$ should be equal to one in the absence of final
state medium effects). Note that $N_{\rm coll}$ is not a directly
measurable quantity, since its determination requires a modelling of
the nucleon distribution inside the nucleus of interest.  The ratio
$R_{AA}$, called the nuclear modification factor, is displayed in the
case of charged hadrons in the figure \ref{fig:raa}, for central
lead-lead collisions at the LHC.
The most obvious feature of the ratio $R_{AA}$ is that it is below
unity over a large momentum range, and slowly reaches one at large
momenta. This means that the spectra of charged hadrons (at high
momentum compared to the QCD scale $\Lambda_{\rm QCD}$) are depleted
in nucleus-nucleus collisions compared to proton-proton collisions,
which can be explained by an increased fragmentation in the presence
of a surrounding medium. Following
Baier-Dokshitzer-Mueller-Peign\'e-Schiff
\cite{Baier:1996kr,Baier:1996sk,Baier:2001yt} and Zakharov
\cite{Zakharov:1996fv,Zakharov:1997uu,Zakharov:1998sv,Zakharov:2000iz,Baier:2000mf},
this medium-induced radiation can be understood semi-quantitatively as
follows. The formation time of a gluon of energy $E_\k$ and transverse
momentum $k_\perp$ is $t_f\approx E_\k/ k_\perp^2$ (this was derived
in eq.~(\ref{eq:tf})). The main difference in the presence of a medium
is that the transverse momentum of the emitted gluon is altered by the
scatterings that occur within its formation time. Assuming that these
scatterings act as random and independent kicks, the increase of the
transverse momentum is a diffusion process in transverse momentum
space. Therefore, the transverse momentum accumulated during the time
$t_f$ is given by $k_\perp^2\approx \widehat{q}\, t_f$, where
$\widehat{q}$ is a constant that quantifies the mean
$\Delta k_\perp^2$ per unit length (it is a quantity that depends on
the properties of the medium: density/temperature, Debye
screening). Consistency between these two relations implies that
$t_f\approx \sqrt{E_\k/\widehat{q}}$. Based on this, the induced
emissions can be divided into three regimes, depending on the energy of the radiated gluon
\begin{itemize}
\item $E_\k <\omega_{BH}$, with
  $\omega_{BH}\equiv \widehat{q}\lambda^2$ ($\lambda$ is the mean free
  path of the emitter in the medium). In this regime, the radiation
  occurs coherently over path lengths shorter than the mean free
  path. Successive collisions, separated by $\lambda$ (therefore there
  are $L/\lambda$ of them), contribute incoherently to the total
  radiation spectrum. In this regime, known as the Bethe-Heitler
  regime, the emission spectrum behaves as $dI/dE\sim E^{-1}$.
\item $\omega_{BH}<E_\k<\omega_c$, with
  $\omega_c\equiv \widehat{q}L^2$ ($L$ is the length traveled by the
  emitter before exiting the medium). In this energy range, the
  formation time is larger than the mean free path, and shorter than
  the medium size. Several scatterings must happen in order to produce
  one emission, which reduces the total yield compared to the
  Bethe-Heitler regime (this relative suppression is known as the
  Landau-Pomeranchuk-Migdal effect). In this regime, the emission
  spectrum behaves as
  \begin{align}
    \frac{dI}{dE}\approx
    \underbrace{\frac{dI}{dE}\Bigg|_{\rm single}}_{\alpha_s\,E^{-1}}\times\underbrace{ \frac{L}{t_f}}_{\sqrt{\omega_c/E}}\sim E^{-3/2}.
  \end{align}
  (For $E_\k< \omega_{BH}$, the factor $L/t_f$ was replaced by the
  constant $L/\lambda$.)
\item $\omega_c < E_\k$. In this regime, the formation time of the
  gluon is larger than the size of the medium, and induced radiation
  is effectively suppressed. One may thus view $\omega_c$ as an upper
  limit of the energy of a radiated gluon (for instance, for
  $\widehat{q}=2$~GeV/fm${}^2$ and $L=5$~fm, this cutoff is
  $\omega_c=50$~GeV). By combining the emission spectra in the three
  regimes, we can obtain the mean radiated energy by a parton:
  $\big<E_\k\big>\propto \alpha_s\omega_c$. In other words, energy
  loss is most often due to a single emission (occurring with probability $\sim\alpha_s$) close to the upper
  limit. Note that a more dilute (or cooler) medium has a smaller
  $\widehat{q}$ and therefore a smaller cutoff energy $\omega_c$, and
  a smaller mean energy loss.
\end{itemize}
Note that two approaches have been used in the literature for
implementing the scattering-induced radiation of a hard parton. One
option is to perform an opacity expansion, starting from a dilute
medium
\cite{Gyulassy:1999zd,Gyulassy:2000fs,Gyulassy:2000er,Vitev:2002pf,Gyulassy:2003mc}. In
this approach, a very small number of scatterings are considered, but
treated with exact kinematics. Another option is to resum multiple
scatterings, in an approximation where the momentum transfer in these
scatterings is soft compared to the parton momentum, more in line with
the BDMPS-Z approach (see for instance
\cite{Salgado:2003gb,CasalderreySolana:2007zz}).

\begin{figure}[htbp]
  \centering
  \includegraphics[width=7cm]{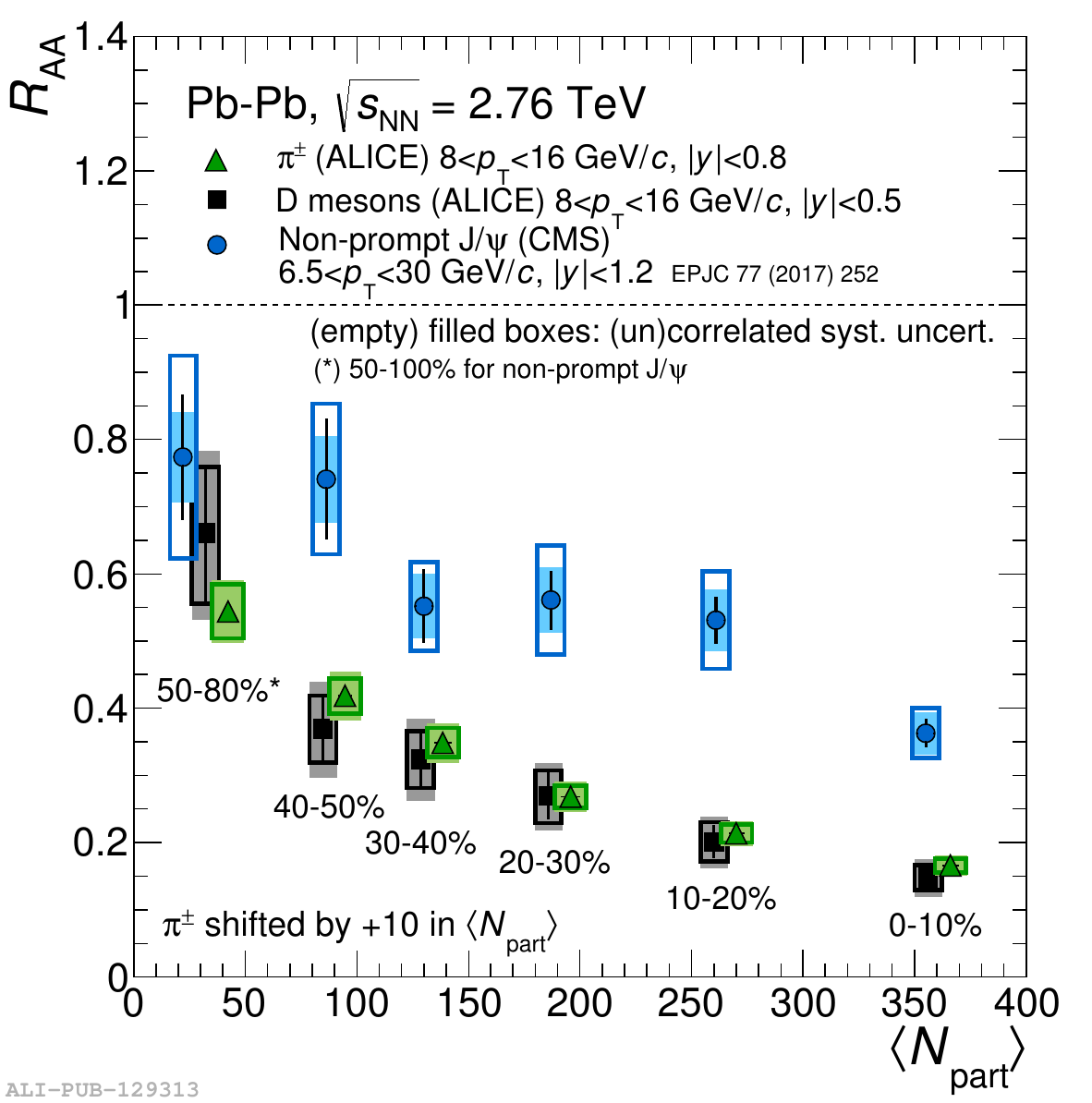}
  \caption{\label{fig:charm-suppression}Open charm and non-prompt
    $J/\Psi$ nuclear suppression factor, compared to that of pions. From \cite{Adam:2015nna}.
  }
\end{figure}

Concrete implementations in the context of heavy ion collisions are of
course more complex than this qualitative discussion
suggests. Firstly, one needs a realistic modeling of the medium and
its evolution, usually taken from hydrodynamics. Another source of
difficulty lies in the fact that the momentum scales characteristic of
the medium (for instance, the Debye screening mass) can be rather soft
at the temperatures reached in heavy ion collisions at present
energies, and in particular not large enough to confidently apply
perturbative QCD at those scales.  This has motivated hybrid
approaches \cite{Casalderrey-Solana:2014bpa} that combine perturbative
techniques for the hard scales and holography-inspired inputs for the
softer medium scales (note that a pure strong coupling approach leads
to an energy loss proportional to $L^3$ instead of $L^2$, and is ruled
out by data).

Let us close this subsection by mentioning that heavy quarks such as
charm are also suppressed in heavy ion collisions, as one can see on
the figure \ref{fig:charm-suppression}.  One can see that $D$ mesons
are as suppressed as the charged pions, suggesting that the
fragmentation of $c$ quarks (at the temperature scales relevant in the
LHC experimental conditions) is comparable to that of light quarks and
gluons. The quark mass starts playing a visible role for $b$ quarks,
since the medium suppression of mesons containing $b$ quarks is
significantly less important. This is consistent with the {\sl dead
  cone effect} \cite{Dokshitzer:2001zm}, a kinematical effect that
prevents gluon radiation inside a cone of opening $m/E$ centered on
the emitting quark of mass $m$. Note that, since the radiative losses
are suppressed by the dead cone effect for heavy quarks, their energy
loss through elastic scatterings (collisional energy loss) become
relatively more important and must be included.

\paragraph{Vacuum antenna pattern} Until this point, we have discussed
the medium modification of the spectra of single hadrons. Another
possibility is to consider similar observables for jets instead of
individual hadrons. One advantage of jets is that jet cross-sections
are in principle perturbative since they do not rely on the details of
the hadronization process. Moreover, jets provide another handle to
probe the loss of energy due to the surrounding medium, since their
opening angle may be chosen at will in their definition.

Before we discuss jet modifications in heavy ion collisions, let us
recall the main feature of gluon radiation from a jet in vacuum.
Compared to radiation by a single parton, the radiation from a jet is
obtained by first considering a parton splitting process $a\to bc$
(where $a,b,c$ could be quarks, antiquarks, gluons, photons). Since
the partons $b,c$ are produced from a common ancestor $a$, their
colors are correlated, and so is the soft radiation they produce.  In
particular, the emission of an additional soft gluon after the
splitting $a\to bc$ depends crucially on whether the new gluon is
emitted inside or outside of the cone formed by the partons $b$ and
$c$. This effect can be understood semi-quantitatively as follows in
the case where the angular opening $\theta_{bc}$ between $b$ and $c$
is small. As we have seen earlier, the formation time of an additional
gluon of energy $E_\k$ at an angle $\theta$ is given by
$t_f^{-1}\approx E_\k \theta^2$. By this time, the pair $bc$ has grown
to a transverse size
\begin{align}
  r_\perp = t_f \theta_{bc}.
\end{align}
Moreover, the gluon wavelength ($E_\k^{-1}$), projected on the plane
orthogonal to the pair momentum, is
$\lambda_\perp \approx (E_\k \theta)^{-1}$.  When the transverse
wavelength of the emitted gluon is smaller than the size of the pair,
the gluon resolves the individual constituents $b$ and $c$ of the pair
and the emission is the sum of the separate emissions from $b$ and
$c$.  Otherwise, the gluon cannot resolve the pair, and sees only its
total charge, which is the charge of the parent $a$.  This condition
reads
\begin{align}
  \frac{1}{E_\k \theta}\lesssim r_\perp\quad
  \Leftrightarrow
  \quad
  \theta\lesssim \theta_{bc}.
\end{align}
In other words, inside the cone, the emission is the incoherent sum of
the emissions from $b$ and $c$, while the emission outside the cone is
equal to that of the total charge, i.e., $a$.  In particular, if the
ancestor is a color neutral object (a virtual photon), then there is no
soft radiation outside the cone of the pair. This property is known as
{\sl angular ordering} in the context of QCD.  It can be used in order
to formulate gluon emission in the form of ``parton cascades'' in
which each quark, antiquark or gluon has a certain probability of
emitting a new gluon. The effect of quantum interferences, that would
in principle preclude such a probabilistic description, is taken into
account by vetoing emissions at large angles.

\begin{figure}[htbp]
  \centering
  \includegraphics[width=8cm]{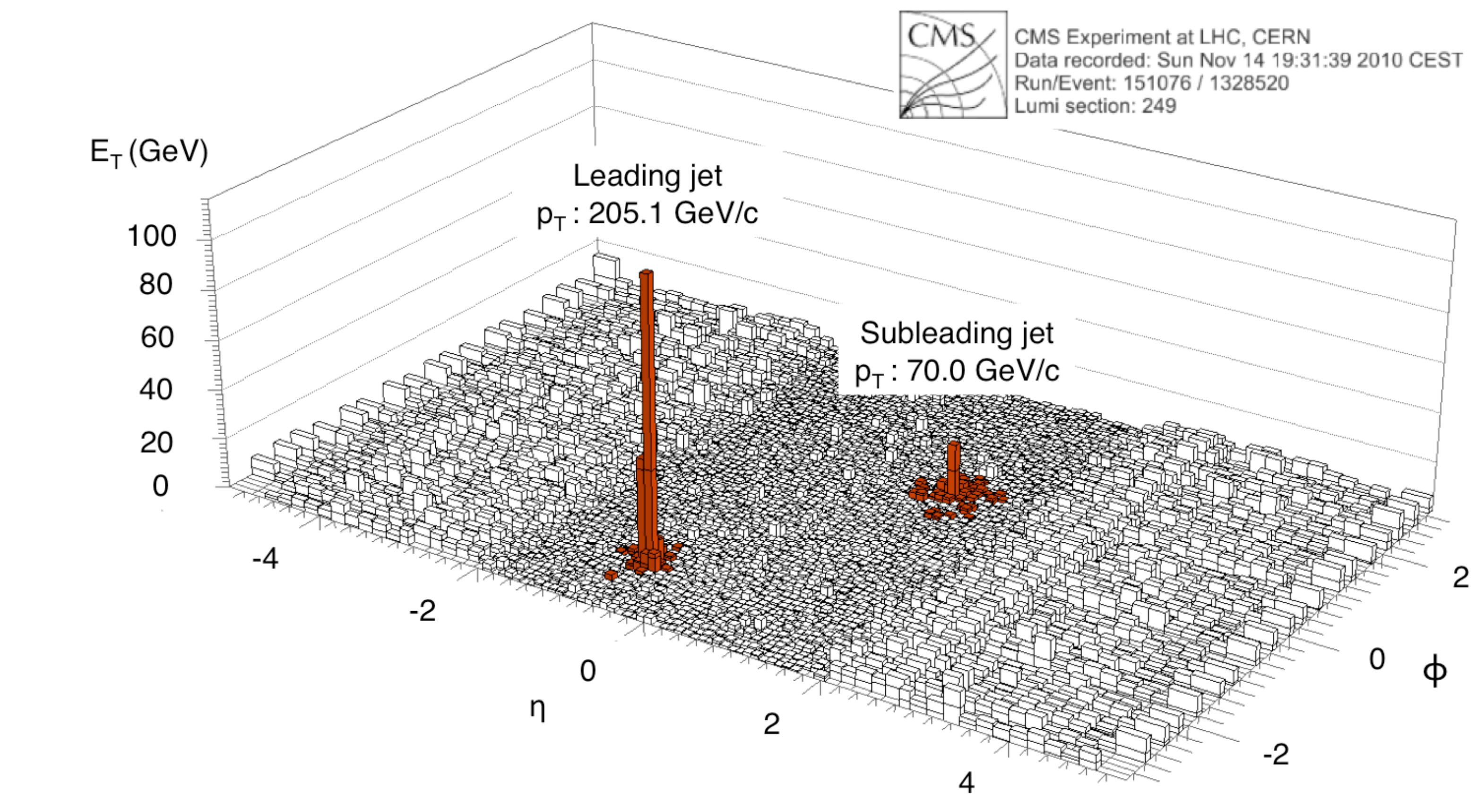}
  \caption{\label{fig:jet-quenching}Event display of a very asymmetrical two-jet event. From \cite{Chatrchyan:2011sx}.}
\end{figure}

\paragraph{In-medium jet energy loss}
A striking observation regarding jets is that
there are events with very imbalanced pairs of jets in heavy ion
collisions, as shown in the figure \ref{fig:jet-quenching}.
In proton-proton collisions, events with two jets in the final states
have jets with nearly balanced transverse momenta, as imposed by
momentum conservation. In the above event display, one jet has almost
three times the energy of the second jet (or, put differently, about
$130$~GeV appear to have been ``lost''), and it is not possible to see
directly on this histogram where the energy has gone (the only
conclusion one may draw by eye is that this energy has been degraded
in the form of soft enough fragments that are lost in the background
of this calorimetric display).

A first medium effect affecting jets is the fact that each parton in
the jet may suffer from medium-induced radiation, in the way discussed
earlier for single particle spectra. The only modification imposed by
the jet definition is that only the radiation that goes outside of the
jet cone must be counted as a loss (thus, this effect is more
pronounced for narrow jets than for wider jets). We have seen in the
previous part that the mean radiative energy loss of a parton is due
to emissions close to the maximum $\omega_c= \widehat{q} L^2$. For
those, the typical gluon emission angle is
$\theta \approx k_\perp/E_\k\approx (L\omega_c)^{-1/2}\ll 1$. The
consequence of this observation is that these rather hard induced
emissions do not alter the energy of a jet for typical jet opening
angles (usually $0.3\le R\le 0.8$), since the radiated gluon stays
within this cone (the jet substructure is altered, but this is not
picked by the jet reconstruction algorithm).  Therefore, the
observed asymmetry between the energy of the pair of jets can be
explained by softer gluons emitted at large angles.

Another important effect affecting a jet as it propagates through a
medium is that the multiple scatterings of the constituents of a pair
of partons eventually lead to the loss of their color coherence
\cite{CasalderreySolana:2012ef,Mehtar-Tani:2013pia,Blaizot:2013vha,Blaizot:2015lma,Qin:2015srf,Milhano:2015mng,Mehtar-Tani:2016aco}. Thanks to this decoherence, the
vacuum-like emissions are no longer forbidden outside of the jet cone.
This happens when the two partons scatter off external color fields
that are uncorrelated, i.e., when the transverse separation $r_\perp$
between them is larger than the coherence length of the color
field. Since $r_\perp$ is proportional to the opening angle of the
pair, the decoherence time is shorter for a pair with a larger opening
angle.  The conclusion of this qualitative argument is that small jets
are more robust against in-medium vacuum-like energy loss than wider
jets.

Moreover, it has been shown that the energy emitted outside of the jet
cone is rapidly degraded into partons whose typical energy is close to
the temperature of the surrounding medium \cite{Blaizot:2012fh,Blaizot:2013hx,Blaizot:2013vha,Caucal:2018dla}. Experimentally, this has
been qualitatively confirmed, since one recovers the missing jet
energy in the form of many softer particles outside of the jet cone.

\paragraph{$\gamma,W^\pm,Z$-jet correlations}
Because of momentum conservation, jets are dominantly produced in
pairs in hard processes, and more rarely (with a suppression of order
$\alpha_s$) in a $3$-jet configuration. After their initial
production, the fate of these jets in the surrounding medium depends
crucially on the location of the production point with respect to the
bulk, and on the direction of motion of the jets. Loosely speaking,
a jet that goes inward looses more energy than a jet that travels
outwards, because it must travel a longer route through the medium. Except
when the production point is very close to the outer boundary of the
medium, both jets loose energy to some degree, making it difficult to
infer event-by-event how much energy was lost by each of them.

A more direct access to the jet energy loss is possible in situations
where a single jet is production in conjunction with another object
that interacts only via electroweak interactions, such as a photon, a
$W^\pm$ or a $Z^0$ boson (for instance, in a process such as
$qg\to q\gamma$). These events are less frequent because of the
electroweak coupling involved at the production vertex, but they offer
the advantage that the weakly interacting object can escape from the
medium without further interactions. Thus, its measurement provides an
unaltered reference for the initial energy of the partner jet.

\section{Thermometric probes}
In this last section, we consider observables that are sensitive to
the local temperature of the medium. In principle, these quantities
could tell if the temperature reached in a heavy ion collision is
above the deconfinement temperature. The actual situation is of course
a bit more complicated, since the temperature of the medium is not
spatially homogeneous, and because the measured quantities result from
the entire history of the system, through which the temperature is
not constant.

\subsection{Electromagnetic radiation from the quark-gluon plasma}
\paragraph{General considerations}
A first quantity which is quite sensitive to the plasma temperature is
the spectrum of photons emitted by the plasma. Let us clarify here a
possible paradox: in a large medium (larger than the photon mean free
path), the electromagnetic radiation would be in thermal equilibrium
with the quarks and the gluons, with a spectrum given by a
Bose-Einstein distribution at the local temperature (but only photons
emitted within one mean free path of the surface would
escape). However, this is not the case in heavy ion collisions, where
the size of the medium is considerably smaller than the mean free path
of the photons. In this case, there is a net production of photons,
and they escape from the medium without further interactions.

Experimentally, the interpretation of photon measurements is quite
challenging, because the detected photons can come from several
sources. It is customary to divide the observed photons into decay
photons (produced from the decay of light hadrons, predominantly
neutral pions) and direct photons (produced directly from partonic
interactions). Direct photons themselves have several sources: some
are produced in hard partonic collisions at the time of the impact of
the two nuclei, some are produced by the pre-equilibrium medium, some
are produced by the interaction of a hard parton and the medium it
traverses \cite{Turbide:2005fk}, some are produced by the QGP and some are produced by the
hot hadron gas after the confinement transition. Among all these
sources, the photons produced by the QGP and by the hot hadron gas are
the most directly sensitive to the properties of the medium produced
in heavy ion collisions, but disentangling them unambiguously from the
overall spectrum is nearly impossible. In the region of low to
intermediate photon energies, the observed spectrum has an exponential
shape in $\exp(-p_\perp/T_{\rm eff})$, suggestive of emissions by a
thermalized medium. However, as we shall see, the parameter
$T_{\rm eff}$ does not have the direct interpretation of the
temperature of the producing medium.
\begin{figure}[htbp]
  \begin{center}
    \includegraphics[width=70mm]{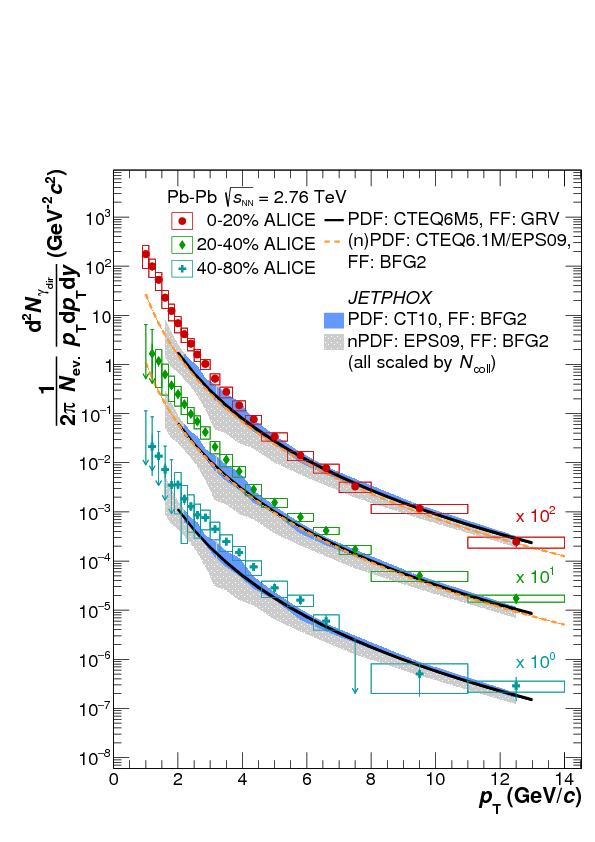}
  \end{center}
  \caption{\label{fig:gam_alice}Direct photon spectrum observed by the
    ALICE experiment, compared to pQCD NLO predictions. Red:
    exponential fit of the excess. From \cite{Adam:2015lda}.
  }
\end{figure}
 
Note also that the produced photons can either be real or virtual. In
the latter case, they can subsequently decay into a lepton pair (this
may be their only decay channel for low invariant masses). Although
the decay into a lepton pair requires another electromagnetic coupling
and thus reduces the yield, the invariant mass of the pair provides
another handle, that may be used to better constrain their possible
source (in particular, a non-zero invariant mass may be used to select
a region where backgrounds are lower).

\paragraph{Thermal radiation from the QGP} The lowest order processes
($qg\to q\gamma$, $q\overline{q}\to g\gamma$
\cite{Baier:1988xv,Altherr:1988bg,Altherr:1992th} and
$q\bar{q}\to \gamma^*$ \cite{McLerran:1984ay}) have been calculated
long ago in an equilibrated quark-gluon plasma. For real photons, the
processes with a quark or antiquark exchanged in the $t$-channel have
a logarithmic singularity, which is cured by resumming the appropriate
quark hard thermal loop \cite{Kapusta:1991qp,Baier:1991em}. However,
it was soon realized that bremsstrahlung processes, formally of higher
order in $\alpha_s$, suffer from a more severe soft singularity when
the photon has a small invariant mass
\cite{Aurenche:1996is,Aurenche:1996sh,Aurenche:1998nw,Aurenche:2002pd}. This
singularity is regularized by the quark in-medium effective mass, but
this leads to an enhancement that promotes bremsstrahlung to the same
order in $\alpha_s$ as the leading processes.

\begin{figure}[htbp]
  \begin{center}
    \includegraphics[width=75mm]{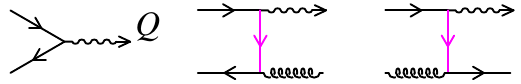}
    \hskip 10mm
    \includegraphics[width=47mm]{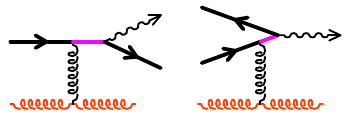}
    \vskip 10mm
  \includegraphics[width=40mm]{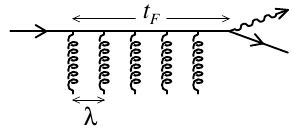}
  \end{center}
  \caption{\label{fig:photons} Top left: Leading Order processes for the
    production of virtual and real photons. Top right: Enhanced
    Next-to-Leading Order bremsstrahlung processes. Bottom: multiple
    scatterings that contribute coherently to the emission of a
    photon.}
\end{figure}

This enhancement also occurs for multiple scattering corrections to
bremsstrahlung \cite{Aurenche:2000gf}. Like in the discussion of
medium induced gluon radiation by a parton, the emission of a photon
by a quark or antiquark is affected by the Landau-Pomeranchuk-Migdal
effect.  The relevant criterion is the comparison between the photon
formation time (i.e., how much time is necessary for the photon
wave-packet to be sufficiently separated from that of the emitter,
which can be estimated to be $t_f^{-1}=E_\k(\k_\perp^2+m^2)/E_\p^2$
where $k$ is the photon momentum, $p$ the quark momentum and $m$ its
in-medium effective mass) and the mean free path $\lambda$ between two
soft (this is sufficient to induce the emission of a photon)
scatterings of a quark in the medium. When $t_f \gtrsim \lambda$,
several scatterings are necessary to induce one emission, which
effectively reduces the photon yield. Note that, in the weak coupling
regime, we have $m\sim gT$ and $\lambda^{-1}\sim g^2 T$ (up to
logarithms). Therefore, the LPM effect plays a role in two cases: for
the production of hard photons emitted at small angle with respect to
the quark, or for the production of soft photons.  The resummation of
these multiple scattering diagrams gives the photon yield at leading
order
\cite{Arnold:2001ba,Arnold:2001ms,Arnold:2002ja,Aurenche:2002wq}. The
next-to-leading order correction has also been calculated more
recently \cite{Ghiglieri:2013gia,Laine:2013vma}.

There have also been attempts to extract the photon production rate of
a quark-gluon plasma in thermal equilibrium from lattice QCD
computations
\cite{Karsch:2001uw,Karsch:2002wv,Ding:2010ga,Ding:2016hua,Ghiglieri:2016tvj,Brandt:2017vgl},
which would in principle be applicable in regimes where the coupling
constant may not be small enough for perturbation theory to be
reliable. However, since this amounts to computing a spectral function
for real energies, there is no direct way to obtain it from an
Euclidean lattice formulation. Instead, one can reach it indirectly by
unfolding (this is an ill-posed problem, that may be attacked with
Bayesian approaches such as the maximal entropy method) the spectral
representation of imaginary time correlation functions.

In the context of an actual heavy ion collision, one has also to face
the fact that the system may not be in local thermal equilibrium. This
is especially true at early times. Firstly, the CGC predicts at the
beginning that there are very few quarks compared to the gluons,
implying that the system is not yet in chemical equilibrium
\cite{Berges:2017eom}. It is possible to handle approximately this
situation by introducing fugacities for the quark and antiquark
distributions, in order to obtain the local photon production rates
for a system where the quarks are underpopulated
\cite{Gelis:2004ep}. Moreover, even when quarks and gluons are present
in the right proportions, the existence of viscous hydrodynamical
corrections implies that their distributions cannot be the equilibrium
ones \cite{Shen:2014nfa,Hauksson:2017udm,Schafer:2019edr}. These deviations should in principle be
taken into account in order to be consistent with the hydrodynamical
framework used to describe the bulk evolution of the system.

\begin{figure}[htbp]
  \begin{center}
    \includegraphics[width=65mm]{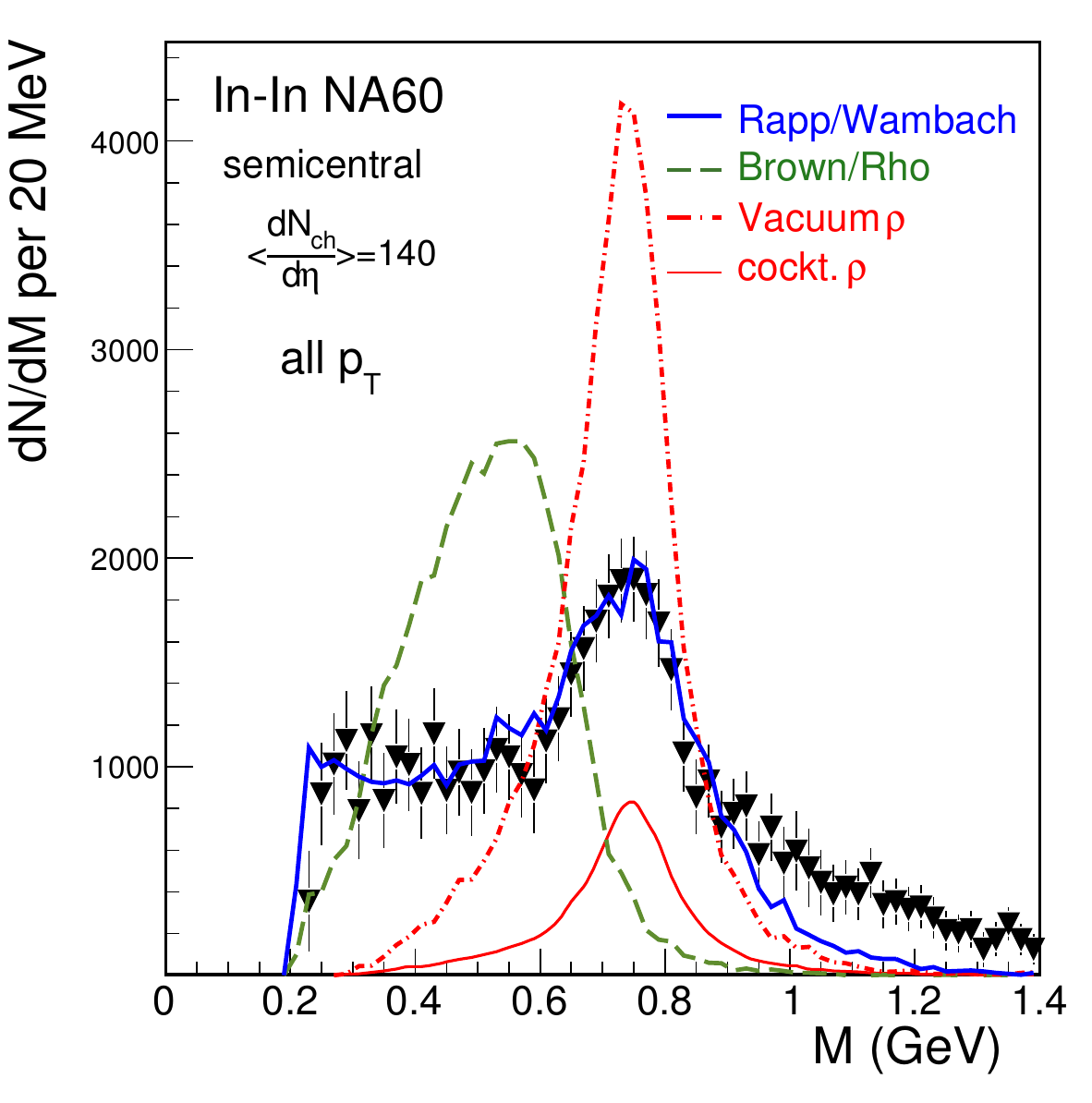}
    \hskip 1mm
    \includegraphics[width=65mm]{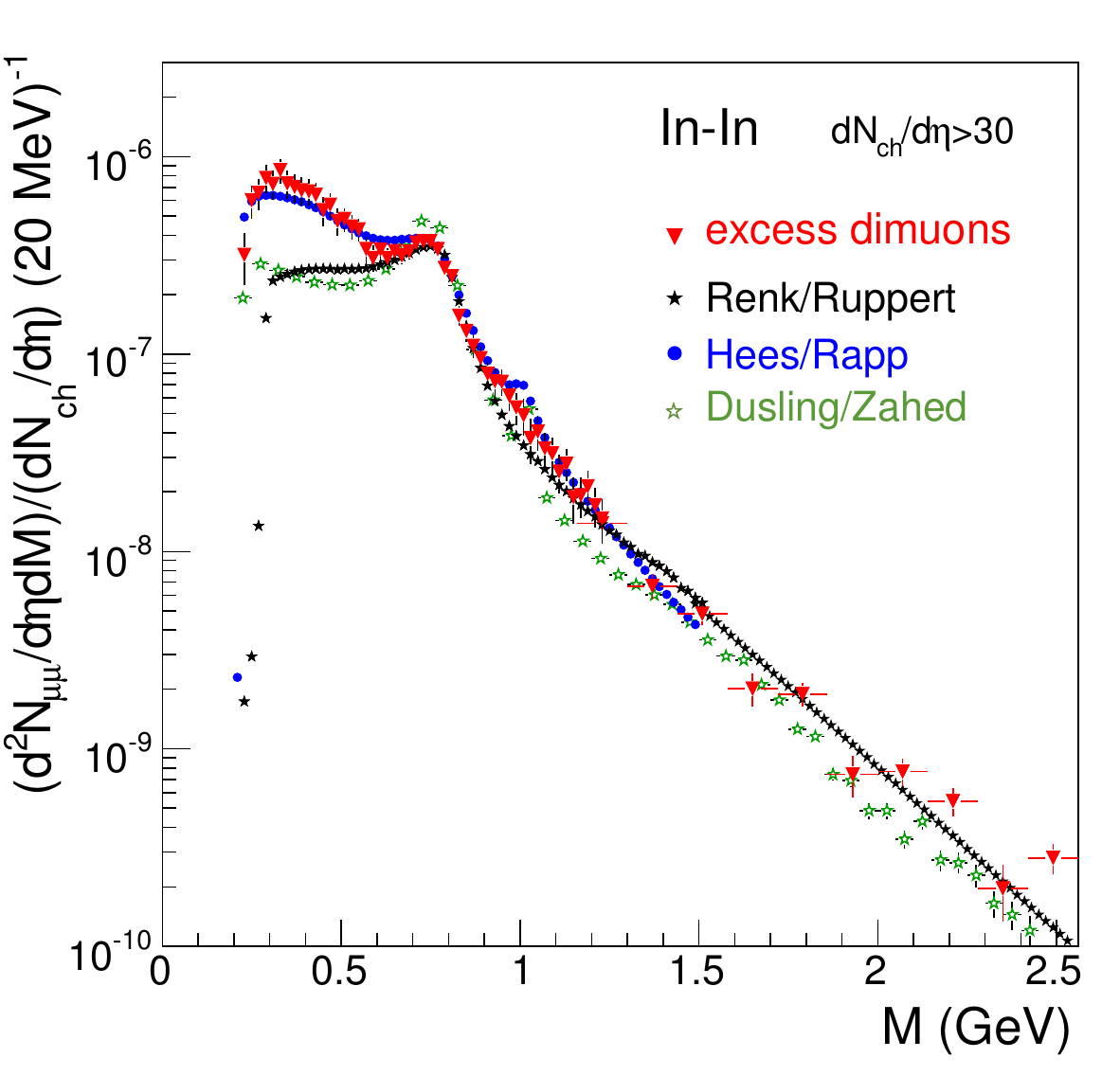}
  \end{center}
  \caption{\label{fig:dileptons}Left: dilepton mass spectrum in the $\rho$ region from the NA60 experiment. Right:  invariant mass spectrum of the excess dimuons, integrated over $p_T$. From \cite{Specht:2010xu}.}
\end{figure}

\paragraph{Thermal radiation from a hot hadron gas} Thermal radiation
from a hot hadron gas is also obtained from the current-current
correlation function, but now it is not accessible to a calculation in
terms of the QCD Lagrangian that has quark and gluon degrees of
freedom. One may instead use an effective Lagrangian that describes
the dynamics and interactions among hadrons at low energy (Such a
Lagrangian in general contains some parameters that are constrained by
empirical data on lifetimes, cross-sections, etc...)
\cite{Gale:1987ki,Rapp:1997fs,Rapp:1999zw,Rapp:2000pe,vanHees:2007th,Rapp:2013nxa,Vujanovic:2013jpa}.
For dileptons of low mass, the spectral function in the vector channel
is dominated by light vector mesons, and is therefore particularly
sensitive to thermal modifications of the $\rho$ meson. The main
thermal effect is a broadening of the $\rho$ peak (while the center of
the peak does not change appreciably), which is even more pronounced
when baryons are taken into account.  Such a broadening seems in quite
good agreement with dilepton measurements in the CERES and NA60
experiments (see Figure \ref{fig:dileptons}).

Let us also mention that, besides effective field theories, a more ab
initio approach has been used recently in order to extract in-medium
spectral functions
\cite{Tripolt:2016cey,Jung:2016yxl,Tripolt:2018jre,Fu:2019hdw,Jung:2019nnr}, based on the
functional renormalization group (FRG) \cite{Berges:2000ew}. The FRG
is a functional equation that tracks the evolution of the quantum
effective action of a theory as one integrates out the quantum
fluctuations in successive layers of momentum, the starting point
being the classical action of the theory (i.e., with no quantum
fluctuations included). In other words, the FRG is an explicit
realization of the renormalization group ``a la Wilson'', where a
theory is coarse-grained to eliminate its details on short distance
scales. In its exact form, there is no practical way to solve the FRG
equation, but it is amenable to tractable equations (at least
numerically) after some kind of truncation has been performed. When
applied to the study of the chiral transition, one observes a
temperature dependence of the chiral condensate in quite good
agreement with lattice QCD computations, and that the $\rho$ and $a_1$
mesons become degenerate at high temperature.
\begin{figure}[htbp]
  \begin{center}
    \includegraphics[width=65mm]{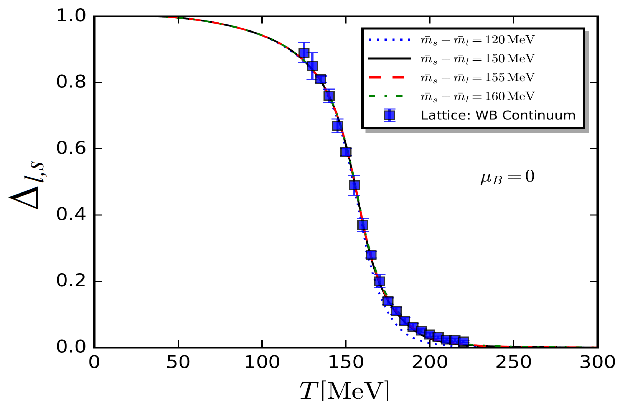}
    \hskip 1mm
    \includegraphics[width=65mm]{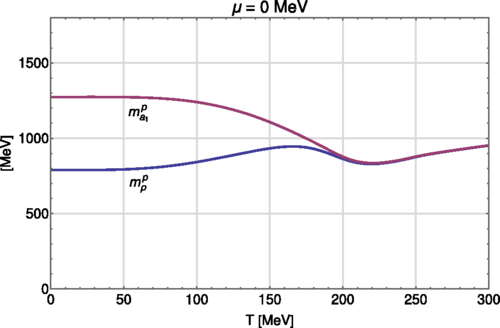}
  \end{center}
  \caption{\label{fig:FRG}Left: FRG computation of the chiral
    condensate as a function of temperature, compared to a lattice QCD
    calculation (From \cite{Fu:2019hdw}). Right: temperature dependence of the masses of the $\rho$
    and $a_1$ mesons, computed in the FRG framework  (From \cite{Jung:2016yxl}).}
\end{figure}

\paragraph{Folding with the medium evolution} The calculations
described above provide a local photon production rate (i.e., the
number of photons produced per unit of time and volume, given the
local temperature of the plasma). To go from there to a photon
spectrum that one may compare with experimental data, this local rate
must be integrated over the entire space-time history of the collision
(in addition, at each space-time point, one must boost the local
spectrum by the $4$-velocity of the plasma at this point). The
important point to keep in mind is that, even if a fit of the
resulting spectrum by an exponential of the form
$\exp(-p_\perp/T_{\rm eff})$ appears to work, the parameter
$T_{\rm eff}$ does not have a direct interpretation as the temperature
of the plasma. In particular, this effective temperature is
blue-shifted by the radial flow of the medium. Moreover, the
integration over time is also more sensitive to the stages of the
evolution where the system spends more time. With dileptons, one may
avoid this blue-shifting effect when looking at the spectrum as a
function of the pair invariant mass since the mass is not affected by
the flow, but the extracted temperature still reflects a spacetime
average rather than an instantaneous temperature. With this caveat in
mind, an exponential mass dependence $\exp(-M/T_{\rm eff})$ of the
dilepton spectrum has been reported by the NA60 experiment, with an
effective temperature $T_{\rm eff}\approx 205\pm 12$~MeV \cite{Specht:2010xu}.

\subsection{Heavy quarkonia in a hot medium}
\paragraph{Qualitative aspects} Bound states made of heavy quarks can
also be viewed as potential ``thermometers'' \cite{Matsui:1986dk}. On
the theory side, the advantage of considering sufficiently heavy
quarks is that they provide a large mass scale (much larger than the
QCD non-perturbative scale $\Lambda_{\rm QCD}$), and this separation
of scales allow the use of effective field theory descriptions such as
non-relativistic QCD. In such a non-relativistic framework, one may
use the concept of interaction potential between a pair of heavy
quarks, in conjunction with a non-relativistic Schr\"odinger
equation. Another theoretical simplification regarding heavy quarks is
that their production happens at very early times (of the order of the
inverse of their mass), and involve parton distributions at reasonably
large momentum fractions, where saturation effects are not important
(this assertion should be contrasted in the case of charm quarks at
the LHC energy -- see \cite{Ma:2015sia,Ma:2018qvc} for a recent study
of $J/\Psi$ production in proton-nucleus collisions; a similar
computation in nucleus-nucleus collisions could be done by solving the
Dirac equation in the glasma color fields, but is considerably more
challenging \cite{Gelis:2005pb,Gelis:2015eua,Tanji:2017xiw}).  Thus, in heavy ion
collisions, one is mostly interested in the subsequent fate of the
produced heavy quarks, rather than the production itself. As far as
experimental measurements are concerned, heavy quarkonia also offer clean
signals via their dilepton decay channel.

Consider for instance a heavy meson $Q\overline{Q}$. Loosely speaking,
when the thermal excitation energy (i.e., the energy gain provided by
the absorption of a gluon from the surrounding thermal bath) is equal
or larger to the binding energy of the quark-antiquark pair, the meson
may be dissociated. An alternate way of describing this phenomenon is by
noting that the interaction potential of the $Q\overline{Q}$ is
affected by Debye screening in the presence of a dense medium. When
the Debye screening length becomes shorter than the size of the
would-be bound state, its dissociation occurs.  Afterwards, the quark
and the antiquark evolve independently in the medium, and the most
likely outcome is that --when the temperature has decreased below the
confinement temperature-- they bind with one of the light quarks or
antiquarks from the surrounding medium, in order to form heavy-light
mesons ($D$ and $B$ mesons). Therefore, in the extreme version of this scenario, the yield
of $Q\overline{Q}$ mesons would be almost completely suppressed and
the produced heavy quarks would all be recovered in the form of
open-flavour mesons. Note that in the case of charm quarks, whose
production is quite abundant in heavy ion collisions at high energy,
the dissociated $c$ and $\overline{c}$ may have a high enough density
for accidental recombinations in $J/\Psi$'s to be important.

\begin{figure}[htbp]
  \centering
  \includegraphics[width=8cm]{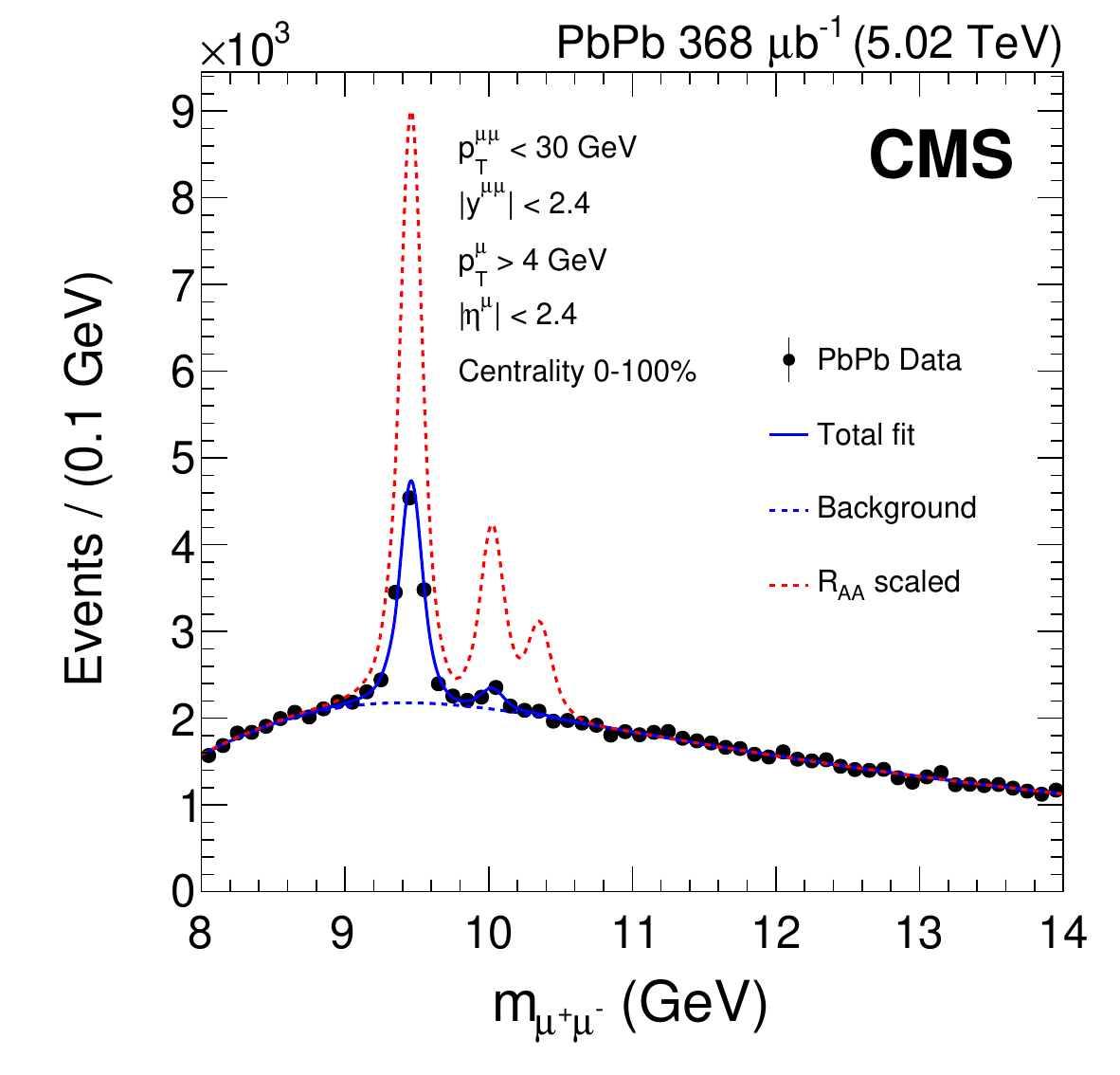}
  \caption{\label{fig:upsilon}Muon-antimuon invariant mass spectrum in the upsilon region. From \cite{Sirunyan:2018nsz}. See also \cite{Chatrchyan:2012lxa,Khachatryan:2016xxp}.}
\end{figure}

Experimentally, a particularly clear observation of this phenomenon
has been performed at the LHC for $b\overline{b}$ mesons, as shown in
the figure \ref{fig:upsilon}.
There, one sees a suppression of the yields of the $\Upsilon''$ and
$\Upsilon'$ mesons, while the ground state $\Upsilon$ has an almost
unmodified yield. Given the above discussion, the interpretation
of this result is that the gluon thermal energy at the temperatures
reached in the collision is sufficient to dissociate the excited
states (because they are more weakly bound) but not enough to alter
the more tightly bound fundamental state.

\paragraph{In-medium singlet heavy quark bound states}
In the vacuum (i.e., at zero temperature), the spectrum of color
singlet $Q\overline{Q}$ states is typically made of a few (meta)stable
states --defined as poles of the propagator in the energy complex
plane-- occupying the low energy region, and a continuum at higher
energies. In the presence of a high temperature medium, several
effects may occur:
\begin{itemize}
\item Some bound states (starting from the high lying ones) may disappear,
\item The lower bound of the continuum of free states may move to a 
  lower energy,
\item The surrounding thermal medium can induce transitions between
  various $Q\overline{Q}$ states, including transitions between
  singlet and octet states.
\end{itemize}
The ab initio approach for studying the in-medium modifications of
heavy quark states consists in calculating the spectral function in
the appropriate channel at non-zero temperature. However, this is not
doable in perturbation theory due to the non-perturbative nature of
bound states. Although lattice QCD is by construction a
non-perturbative approach, spectral functions cannot be computed
directly. Instead, one can calculate an Euclidean propagator
$G(\tau,\p)$ which is related to the corresponding spectral function
$\rho(\omega,\p)$ as follows:
\begin{align}
  G(\tau,\p)
  =
  \int
  d\omega\; \rho(\omega,\p)\, \frac{\cosh(\omega(\tau-1/2T))}{\sinh(\omega/2T)}
  .
\end{align}
A first obvious difficulty for inverting this relationship is that the
propagator is computed only at the finite set of Euclidean times
$\tau$ that exist in the employed lattice setup, while the expected
spectral function depends on a continuous energy $\omega$. But even if
the propagator was known at all the real $\tau$'s in the range
$[0,1/T]$, this inversion is a mathematically ill-posed problem,
because the linear mapping from $\rho$ to $G$ has zero modes (i.e.,
functions $\rho$ that give zero when inserted in the integral in the
previous equation). Thus, even in the ideal situation where the
propagator would be known exactly, the inversion can only be performed
up to a linear combination of these zero modes.

A possible strategy is to remove this ambiguity by imposing additional
(but generic enough so that they do not bias the outcome in unphysical
directions) constraints on the expected spectral function. A minimal
constraint that helps disambiguate the answer is to request the
positivity of the spectral function.  In practice, this can be
implemented by using the {\sl maximal entropy method}, which is a
Bayesian method for finding the most likely spectral function
consistent with the computed values of the propagator and the
additional constraints.  When using this approach, it is necessary to
have very accurate lattice data for a robust extraction of the
spectral function (otherwise, the extracted spectral function may be
dominated by the additional constraints imposed on the
solution). Another limitation is that it is practically impossible to
be sensitive to excited states, as this would require an exponentially
large statistics. With these caveats in mind, the general trend
observed for heavy bound states is a sequential melting of states,
starting with the high lying ones, and a trend towards negative
medium-induced mass shifts \cite{Burnier:2013nla,Aarts:2014cda,Ikeda:2016czj,Kim:2018yhk}.

An alternative to Bayesian methods is to model the spectral functions
$\rho(\omega,\p)$ with a few free parameters and to perform a standard
fit to adjust these parameters in order to reproduce the computed
propagator $G(\tau,\p)$.  Besides the location and width of the lowest
lying peak, the model may contain parameters that describe the
transport properties of heavy quarks, or excited states. However, one
should keep in mind that the propagator may be very weakly sensitive
to these additional features of the spectral function
\cite{Aarts:2002cc}, and that an unrealistic modeling may introduce a
strong model dependence on the outcome. A model of the spectral
function may be obtained from effective field theory descriptions,
such as non-relativistic QCD (NRQCD -- obtained from QCD by
integrating out the heavy quark mass scale $m_{_Q}$) or even potential
non-relativistic QCD (pNRQCD -- obtained from NRQCD by further
integrating out the softer scale $m_{_Q}v$, where $v$ is the heavy
quark velocity)
\cite{Brambilla:1999xf,Brambilla:2004jw,Brambilla:2008cx}, that rely
on the heavy mass of the quarks, both compared to the QCD
non-perturbative scale $\Lambda_{\rm QCD}$ and to the typical energy
scale of the surrounding medium.

In fact, the pNRQCD effective theory also provides a proper connection
between QCD and the non-relativistic Schr\"odinder equation used in
order to study the $Q\overline{Q}$ bound states (see for instance
\cite{Laine:2006ns,Beraudo:2007ky}). Indeed, one of the parameters in
pNRQCD is the $Q\overline{Q}$ interaction potential (that should in
principle be obtained by a matching to the underlying field theory),
and its equation of motion in the singlet sector is a Schr\"odinder
equation whose discrete energy levels correspond to the
$Q\overline{Q}$ singlet bound states.
\begin{figure}[htbp]
  \begin{center}
    \includegraphics[width=65mm]{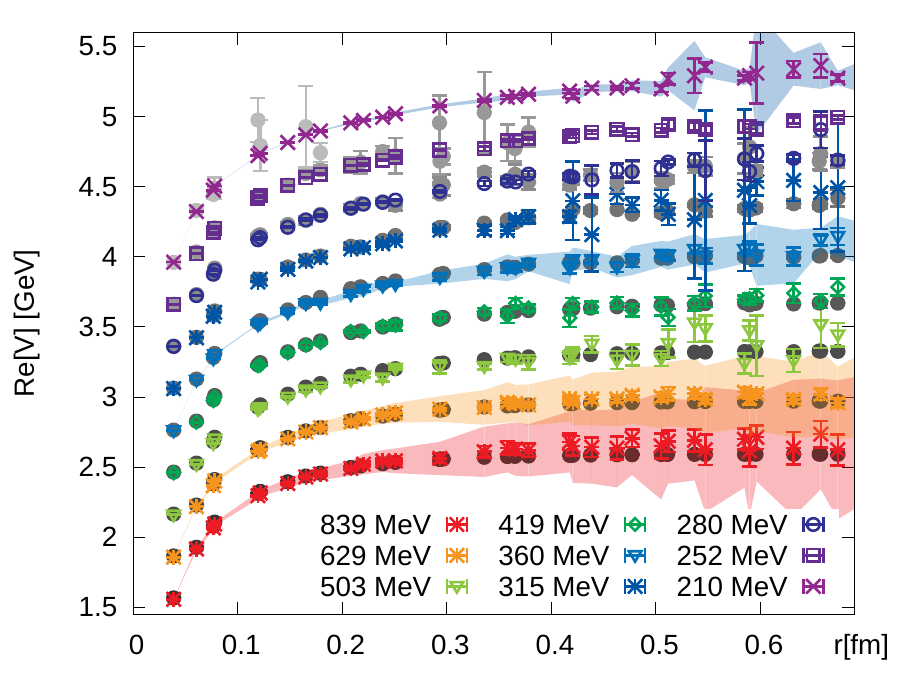}
    \includegraphics[width=65mm]{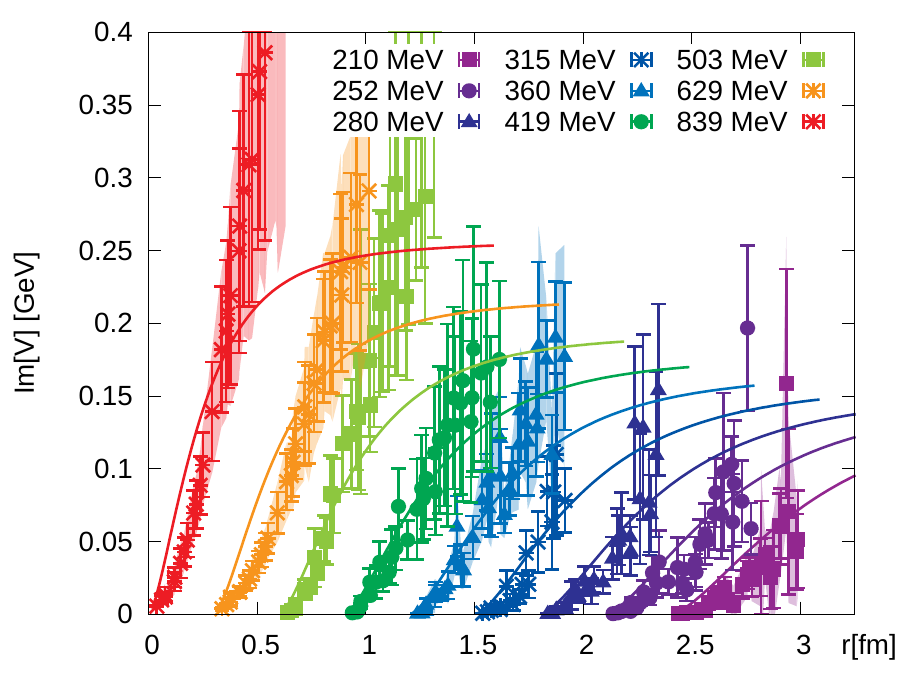}
  \end{center}
  \caption{\label{fig:Qpotential}Real and imaginary parts of the
    singlet $Q\overline{Q}$ potential at various temperatures. From
    \cite{Burnier:2014ssa}. See also \cite{Lafferty:2019jpr}.}
\end{figure}
The singlet potential $V_s(r)$ can be obtained as 
\begin{align}
  V_s(r)=\lim_{t\to \infty}\frac{\partial_t W(r,t)}{W(r,t)},
\end{align}
where $W(r,t)$ is a rectangular Wilson loop of spatial extent $r$ and
temporal extent $t$. Note however that $t$ is here the Minkowski time,
and a direct evaluation of this quantity in lattice QCD is therefore
not possible. A possible strategy is to start from an Euclidean
rectangular Wilson loop, whose associated spectral function can be
used to express $V_s(r)$, thus allowing to constrain the potential via
Bayesian methods as shown in Figure \ref{fig:Qpotential}. Note that
the real part of this potential behaves similarly but is not identical to the
potential sometimes inferred from the logarithm of the free energy of a
singlet $Q\overline{Q}$ pair \cite{Borsanyi:2015yka}. (The imaginary part, due to Landau
damping and transitions from singlet to octet states, is not present
in the potential defined from the free energy.)

\paragraph{In-medium dynamical evolution}
In the presence of a high temperature medium, not only the spectrum of
singlet quarkonia states is modified, but the surrounding medium can
also induce transitions between various singlet states, and between
singlet and octet states. In equilibrium, there are as many
transitions in either direction, and the density matrix of the system
is time independent.  The situation is far more complicated
out-of-equilibrium, since the density matrix is now time
dependent. Various forms of kinetic or stochastic equations have been
employed to describe the evolution of heavy quarks and quarkonia
embedded in a medium \cite{Moore:2004tg,Ke:2018tsh,Yao:2018zrg}.

The main physical effect one would like to study here is the possible
recombination of the heavy quarks and antiquarks, that may have a
sizeable probability of close encounter when their density is large
enough (which appears to be the case for charm quarks). When this is
the case, the final yield of quarkonia bound states is enhanced
compared to what one would get with the assumption that all the bound
states that are dissociated go into open-heavy flavor mesons
\cite{Thews:2000rj,Blaizot:2015hya,Du:2017qkv,Yao:2017fuc,Yao:2018sgn}.
\begin{figure}[htbp]
  \begin{center}
    \includegraphics[width=80mm]{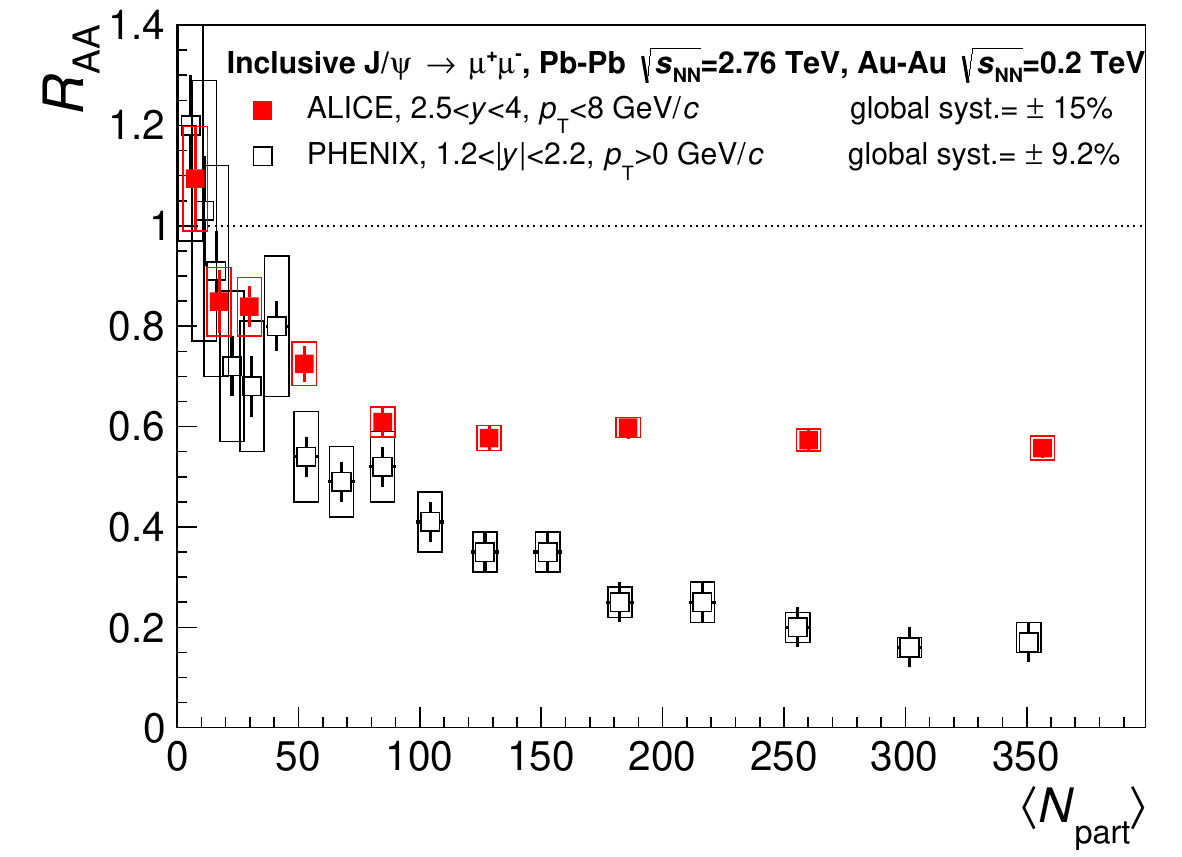}
  \end{center}
  \caption{\label{fig:reco}Nuclear modification factor of $J/\psi$ as
    a function of the number of participants, for RHIC and LHC
    energies. From \cite{Adam:2015isa}.}
\end{figure}
This phenomenon can be seen for charm quarks by comparing the nuclear
modification factor of $J/\psi$ at RHIC and LHC energies, as shown in
Figure \ref{fig:reco}. On this plot, one sees that $J/\psi$'s are less
suppressed in central collisions at higher energy, which can be
explained by the more abundant production of $c$ and $\overline{c}$
quarks at the LHC, which favors their recombination into charmonium.

More recently, there has been an effort to derive these description
from a more fundamental starting point, shedding some light on the
sequence of approximations necessary to obtain them. The strategy,
based on the framework of {\sl open quantum systems}
\cite{Akamatsu:2014qsa,Kajimoto:2017rel,Yao:2018nmy,Brambilla:2017zei,Brambilla:2016wgg,Blaizot:2018oev},
consists in starting from the system made of the $Q\overline{Q}$ pairs
and their environment.  This system is closed and evolves unitarily,
with a density matrix $\rho$ that obeys the von Neumann equation (also
known as the quantum Liouville equation),
\begin{align}
  \frac{d\rho}{dt}=-i\,[H,\rho].
\end{align}
The density matrix $\rho$ contains a lot of information regarding the
surrounding medium in which the heavy quarks are embedded. A {\sl reduced
density matrix} describing specifically the heavy quark degrees of
freedom can be obtained by performing a partial trace over environment
degrees of freedom, $\rho_{_Q}\equiv {\rm tr}_{\rm
  env}(\rho)$. However, after doing this, the evolution equation for
the reduced matrix density is no longer a von Neumann equation, and is
in general dissipative and non unitary (this is just a consequence of
the fact that the subsystem made of the heavy quarks is an open
system). With the assumption that the environment relaxes more quickly
than the heavy quarks, it has been shown that $\rho_{_Q}$ obeys a
Lindblad equation. Under the additional assumption that the transitions
between the various $Q\overline{Q}$ states are faster than their
adjustment to changes of the environment, this Lindblad equation can
lead to a Boltzmann equation (and a rate equation is the momenta are
integrated out).

\section{Conclusions, open questions for the future}
What should first come to mind after this brief survey of the
theoretical aspects of heavy ion collisions is the difficulty of
describing a large, time-dependent, non-equilibrated system in terms
of a (rather complicated) underlying microscopic theory (QCD). Since
an ab-initio description in terms of QCD of these collisions is not
practically feasible, most approaches are based on effective
descriptions that capture the relevant dynamics at scales larger than
the typical QCD scales. Thus, instead of a unique theory from which
everything would be derived, theoretical works in this field use
instead a large variety of tools that are more or less connected to QCD:
\vglue 10mm
\begin{center}
\xymatrix{
    *+[F-]\txt{{\colorb\bf Lattice QCD}} 
    &
    *+[F-]\txt{{\colorc\bf Perturbative QCD}}
    &
    *+[F-]\txt{{\colord\bf CGC}}
    \\
    *+[F-]\txt{{\colore\bf Hydrodynamics}}
    &
    *+[F-,]{\txt{{\Large\bf QCD}\\{\ }\\${\cal L}=-\frac{1}{4}{\colorb F^2} +{\colora\overline\psi}(i{\colorb\slD}-{\colora m}){\colora\psi}$}}
    \ar[dl]\ar[ul]\ar[dr]\ar[ur]\ar[r]\ar[u]\ar[l]\ar[d]
    &
    *+[F-]\txt{{\colord\bf NRQCD}}
    \\ 
     *+[F-]\txt{{\colore\bf Kin. Theory}} 
    & 
    *+[F-]\txt{{\colora\bf AdS/CFT}}
    &
    *+[F-]\txt{{\bf\colord $\chi$PT}}
}
\end{center}
\vskip 1mm Although we have not spent much time discussing this, we
should also stress the fact that many observables in heavy ion
collisions depend on a number of mundane aspects of low energy nuclear
physics, namely the shape and size of the nuclei, and the distribution
of the nucleons inside a nucleus and the fluctuations thereof. These
properties, that are not the main targets of the heavy-ion collision
program, play nevertheless an important role when trying to uncover
some property of QCD from experimental data.

By a combination of experimental and theoretical efforts, many
properties of the quark-gluon plasma have been uncovered:
\begin{itemize}
\item the QGP is a nearly perfect fluid,
\item its shear viscosity to entropy ratio is in the range
  $[1,2.5]$ (in units of $\hbar/4\pi$), making it the substance with
  the smallest ratio so far,
\item its equation of state is consistent with lattice QCD
  expectations, and with the deconfinement of the color degrees of
  freedom,
\item the yield of ``light'' partons, including charm quarks, is
  significantly suppressed compared to rescaled proton-proton
  collisions,
\item the suppression of bottom quarks is less pronounced, in
  agreement with theoretical expectations (dead-cone effect due to the
  mass of the emitter),
\item the studies of energy loss can now be supplemented by direct
  observations of reconstructed jets. This has allowed to determine
  that a large amount of energy is radiated by soft emissions at large
  angle,
\item a sequential pattern has been observed in the disappearance of
  $b\overline{b}$ bound states, consistent with the theoretical
  understanding of the dissociation phenomenon,
\item at the highest energies, the production of charm quarks is
  copious enough to lead to the formation of $J/\psi$ bound states by
  recombination of uncorrelated quarks and antiquarks.
\end{itemize}
The picture that emerges from these observations is that the matter
produced in heavy ion collisions is a very ``opaque'', strongly
interacting, fluid, in rather sharp contrast with the ethereal
quark-gluon {\sl plasma} that was the common point of view before the
RHIC experiment. Despite many progresses, it is also clear that
extracting the underlying QCD properties from the outcome of heavy ion
collisions is extremely difficult, since in several instances the
comparisons with QCD have remained rather qualitative although the
experimental measurements were quite detailed. Another source of
complication is that some of these studies are done by comparing the
outcome of nucleus-nucleus collisions with that of rescaled
proton-nucleus or proton-proton collisions.  However, high energy
proton-nucleus and proton-proton collisions have turned out to display
some features close to those observed in nucleus-nucleus collisions,
casting some doubts on their use as ``references'' to compare with in
order to pinpoint effects specific to the quark-gluon plasma. If flow
is confirmed to occur even in proton-proton collisions, one will have
to learn to live without such a reference for certain observables.

It is of course hard to predict where the next advances will happen,
but given the areas that have received most attention in the past
years, some improvements are probably within reach in a reasonable
future in the following directions:
\begin{itemize}
\item Determine the temperature dependence of the shear viscosity,
\item Obtain a better determination of the bulk viscosity,
\item Better disentangle the mechanisms of energy loss, especially in the case of jets,
\item Characterize {\sl when} heavy quark bound states are formed,
\item estimate the initial temperature from thermal photons and the melting of quarkonia,
\item Clarify to what extent the concept of flow applies to the system
  formed in proton-proton collisions. This entails a more robust
  control over the state of the system immediately after the
  collision, in order to disentangle initial flow from the flow
  hydrodynamically generated later on.
\end{itemize}
For this to be possible, besides some improvements to the calculation
of the elementary relevant phenomena, an important aspect (and
difficulty) is to merge as seamlessly as possible tools that have
originally been developed independently. This is especially true for
observables that depend on the interactions between some probe and the
surrounding medium, for which it is crucial to use a modeling of the
background and its evolution which is as realistic as possible.

\paragraph{Acknowledgements:} The author is supported by the Agence
Nationale de la Recherche through the project 11-BS04-015-01.


\end{document}